\documentclass[10pt,times,letter]{article}

\usepackage{bm}

\usepackage{amsmath, amsfonts,amsthm,amssymb,graphicx}
\usepackage{dsfont}

\usepackage{color}

\newcommand{\Px}{ \mathbb{P} }
\newcommand{\Qx}{ \mathbb{Q} }

\newcommand{\Ex}{ \mathbb{E} }

\newcommand{\A}{{\cal A}}

\newcommand{\D}{\mathrm{d}}
\newcommand{\too}{-\!\!\!\to}

\newcommand{\G}{\mathcal{G}}
\newcommand{\R}{\mathbb{R}}

\usepackage{natbib}

\definecolor{Red}{rgb}{1.00, 0.00, 0.00}

\definecolor{DRed}{rgb}{0.5, 0.00, 0.00}

\definecolor{Blue}{rgb}{0.00, 0.00, 1.00}

\definecolor{Green}{rgb}{0.0, 0.4, 0.0}

\usepackage[colorlinks=true,breaklinks=true,bookmarks=true,urlcolor=blue,
     citecolor=blue,linkcolor=blue,bookmarksopen=false,draft=false]{hyperref}

\newtheorem{theorem}{Theorem}[section]
\newtheorem{definition}{Definition}[section]

\newtheorem{proposition}[theorem]{Proposition}
\newtheorem{remark}[theorem]{Remark}
\newtheorem{lemma}[theorem]{Lemma}

\begin{document}




\title{Robust Optimization of Credit Portfolios}

\author{
Lijun Bo \thanks{Email: lijunbo@ustc.edu.cn, School of Mathematical Sciences, University of Science and Technology of China, Hefei, Anhui
Province, 230026, China.} \and
Agostino Capponi \thanks{E-mail: ac3827@columbia.edu, Department of Industrial Engineering and Operations Research, Columbia University, New York, 10027, NY, USA. }
}


\maketitle

\begin{abstract}%
We introduce a dynamic credit portfolio framework where optimal investment strategies are robust against misspecifications of the reference credit model. The risk-averse investor
models his fear of credit risk misspecification by considering a set of plausible alternatives whose expected log likelihood ratios are penalized.
We provide an explicit characterization of the optimal robust bond investment strategy, in terms of default state dependent value functions associated with the max-min robust
optimization criterion. The value functions can be obtained as the solutions of a recursive system of HJB equations. We show that each HJB equation is equivalent to a suitably truncated
equation admitting a unique bounded regular solution. The truncation technique relies on estimates for the solution of the master HJB equation that we establish. 
\end{abstract}%




%

\section{Introduction.}\label{intro} 
Portfolio optimization problems rely on models of asset price dynamics whose probabilistic behavior is imprecisely known.
Although a great deal of effort is devoted to model calibration, the limited number of available observations as
well as the perturbing noise often result in parameter estimates subject to estimation errors. As a result, the investor always fears model misspecifications of the transition law
governing the joint dynamic evolution of default probabilities. Since he is unable to detect the true underlying model, he wants to design decision rules which are robust to model errors, i.e. take into consideration parameter uncertainty into his optimization procedure. His objective is to construct portfolio strategies which, besides working well when the model describing the price dynamics is correctly specified, also perform reasonably well in the case when the model is misspecified.

Depending on the approach used to perturb the actual underlying model, there can be different robust control formulations.
The Bayesian approach, pioneered by \cite{Gilboa}, models ambiguity aversion through the formulation of multiple priors preferences. This approach
has been extended to a dynamic setting by \cite{Epstein}, where priors are updated over time. The other approach, pioneered by \cite{AndersonSarg} (see also \cite{Anderson} for a related study), formulates the robust decision making problem using a penalty function for model misspecifications. \cite{Mahenhout} builds on this framework by considering a diffusion model with uncertainty in the equity risk premium. \cite{MahenhoutJT} further extends the framework in \cite{Mahenhout} by considering stochastic investment opportunities. \cite{Liu} extends the analysis in \cite{Mahenhout} to a jump-diffusion model, where the investor knows the diffusion component but is uncertain about the size and frequency of jumps. \cite{JinZhang} extend the work of \cite{Liu} to the case of multiple assets. We also refer to \cite{HansenNoa} for a survey of various mathematical formulations to achieve robustness. 

The objective of our paper is to study the impact of credit risk model misspecification on optimal investment strategies. Previous studies on optimal credit portfolios assume the underlying model governing default intensities and contagion risk to be known. This is the case in the early work of \cite{BielJang}, in the first passage time framework of \cite{KS}, as well as in the default contagion model by \cite{KSAll}. The credit model is also assumed to be known in the work of \cite{CapFig} and \cite{CappLopezhidd}, where Markov-modulated dynamics drive the behavior of default intensities and security prices. \cite{BoCappMF} construct a credit default swaps portfolio framework, but assume that changes in default intensities of obligors in reaction to default events are known.

Empirical evidence, however, suggests that actual default intensities and default correlations are difficult to estimate. This is because defaults are rare events, and most of the firms whose securities are traded in the market have never defaulted or rarely experienced severe financial distress.\footnote{For instance, the typical cohort approach used by Moody{'}s and Standard and Poor{'}s is well known to underestimate default risk, and has led to considering alternative approaches such as the continuous-time analysis of rating transitions proposed by \cite{Landodrift}.}

Our study is the first to develop a dynamic credit portfolio optimization framework which accounts for robust decision rules against misspecifications of the model for the actual
default intensity process. We consider a portfolio consisting of defaultable, coupon paying bonds. As in \cite{BoCappMF}, we model default contagion via an interacting intensity model. Different from their work, the investor now protects himself against ambiguity aspects of the reference credit model. He considers it to be the best statistical characterization of the data, but models his fear of credit model misspecification by considering a set of plausible alternatives whose expected log likelihood ratios (i.e., whose relative entropies) are penalized. As in \cite{AndersonSarg} (see also \cite{HansenNoa}), we restrict attention to perturbations that are absolutely continuous over finite intervals, as these are statistically difficult to distinguish from the reference model. 
A related study to ours is \cite{JaiSig}, who consider a hybrid credit model where default is modeled as the first jump of a Poisson process after the credit worthiness index
of a company has a crossed a certain barrier. Under this default model and accounting for model uncertainty, they study robust indifference pricing of defaultable bonds and CDSs.
As in our study, they penalize deviations from the reference measure using an entropic penalty function. Their robust formulation allows to explain the main drawback of structural models,
namely the underestimation of short-term credit spreads.

We next list our main technical contributions. We develop an explicit characterization of the optimal robust bond investment strategy. This is obtained by recovering an analytical expression for the vector of optimal feedback functions, given as the product of two terms, the inverse of a matrix measuring the
bond depreciations at the default events, and a vector associated with the worst-case probability measure. Due to the presence of default contagion, the value function associated with the max-min robust optimization criterion
depends on the default state. More specifically, we show that it corresponds to the solution of a recursive system of nonlinear HJB equations.
We remark that the recursive decomposition of a global optimal investment problem has also been considered by \cite{JiaoKhaPham}. Their approach consists in first defining the sub-control problems in the reference market filtration exclusive of default event information, and then connecting them by assuming the existence of a conditional density on the default times.
Despite this similarity, there exist significant differences between ours and their approach. We consider the wealth dynamics under the enlarged market filtration inclusive of default events and do not perform any decomposition of the control problem at the level of the stochastic differential equation.
It is only after deriving the HJB equations that the recursive dependence between ODEs associated with different default states naturally arises. Their approach instead exploits the exponential utility preference function of the investor and reduce the optimal investment problem to solving a recursive system of backward stochastic differential equations with respect to the default-free market filtration.

In our control problem, both the worst-case measure and the optimal feedback functions are coupled with the HJB equations. By exploiting the property of a carefully identified smooth and
increasing transform, we are able to prove existence and uniqueness of a global classic solution to each equation. This is achieved by showing the equivalence of each HJB equation to a
truncated equation admitting a unique bounded regular solution. The truncation is defined in terms of estimates established for the solution of the
original HJB equation. {The study of smooth solutions to HJB equations with unbounded control space has also been considered by \cite{FeGaGo} and \cite{Pham}. Therein, they
consider one default-free stock and an infinite-horizon framework in order to study the regularization of the viscosity solution of the HJB equation, and its correspondence with the value
function of the control problem.}

We perform a numerical analysis of the robust strategies and value function. The investor allocates higher fraction of wealth to the risky bond if the reference default intensity increases. However, he faces a trade-off between investing more in risky securities to capture default risk premium and reducing his long investment to avoid losses when the bond defaults. Indeed, his risk aversion leads him to divert wealth from the riskier to the safer bond when the default risk becomes sufficiently high. Model uncertainty reduces the utility achievable by the investor. In particular, it leads him to reduce the demand for risky bonds if he is more tolerant against model misspecifications. In this case, the worst-case default intensity gets higher and the investor derives smaller utility by implementing his robust credit strategy. The investor's decisions are more sensitive to penalty for mispecification in the current
default state, but also take into account model uncertainty in future states reached when an additional obligor defaults.

The rest of the paper is organized as follows. Section \ref{sec:model} introduces the model. Section \ref{sec:CDSdyn} derives the master HJB equation associated with the robust control problem. Section \ref{sec:optstrategies} derives the robust
bond investment strategy. Section \ref{sec:HJB} analyzes the HJB equation. Section \ref{sec:verification} proves a verification theorem. Section \ref{sec:numanalysis} performs a numerical analysis. Section \ref{sec:concl} concludes. Technical proofs of auxiliary results are delegated to the appendix, while proofs of the main results are given in the main body of the paper.


\section{The Model}\label{sec:model}
We use three probability measures in the specification of our model, which are equivalent to each other: {\sf(I)} $\Px$ represents the reference measure, i.e. the one associated with the best description of the actual default intensity process available to the investor, {\sf(II)} $\tilde{\Px}$ corresponds to an alternative model chosen by the investor who wants to protect himself against misspecifications of the reference measure, and {\sf(III)} the risk-neutral measure $\Qx$ which is the measure under which prices of fixed income securities are observed. The investor is uncertain about the actual default intensities,
but is assumed to be certain about the pricing measure $\Qx$. We will elaborate more on the relations between these measures as well as on the rational behind such a model of uncertainty later in the section. We provide basic notation and definitions used throughout the paper in section \ref{sec:notations}. We give the default model in section \ref{sec:default}. We describe the portfolio securities in section \ref{sec:securities}. We formulate the robust control problem in section \ref{sec:formulation}.

\subsection{Notation} \label{sec:notations}
Let $\mathcal{S}:=\{0,1\}^M$.
Throughout the paper, the vector ${\bm z}=(z_1,\ldots,z_M) \in \mathcal{S}$ captures the default state of the portfolio, with $z_i=0$ if the obligor $i$ has not defaulted and $z_i=1$ if he has defaulted. For ${\bm z}\in{\cal S}$ such that $z_j=0$, we use
\begin{equation}
{\bm z}^j:=(z_1,\ldots,z_{j-1},1,z_{j+1},\ldots,z_M),\ \ \ \ j=1,\ldots,M,
\label{eq:zjdef}
\end{equation}
to denote the vector obtained from ${\bm z}$ by setting its $j$-th component to one. Let $m = 1,\ldots,M$ and $j_1, j_2, \ldots, j_m \in \{1,\ldots,M\}$, be $m$ distinct integers. Given ${\bm z}\in{\cal S}$ such that $z_{j_1} = z_{j_2} = \cdots = z_{j_m} = 0$, we use ${\bm z}^{j_1,\ldots,j_m}:=\left(\left({\bm z}^{j_1}\right)^{\ldots}\right)^{j_m}$
to denote the vector obtained from ${\bm z}$ by setting its components $j_1,j_2,\ldots j_m$ to one. In other words, ${\bm z}^{j_1,\ldots,j_m}$ denotes a default state where the names $j_1,j_2,\ldots,j_m$ have defaulted. {In particular, ${\bm z}^{j_1,\ldots,j_m}={\bm z}$ if $m=0$.} For brevity of notation, we will use
\begin{equation}
f_{j_1,\ldots,j_m}(\cdot):=f_{{\bm 0}^{j_1,\ldots,j_m}}(\cdot),
\label{eq:fz}
\end{equation}
where {${\bm0}=(0,\ldots,0)$ denotes the zero vector}, and $f_{\bm z}(\cdot)$ is an arbitrary measurable function depending on the default state ${\bm z}\in{\cal S}$.
Moreover, if $j \neq j_1,j_2,\ldots,j_m$, we set
\begin{equation}
g_{j,j_1,\ldots,j_m}(\cdot) := g_{j,{\bm 0}^{j_1,\ldots,j_m}}(\cdot),
\label{eq:gjnot}
\end{equation}
for any measurable function $g_{j,{\bm z}}(\cdot)$ depending on the default state ${\bm z}\in{\cal S}$ and the index $j$ of the obligor.

\subsection{Default Model} \label{sec:default}
We model default risk through an interacting intensity model. Models of this type are well suited for modeling default contagion. We also refer the reader to \cite{FreyBackhaus04} and \cite{JY} for additional details.

We consider $M \geq 2$ obligors subject to default risk. The default state is described by an $M$-dimensional default indicator process ${\bm Z}=(Z_1(t),\ldots,Z_M(t))_{t\geq0}$ supported by a probability space $(\Omega,\G,{\Px})$. Here, $\Px$ denotes the probability measure associated with the reference model corresponding with the best description of the data generating process available to the investor.
We denote by $\Ex^{\Px}$ the expectation operator w.r.t. $\Px$. The state space of the default indicator process ${\bm Z}$ is given by $\mathcal{S}=\{0,1\}^M$, where $Z_i(t)=1$ if the name $i$ has defaulted by time $t$ and $Z_i(t)=0$ otherwise. The default time of the $i$-th name is given by
\[
\tau_i :=\inf\{t\geq0;\ Z_i(t)=1\},\ \qquad i=1,\ldots,M.
\]
Hence, we have {$Z_i(t)={\bf1}_{\tau_i\leq t}$}, where $t\geq0$. Here, ${\bf1}_A$ denotes the indicator of the event $A$. The default indicator process ${\bm Z}$ is assumed to follow a continuous-time Markov chain on $\mathcal{S}$, where ${\bm Z}(t)$ transits to a neighbouring state ${\bm Z}^i(t)$ at rate ${\bf1}_{\{Z_i(t)=0\}}{h^{\Px}_{i,{\bm Z}(t)}(t)}$. Here, for $i=1,\ldots,M$, {$h^{\Px}_{i,{\bm z}}(t)$ is a continuous function in $t\geq0$}, for each ${\bm z}\in \mathcal{S}$. {We assume strictly positive default intensities satisfying $\inf_{t\geq0}h^{\Px}_{i,{\bm z}}(t)>0$.}

The market filtration is given by $\G_t =\sigma({\bm Z}(u);\ u\in[0,t])$, $t\geq0$, augmented with its null sets so to satisfy the usual conditions of completeness and right continuity; see Section 2.4 of \cite{Belanger}.
Using the Dynkin's formula (see (10.13) in \cite{RogersWilliams}, pag. 254), 
we have
\begin{eqnarray}\label{eq:Q-default-martingle}
\xi_i^{\Px}(t) &:=& Z_i(t) - \int_0^t(1-Z_i(u))h_{i,{\bm Z}(u)}^{\Px}(u)\D u,\ \ \ \ t\geq0
\end{eqnarray}
is a $(\Px,(\G_t)_{t\geq0})$-martingale.

\subsection{The Portfolio Securities} \label{sec:securities}
The portfolio of securities at disposal of the investor are:

\begin{itemize}
\item {\bf Money market account.} The value of one share at time $t$ is denoted by $B(t)$, and accrues interest at a constant rate $r>0$ so that $B(t) = e^{rt}$, {$t\geq0$}. We set $B(0)=1$.
\item {\bf Risky bonds.} We consider $M$ risky bonds referencing obligors whose default times are modeled as described in section \ref{sec:default}.
Different from a primary asset such as the stock where one can directly assume a convenient price process under the reference probability measure, fixed income securities are claims depending on the occurrence of a credit event. Consequently, as for any traded derivative contract, the bond price is equal to the expected discounted value of the credit contingent dividend process under the risk-neutral measure $\Qx$. It is important to distinguish $\Qx$ from the reference (subjective) probability measure $\Px$ of the investor. Bond prices are determined by the market and not by a single investor.

The dividend process of the $i$-th bond with maturity $T_i$ is given by
\begin{eqnarray}\label{eq:dividend}
D_i(t) &:=& \int_0^t (1-Z_i(u)) C_i \D u + \int_0^t R_i \D Z_i(u) + (1-Z_i(T_i)) {\bf 1}_{t\geq T_i},\ \ \ \ \ t\geq0.
\end{eqnarray}
Above, $C_i\geq0$ is the continuously paid coupon rate and thus $C_i \int_0^t(1-Z_i(u))\D u$ is the cumulative payment of the $i$-th bond before obligor $i$ defaults. $R_i\in{[0,1)}$ is the constant recovery rate paid at default time $\tau_i$. The quantity $(1-Z_i(T_i)) {\bf 1}_{t\geq T_i}$ is the unit notional payment received by the bond holder at the maturity time $T_i$ if the obligor $i$ has not defaulted.

Denote by $h_{i,{\bm Z}(t)}(t)$ the positive risk-neutral default intensity of obligor $i$ at time $t$. To guarantee that $\Qx$ is well defined, we assume that $h_{i,{\bm z}}(t)$ is continuous in $t$ for each default state ${\bm z}\in{\cal S}$.
Then the following process is a $(\Qx,(\G_t)_{t\geq0})$-martingale:
\begin{eqnarray}\label{eq:default-mart-Q}
\xi_i(t) := Z_i(t) - \int_0^{t} (1-Z_i(u))h_{i,{\bm Z}(u)}(u)\D u,\ \ \ \ \ \ t\geq0.
\end{eqnarray}
{Lemma~\ref{lem:uniqe-Q} in Appendix~\ref{sec:appHJB} shows that, in the market without model uncertainty, the above risk-neutral default intensity is uniquely determined under a mild invertibility condition on the matrix of bond depreciations (see Lemma~\ref{lem:uniqe-Q} for the precise statement). By the Second Fundamental Theorem of Asset Pricing (see for example Theorem 1.2 in \cite{bia}), this implies that the market model consisting of bank account and risky bonds is complete.}

The time-$t$ price of the $i$-th bond is given by
\begin{eqnarray}\label{eq:Phi0}
P_i(t) &:=& (1-Z_i(t))\Ex_t\left[ \int_t^{T_i}e^{-\int_t^ur\D s} \D D_i(u)  \right],\ \ \ \ \ \ \ \ t\in[0,T_i],
\end{eqnarray}
where the expectation $\Ex_t[\cdot]:=\Ex[\cdot|\G_t]$ is under the risk-neutral measure $\Qx$ and conditional on the current information set. Moreover, the price formula \eqref{eq:Phi0} shows that on $\tau_i\leq t$, the price of the $i$-th bond is given by $0$, while on $\tau_i>t$, it is
\begin{eqnarray}\label{eq:Phi1}
P_i(t) &=& \Ex_t\left[C_i\int_t^{T_i}e^{-\int_t^ur\D s} (1-Z_i(u)) \D u  \right] + R_i\Ex_t\left[Z_i(T_i)e^{-\int_t^{T_i} r(1-Z_i(u))\D u}\right]\nonumber\\
&&+\Ex_t\left[(1-Z_i(T_i))e^{-\int_{t}^{T_i}r\D s}\right].
\end{eqnarray}
%

\end{itemize}

\subsection{Robust Control Formulation}\label{sec:formulation}
This section describes the robust portfolio optimization problem of the investor. Our investor dynamically allocates her wealth into the money market account and the $M$ risky bond securities. {Hereafter, let $0<T<\wedge_{i=1}^MT_i$ be the terminal horizon, i.e. the investment horizon is smaller than the maturities of the bond securities.}

For $i=1,\ldots,M$, and {$t\in[0,T]$}, denote by $\phi_i(t)$ the number of shares of the $i$-th risky bond that the investor buys $(\phi_i(t)>0)$ or sells $(\phi_i(t)<0)$ at time $t$.  A short credit position is implemented by short-selling a bond, while a long credit position is implemented by purchasing the bond security. In the latter case, the investor pays the bond price and receives a stream of coupons until default occurs. Moreover, he receives the recovery rate when default happens.

We use $\phi_{B}(t)$ to denote the number of shares held in the money market account at time $t$. The process $\bar{\bm\phi}=(\bm\phi(t),\phi_B(t))_{t\in[0,T]}$ with ${\bm\phi}(t)=(\phi_i(t))_{i=1,\ldots,M}$ is called a portfolio process. The wealth process associated with the portfolio $\bar{\bm\phi}=({\bm\phi}(t),\phi_B(t))_{t\in[0,T]}$, denoted by $V_t^{\bar{\bm\phi}}$, is given by
\begin{equation}\label{eq:wealth-def}
V_t^{\bar{\bm\phi}} = \sum_{i=1}^M\phi_{i}(t)P_i(t)  + \phi_B(t) B(t).
\end{equation}
Actual default intensities and default correlations are notoriously difficult to estimate given that default events happen rarely. For instance, \cite{Duffie6} find that a proportional-hazards form for the default intensity consisting of four macroeconomic and firm-specific covariates (the firm{'}s distance to default, the firm{'}s trailing one-year stock return, the three-month Treasury bill rate, and the trailing one-year return on the DJIA)
is unable to fit empirically estimated historical default correlations. 
On the other hand, risk-neutral default intensities can be more
accurately estimated, especially after the growth of the liquid CDS market. On the basis of these considerations, we only consider uncertainty in the actual default intensities and
assume risk-neutral default intensities to be perfectly known.

Given his limited ability to assess the likelihood of default events and their correlation, the investor considers alternative models to protect himself against possible model misspecifications. Each alternative model is defined by an equivalent probability measure $\tilde{\Px}{\sim{\Px}}$ on $\G_t$ specified via the Radon-Nikodym derivative $\eta_t^{{\bm\vartheta}}$ given by
\begin{eqnarray}\label{eq:robust-density}
\eta_t^{{\bm\vartheta}} = {\cal E}\left(\sum_{i=1}^M\int_0^{\cdot}\big(\vartheta_i(u-)-1\big)\D\xi_i^{\Px}(u)\right)_t.
\end{eqnarray}
In the above expression, $\vartheta_i(t)=\vartheta_{i,{\bm Z}(t)}(t)$, i.e. $\eta_t^{{\bm\vartheta}}$ changes the investor{'}s probability assessment of reference default intensities and default correlations.
\begin{remark}
{\it The formulation of the robust decision making problem requires the introduction of three equivalent probability measures $\Px$, $\tilde{\Px}$ and $\Qx$.
While bond prices are observed under the risk neutral measure $\Qx$, the investor wishes to optimize his expected utility from terminal wealth under the worst-case measure, i.e. under the worst-case alternative model that he considers.}
\end{remark}

Under the alternative measure $\tilde{\Px}$, the default intensity of obligor $i$ becomes $h^{\tilde{\Px}}_{i,{\bm Z}(t)}(t):=\vartheta_i(t)h_{i,{\bm Z}(t)}^{\Px}(t)$. {In light of \eqref{eq:robust-density},} the investor operates under the reference model by choosing $\vartheta_i(t) = 1$, for each $i=1,\ldots,M$, and selects other models by choosing $\vartheta_i(t) \neq 1$.
If for a name $i$ and time $t$, $\vartheta_i(t) < 1$, it means that the investor is more optimistic than the history on the credit quality of the obligor $i$. Viceversa, if {$\vartheta_i(t) > 1$}, it means that he is more pessimistic and believes that the credit quality of $i$ is worse than what predicted by his estimation method. Notice that the following process is a $(\tilde{\Px},(\G_t)_{t\in[0,T]})$-martingale:
\begin{eqnarray}\label{eq:default-mart-Ptilde}
\xi_i^{\tilde{\Px}}(t) &:=& Z_i(t) - \int_0^{t} (1-Z_i(u)) h_{i,{\bm Z}(u)}^{\tilde{\Px}}(u)\D u,\ \ \ \ \ \ {t\in[0,T]}.
\end{eqnarray}

Since $\Px$ is statistically the best representation of existing data, the investor penalizes his choice of $\tilde{\Px}$ according to how much it deviates from the reference measure $\Px$. The distance measure is captured by the relative entropy ${\cal H}_t(\tilde{\Px}|\Px)$. The latter is defined as the expectation, under the probability measure $\tilde{\Px}\sim\Px$ {on $\G_t$}, of the logarithm of the Radon-Nikodym derivative at time $t$ given by \eqref{eq:robust-density}. We can rewrite $\eta_t^{{\bm\vartheta}}$ using It\^o's formula as
\begin{eqnarray}\label{eq:solution-eta-var}
\eta_t^{{\bm\vartheta}} = \exp\left\{\sum_{i=1}^M\int_0^t\log(\vartheta_i(u-))\D Z_i(u)
-\sum_{i=1}^M\int_0^{t\wedge\tau_i}\big(\vartheta_i(u)-1\big)h_{i,{\bm Z}(u)}^{\Px}(u)\D u\right\}.
\end{eqnarray}

We denote by ${\cal V}_{0}$ the space of all $(\G_t)_{t\in[0,T]}$-adapted positive processes ${\bm\vartheta}=({\bm\vartheta}(t))_{t\in[0,T]}$ so that the density process $\eta^{{\bm\vartheta}}=(\eta_t^{{\bm\vartheta}})_{t\in[0,T]}$ is a $(\Px,(\G_t)_{t\in[0,T]})$-martingale if the initial time is $0$. Similarly, we use ${\cal V}_t$ to represent the counterpart if the initial time is $t\in[0,T]$. Notice that the log Radon-Nikodym derivative process under $\tilde{\Px}$ is given by
\begin{eqnarray*}
\log\big(\eta_t^{{\bm\vartheta}}\big) = \sum_{i=1}^M\int_0^t\log(\vartheta_i(u-))\D{\xi}_i^{\tilde{\Px}}(u)
+ \sum_{i=1}^M\int_0^{t\wedge\tau_i}\big[\vartheta_i(u)\log(\vartheta_i(u))-\vartheta_i(u)+1\big]h_{i,{\bm Z}(u)}^{\Px}(u)\D u.
\end{eqnarray*}
Using \eqref{eq:default-mart-Ptilde}, this leads to the following expression of the relative entropy:
\begin{eqnarray}\label{eq:relative-entropy}
{\cal H}_t(\tilde{\Px}|\Px) := \Ex^{\tilde{\Px}}\left[\log\big(\eta_t^{{\bm\vartheta}}\big)\right] =\Ex^{\tilde{\Px}}\left[\sum_{i=1}^M\int_0^{t\wedge\tau_i}\big[\vartheta_i(u)\log(\vartheta_i(u))-\vartheta_i(u)+1\big]h_{i,{\bm Z}(u)}^{\Px}(u)\D u\right].
\end{eqnarray}
Moreover, over a sufficiently small time interval the relative entropy admits the following limit
\begin{eqnarray}\label{eq:limit-entropy}
\lim_{\delta\too0}\frac{1}{\delta}\Ex_t^{\tilde{\Px}}\left[\log\frac{\eta_{t+\delta}^{{\bm\vartheta}}}{\eta_t^{{\bm\vartheta}}}\right]
&=&\sum_{i=1}^M(1-Z_i(t))\big[\vartheta_i(t)\log(\vartheta_i(t))-\vartheta_i(t)+1\big]h_{i,{\bm Z}(t)}^{\Px}(t)
=:\sum_{i=1}^M\tilde{p}_t^{i,{\bm\vartheta}},
\end{eqnarray}
since ${\bm\vartheta}(t)$ is $\G_t$-adapted. Here $\Ex_t^{\tilde{\Px}}:=\Ex^{\tilde{\Px}}[\cdot|\G_t]$.

We follow \cite{Anderson} and assume that the investor chooses a robust portfolio strategy which is the best choice in some worst-case model. We consider a rational risk-averse investor who wants to maximize his power utility from terminal wealth, i.e. $U:[0,\infty)\too[0,\infty)$ is given by $U(v)=\frac{v^{\gamma}}{\gamma}$, where $\gamma\in(0,1)$ is the risk-aversion parameter. The investor maximizes his utility function, adjusted for model ambiguity, by choosing an optimal admissible allocation strategy (the precise definition of admissibility will be given in the next section) over the risky bond instruments. Concretely, we define the value function
\begin{eqnarray}\label{eq:robust-problem}
w_{\bm z}(t,v) := \sup_{{\bm\phi}\in\tilde{\cal U}_t}\inf_{{\bm\vartheta}\in{\cal V}_t}\Ex^{\tilde{\Px}}_{t,v,{\bm z}}\left[U\big(V_T^{\bar{\bm\phi}}\big)
+\sum_{i=1}^{M}\int_t^T\frac{\tilde{p}_u^{i,{\bm\vartheta}}}{\Upsilon_{i,{\bm Z}(u)}(u,V_u^{\bar{\bm\phi}})}\D u\right],
\end{eqnarray}
where the $\tilde{\Px}$-conditional expectation $\Ex^{\tilde{\Px}}_{t,v,{\bm z}}[\cdot]:=\Ex^{\tilde{\Px}}[\cdot|V_t=v,{\bm Z}(t)={\bm z}]$ for $(t,v,{\bm z})\in[0,T]\times\R_+\times{\cal S}$, the penalty rate $\tilde{p}_t^{i,{\bm\vartheta}}$, $t\in[0,T]$, ($i=1,\ldots,M$) is defined by \eqref{eq:limit-entropy} and $\Upsilon_{i,{\bm z}}(t,v)$ denotes the preference parameter governing aversion to uncertainty with respect to the reference default intensity of obligor $i$.
For $i=1,\ldots,M$, and ${\bm z}\in{\cal S}$, this is assumed to be of the following form:
\begin{eqnarray}\label{eq:weights}
\Upsilon_{i,{\bm z}}(t,v) = \frac{\mu_{i,{\bm z}}(t)}{U(v)},\ \ \ \ \ \ (t,v)\in[0,T]\times\R_+,
\end{eqnarray}
where $\mu_{i,{\bm z}}(t)$ satisfies $\inf_{t\in[0,T]}\mu_{i,{\bm z}}(t)>0$, is a continuous function in time $t$, and is allowed to depend on the current default state. {As in \cite{Mahenhout}, \cite{MahenhoutJT} and \cite{Liu}, we are imposing a homothetic specification of ambiguity aversion which renders the problem tractable and allows to separate the Hamiltonian into a term depending on the level of wealth and another term which only depends on time (see Eq.~\eqref{eq:H} for the detailed expression). Under this homotheticity form, we are able to derive a closed-form representation of the worst-case measure and of the optimal strategies. Under the above specification of ambiguity, the larger the values of these functions and the less a given deviation from the reference model is penalized. Symmetrically, the less confidence the investor has in the reference default model, and the more the worst-case model will deviate from the reference model. Although this homothetic specification is made for analytical convenience, it has been argued in \cite{Mahenhout} that it is convenient for calibration purposes 
and for economic reasons because it facilitates the construction of a representative agent, see section 2 in his paper.} The specification in Eq.~\eqref{eq:weights} also allows distinguishing  the degree of uncertainty in the different information sources. For instance, if the source of information used to estimate the physical default intensity of name $i$ in the default state ${\bm z}\in{\cal S}$ is very reliable, the investor has more faith in the reference model, hence he would choose high values of the penalty (low values of $\mu_{i,{\bm z}}(t)$). If instead the investor has unreliable information to perform such an estimation, he is more robust and has less faith in the reference model, hence he would penalize less (higher values of $\mu_{i,{\bm z}}(t)$) deviations from the reference model. 

\section{Dynamics Programming Formulation}\label{sec:CDSdyn}
The objective of this section is to derive the HJB equation associated with the robust control problem in Eq.~\eqref{eq:robust-problem}.
For this, we need the dynamics of the wealth process which in turns depends on the dynamics of the price process $P_i(t)$ of the $i$-th bond security. We give the exact price dynamics of the bond price process $P_i(t)$, $t\in[0,{T_i}]$, in section \ref{sec:pricedyn}. We obtain the master HJB equation in section \ref{sec:HJB1}.

\subsection{Price Dynamics of Risky Bonds} \label{sec:pricedyn}

For $i=1,\ldots,M$, using the price representation \eqref{eq:Phi1}, we can rewrite it as
\begin{eqnarray}\label{eq:cds-price}
P_i(t) = (1-Z_i(t))F_{i,{\bm Z}(t)}(t),\ \ \ \ \ \ \ \ t\in[0,T_i],
\end{eqnarray}
where $F_{i,{\bm z}}(t)$, $(t,{\bm z})\in[0,T_i]\times{\cal S}$, denotes the pre-default price function given by
\begin{eqnarray}\label{eq:pre-default-cds-price}
F_{i,{\bm z}}(t)&=& R_i F^a_{i,{\bm z}}(t)+ C_i F^b_{i,{\bm z}}(t)+F^c_{i,{\bm z}}(t).
\end{eqnarray}
Here the decomposed price functions are given by
\begin{eqnarray}
F^a_{i,{\bm z}}(t) &:=& \Ex\left[Z_i(T_i)e^{-\int_t^{T_i} r(1-Z_i(u))\D u}\Big|{\bm Z}(t)={\bm z}\right],\nonumber\\
F^b_{i,{\bm z}}(t) &:=& \Ex\left[\int_t^{T_i}e^{-\int_t^u r\D s}(1-Z_i(u))\D u \bigg|{\bm Z}(t)={\bm z}\right],\\
F^c_{i,{\bm z}}(t) &:=& \Ex\left[(1-Z_i(T_i))e^{-\int_{t}^{T_i}r\D u}\Big|{\bm Z}(t)={\bm z}\right].\nonumber
\label{eq:Ffun}
\end{eqnarray}
{Obviously, $F_{i,{\bm z}}(T_i)=R_iz_i+1-z_i$. Hence $P_i(T_i)=1-Z_i(T_i)$ for $i=1,\ldots,M$.} We next give an auxiliary lemma giving the $\tilde{\Px}$-dynamics of the dividend adjusted
bond price process, later used to derive the dynamics of the wealth process under the measure $\tilde{\Px}$.
\begin{lemma}\label{lem:P+D}
Under $\tilde{\Px}$, for each $i=1,\ldots,M$, we have the following dynamics: {for $t\in[0,T]$,}
\begin{eqnarray}\label{eq:P+D}
\frac{\D\big(P_i(t)+D_i(t)\big)}{P_i(t-)} &=& \left[r-(1-Z_j(t))\big(h_{j,{\bm Z}(t)}(t)-\vartheta_j(t)h^{\Px}_{j,{\bm Z}(t)}(t)\big)\right] \D t+\sum_{j=1}^M G_{i,j,{\bm Z}(t-)}(t)\D \xi_j^{\tilde{\Px}}(t).
\end{eqnarray}
We recall that the $\tilde{\Px}$-martingale $\xi_j^{\tilde{\Px}}(t)$ in the above expression is given by \eqref{eq:default-mart-Ptilde}, and for $i,j=1,\ldots,M$, we define the functions
\begin{eqnarray}\label{eq:Phi}
G_{i,j,{\bm z}}(t) &:=& \frac{F_{i,{\bm z}^j}(t)}{F_{i,{\bm z}}(t)}-1,\ \ \ \ \ \ (t,{\bm z})\in[0,T]\times{\cal S}.
\end{eqnarray}
\end{lemma}

\noindent{\it Proof.}\quad From Lemma~\ref{thm:mul-cds-price}, it follows that
\begin{eqnarray*}
P_i(t) = P_i(0) + \int_0^t (1-Z_i(u))\big[rP_i(u)-C_i\big]\D u-R_i Z_i(t)
+\int_0^tP_i({u-})\sum_{j=1}^M G_{i,j,{\bm Z}(u-)}(u)\D \xi_j(u).
\end{eqnarray*}
Take the dividend given by \eqref{eq:dividend} into account, and notice that {$T\in(0,\wedge_{i=1}^MT_i)$}. Then we have ${\bf1}_{t\geq T_i}=0$ for all $t\in[0,T]$. Hence
\begin{eqnarray*}
P_i(t)+D_i(t) = P_i(0)+D_i(0) + r \int_0^t P_i(u)\D u+\int_0^tP_i(u-)\sum_{j=1}^M G_{i,j,{\bm Z}(u-)}(u)\D \xi_j(u).
\end{eqnarray*}
Then the $\tilde{\Px}$-dynamics \eqref{eq:P+D} follows from \eqref{eq:default-mart-Q} and the relation
\begin{eqnarray}\label{eq:default-mart-PQ}
\xi^{\tilde{\Px}}_j(t)=\xi_j(t) + \int_0^{t} (1-Z_j(u))\big(h_{j,{\bm Z}(u)}(u)-\vartheta_j(u)h^{\Px}_{j,{\bm Z}(u)}(u)\big)\D u.
\end{eqnarray}
This completes the proof of the lemma. \hfill$\Box$

We can derive recursive explicit expressions for the pre-default price functions $F_{i,{\bm z}}(t)$. This in turn yields explicit recursive expressions for the function $G_{i,j,{\bm z}}(t)$, and consequently for the dynamics of the bond price process $P_i(t)$.
We give the detailed expressions in Appendix~\ref{appen:express-pre-fcn}.

\subsection{The HJB Equation} \label{sec:HJB1}
This section derives the HJB equation associated with the robust control problem. We require the portfolio process $\bar{\bm\phi}$ to be $(\G_t)_{t\in[0,T]}$-predictable. A $(\G_t)_{t\in[0,T]}$-{predictable} portfolio process $\bar{\bm\phi}=({\bm\phi}(t),\phi_B(t))_{t\in[0,T]}$ is said to be {\it self-financing} if $V_t^{\bar{\bm\phi}} = V_0^{\bar{\bm\phi}} + {\Theta}_t^{\bar{\bm\phi}}$, {where the time-$t$ wealth $V_t^{\bar{\bm\phi}}$ is defined in Eq.~\eqref{eq:wealth-def}}, and the time-$t$ gains process is given by
\begin{equation}\label{eq:gainsp}
{\Theta}_t^{\bar{\bm\phi}} = \sum_{i=1}^M\int_0^t \phi_{i}({u})\D(P_i(u)+D_i(u)) + \int_0^t\phi_{B}(u) \D B(u).
\end{equation}
Above, $D_i=(D_i(t))_{t\geq0}$ is the dividend process of the $i$-th risky bond given by \eqref{eq:dividend}.

For $t\in[0,T]$, we use $\tilde{\pi}_B(t):=\frac{\phi_{B}(t)B_t}{V_{t-}^{\bar{\bm\phi}}}$ to denote the proportion of wealth invested in the money market account. Similarly, $\tilde{\pi}_i(t):=\frac{\phi_i(t)P_i(t-)}{V_{t-}^{\bar{\bm\phi}}}$, $i=1,\ldots,M$, denotes the the proportion of wealth invested in the $i$-th risky bond. From the above definition and using \eqref{eq:Phi0}, it can be easily seen that $\tilde{\pi}_i(t)=(1-Z_i(t-))\tilde{\pi}_i(t)$ for $t\in[0,T]$ . Further, from \eqref{eq:wealth-def}, it follows that
\begin{eqnarray}\label{eq:pisum=1}
\sum_{i=1}^M \tilde{\pi}_i(t) + \tilde{\pi}_B(t) =1.
\end{eqnarray}
We can then define the space of admissible strategies.
\begin{definition}\label{def:control-U}
{\it Let $t\in[0,T]$. The $t$-admissible control set $\tilde{\cal U}_t=\tilde{\cal U}_t(v,{\bm z})$, $(v,{\bm z})\in{\R_+}\times{\cal S}$, is a class of $(\G_t)_{t\in[0,T]}$-predictable locally bounded feedback strategies $\tilde{\pi}(u)= (\tilde{\pi}_i(u))_{i=1,\ldots,M}$ given by
\begin{eqnarray}\label{eq:pit-}
\tilde{\pi}_i(u) ={\pi}_{i,{\bm Z}(u-)}\big(u,V_{u-}^{\tilde{\bm\pi}}\big),\ \ \  i=1,\ldots,M, \ \ \ \ \ u\in[t,T].
\end{eqnarray}
The feedback function $\pi_{i,{\bm z}}(\cdot)$ is locally bounded on $[0,T]\times\R_+$ for $i=1,\ldots,M$ and ${\bm z}\in{\cal S}$, and $\tilde{\pi}_i(u)=(1-Z_i(u-)){\pi}_{i,{\bm Z}(u-)}(u,V_{u-}^{\tilde{\bm\pi}})$ for $u\in[t,T]$.
The relative wealth process is of the form given by
$$
\int_t^T \frac{\D V_u^{\tilde{\bm\pi}}}{V_{u-}^{\tilde{\bm\pi}}}  = \int_t^T \alpha(u) \, \D u + \sum_{j=1}^M \int_t^T \beta_j(u)\D Z_j(u),
$$
where $V_t^{\tilde{\bm\pi}}= v \in \R_+$, and ${\bm Z}(t)={\bm z}\in{\cal S}$. Above, $\alpha$ is an adapted process with well defined integral $\int_t^T \alpha (u) \, \D u$, and $\beta_j$
is a predictable process and bounded away from $-1$ and $\infty$ for all $j = 1,\dots,M$. Moreover, we define ${\cal U}_t$ to be the $t$-admissible set of locally bounded feedback function vectors ${\bm\pi}=(\pi_{i,{\bm z}}(\cdot))_{i=1,\ldots,M, {\bm z}\in{\cal S}}$.}
\end{definition}

The following lemma gives the dynamics of the wealth process:
\begin{lemma}\label{lem:wealth-dynamics}
Let $\tilde{\bm\pi}\in\tilde{\cal U}_0$ {and $t\in[0,T]$}. Then the $\tilde{\Px}$-wealth dynamics is given by, $V_0^{\tilde{\bm\pi}}= v>0$, and
\begin{eqnarray}\label{eq:wealth}
\frac{\D V_t^{\tilde{\bm\pi}}}{V_{t-}^{\tilde{\bm\pi}}} &=&r \D t - \sum_{j=1}^M\left(\sum_{i=1}^M\tilde{\pi}_i(t)G_{i,j,{\bm Z}(t-)}(t)\right)(1-Z_j(t-))\big(h_{j,{\bm Z}(t-)}(t)-\vartheta_j(t)h^{\Px}_{j,{\bm Z}(t-)}(t)\big)\D t\nonumber\\
&&+\sum_{j=1}^M\left(\sum_{i=1}^M\tilde{\pi}_i(t)G_{i,j,{\bm Z}(t-)}(t)\right)\D \xi_j^{\tilde{\Px}}(t),
\end{eqnarray}
where, for $i,j=1,\ldots,M$, the function $G_{i,j,{\bm z}}(t)$ with $(t,{\bm z})\in[0,T]\times{\cal S}$, is defined by \eqref{eq:Phi}.
\end{lemma}

\noindent{\it Proof.}\quad  For $\tilde{\bm\pi}\in\tilde{\cal U}_0$, using \eqref{eq:gainsp}, it follows that
\begin{eqnarray*}
\D V_t^{\tilde{\bm\pi}} &=& \sum_{i=1}^M \phi_{i}(t)\D(P_i(t)+D_i(t)) + \phi_{B}(t) \D B(t)
=V_{t-}^{\tilde{\bm\pi}}\sum_{i=1}^M \tilde{\pi}_i(t)\frac{\D(P_i(t)+D_i(t))}{P_i(t-)} +  V_{t-}^{\tilde{\bm\pi}}\tilde{\pi}_B(t)r\D t.
\end{eqnarray*}
It follows from Lemma~\ref{lem:P+D} that
\begin{eqnarray*}
\D V_t^{\tilde{\bm\pi}}&=&V_{t-}^{\tilde{\bm\pi}}\sum_{i=1}^M \tilde{\pi}_i(t)\frac{\D(P_i(t)+D_i(t))}{P_i(t-)} +  V_{t-}^{\tilde{\bm\pi}}\tilde{\pi}_B(t)r\D t\nonumber\\
&=&V_{t-}^{\tilde{\bm\pi}}\sum_{i=1}^M \tilde{\pi}_i(t)\left[r-\sum_{j=1}^M G_{i,j,{\bm Z}(t-)}(t)(1-Z_j(t-))\big(h_{j,{\bm Z}(t-)}(t)-\vartheta_j(t)h^{\Px}_{j,{\bm Z}(t-)}(t)\big)\right]\D t\nonumber\\
&&+  V_{t-}^{\tilde{\bm\pi}}\tilde{\pi}_B(t)r\D t+V_{t-}^{\tilde{\bm\pi}}\sum_{j=1}^M \left(\sum_{i=1}^M \tilde{\pi}_i(t)G_{i,j,{\bm Z}(t-)}(t)\right)\D\xi_j^{\tilde{\Px}}(t).
\end{eqnarray*}
Then the $\tilde{\Px}$-wealth dynamics \eqref{eq:wealth} follows from the condition \eqref{eq:pisum=1}. This completes the proof of the lemma. \hfill$\Box$

The wealth dynamics \eqref{eq:wealth} may be intuitively interpreted as follows. The investor accrues instantaneous interest rate $r$ on his wealth. When a credit event occurs, his wealth level is updated to reflect changes in the mark-to-market value of his bond position resulting from contagion effects induced by the defaulted name on the surviving entities.
\begin{remark}\label{rem:wealth}
{\it Using Definition \ref{def:control-U} of admissible controls, we have $\tilde{\bm\pi}(u)={\bm\pi}_{{\bm Z}(u-)}(u,V_{u-}^{t,v,{\bm\pi}})$ for $u\in[0,T]$. Then, we can rewrite the wealth dynamics \eqref{eq:wealth} as follows: {for $t\in[0,T]$,}
\begin{eqnarray}\label{eq:wealth0}
\frac{\D V_t^{{\bm\pi}}}{V_{t-}^{{\bm\pi}}} &=&r \D t - \sum_{j=1}^M\left(\sum_{i=1}^M\pi_{i,{\bm Z}(t-)}(t,V_{t-}^{{\bm\pi}})G_{i,j,{\bm Z}(t-)}(t)\right)(1-Z_j(t-))\big(h_{j,{\bm Z}(t-)}(t)-\vartheta_j(t)h^{\Px}_{j,{\bm Z}(t-)}(t)\big)\D t\nonumber\\
&&+\sum_{j=1}^M\left(\sum_{i=1}^M\pi_{i,{\bm Z}(t-)}(t,V_{t-}^{{\bm\pi}})G_{i,j,{\bm Z}(t-)}(t)\right)\D \xi_j^{\tilde{\Px}}(t).
\end{eqnarray}}
\end{remark}

Next, we derive the master HJB equation {using \eqref{eq:limit-entropy} and \eqref{eq:wealth0}}. We first start by heuristic arguments and then give a rigorous proof in the verification theorem. {Recall the robust control optimization criterion given by \eqref{eq:robust-problem}}. If $w_{\bm z}(t,v)$ is $C^1$ in $t$, and is $C^2$ in $v$ for each ${\bm z}\in{\cal S}$, using the dynamic programming principle, {we expect that the master HJB equation associated with the robust control problem \eqref{eq:robust-problem}} is given by
\begin{eqnarray}\label{eq:HJB}
&&\sup_{{\bm\pi}\in{\cal U}_0}\inf_{{\bm\vartheta}\in{\cal V}_0}\Bigg\{\left(\frac{\partial }{\partial t}+{\cal L}^{{\bm\pi},{\bm\vartheta}}\right)w_{\bm z}(t,v) +\sum_{i=1}^M\frac{(1-z_i)\big[\vartheta_{i,{\bm z}}\log(\vartheta_{i,{\bm z}})-\vartheta_{i,{\bm z}}+1\big]h_{i,{\bm z}}^{\Px}(t)}{\Upsilon_{i,{\bm z}}(t,v)}\Bigg\}=0
\end{eqnarray}
with terminal condition $w_{\bm z}(T,v) = U(v)$. Here the operator ${\cal L}^{{\bm\pi},{\bm\vartheta}}:={\cal L}_{c}^{{\bm\pi}}+{\cal L}_{J}^{{\bm\pi},{\bm\vartheta}}$ acts on any function $\varphi_{\bm z}(t,v)$ which is $C^1$ in $(t,v)$, as follows:
\begin{eqnarray}\label{eq:operatorsL}
{\cal L}_{c}^{{\bm\pi}}\varphi_{\bm z}(t,v)&:=& v\frac{\partial \varphi_{\bm z}(t,v)}{\partial v}\left[r-\sum_{j=1}^M\left(\sum_{i=1}^M\pi_{i}{(1-z_i)} G_{i,j,{\bm z}}(t)\right)(1-z_j)h_{j,{\bm z}}(t)\right],\\
{\cal L}_J^{{\bm\pi},{\bm\vartheta}}\varphi_{\bm z}(t,v)&:=& \sum_{j=1}^M\left[\varphi_{{\bm z}^{j}}\bigg(t,v+v\bigg(\sum_{i=1}^M\pi_{i}{(1-z_i)} G_{i,j,{\bm z}}(t)\bigg)\bigg)-\varphi_{\bm z}(t,v)\right](1-z_j)\vartheta_{j,{\bm z}}{h}_{j,{\bm z}}^{\Px}(t),\nonumber
\end{eqnarray}
where, for $i=1,\ldots,M$ and ${\bm z}\in{\cal S}$, $\pi_{i}=\pi_{i,{\bm z}}(t,v)$ on $(t,v)\in[0,T]\times\R_+$ so that ${\bm\pi}=(\pi_{i})_{i=1,\ldots,M}\in{\cal U}_0$.

\section{Optimal Feedback Functions} \label{sec:optstrategies}
This section rigorously analyzes the optimal feedback functions. Section \ref{sec:worstcm} derives the HJB equation associated with the worst-case measure. We then analyze the optimal feedback functions under such a measure in section \ref{sec:worstopt}.

\subsection{{Worst-Case Measure}} \label{sec:worstcm}
This section gives the explicit form of the HJB equation associated with the worst-case measure, i.e. it analyzes the inner minimization problem in Eq.~\eqref{eq:HJB}. To this purpose, we first use separation of variables and propose the following decomposition of the value function:
\begin{eqnarray}\label{eq:decom-value}
w_{\bm z}(t,v) = U(v) B_{\bm z}(t),
\end{eqnarray}
where $B_{\bm z}(t)$ is a positive $C^1$-function in $t\in[0,T]$ for each ${\bm z}\in{\cal S}$.
Using the expressions for the operators in \eqref{eq:operatorsL} and the decomposition~\eqref{eq:decom-value}, we define the following Hamiltonian by
\begin{eqnarray}\label{eq:H}
H_{\bm z}^{{\bm\pi},{\bm\vartheta}}(t,v) &:=&{\cal L}^{{\bm\pi},{\bm\vartheta}}w_{\bm z}(t,v) +\sum_{i=1}^M\frac{(1-z_i)\big[\vartheta_{i,{\bm z}}\log(\vartheta_{i,{\bm z}})-\vartheta_{i,{\bm z}}+1\big]h_{i,{\bm z}}^{\Px}(t)}{\Upsilon_{i,{\bm z}}(t,v)}\nonumber\\
&=& \gamma U(v)B_{\bm z}(t)\left[r -\sum_{j=1}^M\Gamma_{j,{\bm z}}^{{\bm\pi}}(t)(1-z_j)h_{j,{\bm z}}(t)\right]\nonumber\\
&&+U(v)\sum_{j=1}^M\big[B_{{\bm z}^j}(t)\big(1+\Gamma_{j,{\bm z}}^{{\bm\pi}}(t)\big)^{\gamma} - B_{\bm z}(t)\big](1-z_j)\vartheta_{j,{\bm z}}h_{j,{\bm z}}^{\Px}(t)\nonumber\\
&&+U(v)\sum_{j=1}^M\frac{(1-z_j)\big[\vartheta_{j,{\bm z}}\log(\vartheta_{j,{\bm z}})-\vartheta_{j,{\bm z}}+1\big]h_{j,{\bm z}}^{\Px}(t)}{\mu_{j,{\bm z}}(t)}.
\end{eqnarray}

Above, we have used the following linear transformation of the feedback function: for $(t,{\bm z})\in[0,T]\times{\cal S}$, and ${\bm\pi}\in{\cal U}_0$,
\begin{eqnarray}\label{eq:linear}
\Gamma_{j,{\bm z}}^{{\bm\pi}}(t) &:=& \sum_{i=1}^M\pi_i(1-z_i)G_{i,j,{\bm z}}(t),\ \ \ \ \ \ j=1,\ldots,M.
\end{eqnarray}
Then, for $j=1,\ldots,M$, and ${\bm z} \in \cal{S}$,
the solution of the first-order condition $\frac{\partial H_{\bm z}^{{\bm\pi},{\bm\vartheta}}(t,v)}{\partial\vartheta_{j,{\bm z}}}=0$ is given by
\begin{eqnarray}\label{eq:werst-controlj}
\vartheta_{j,{\bm z}}^{*,{\bm\pi}}(t) &=& \exp\bigg\{-\mu_{j,{\bm z}}(t)\left[B_{{\bm z}^j}(t)\big(1+\Gamma_{j,{\bm z}}^{{\bm\pi}}(t)\big)^{\gamma} - B_{\bm z}(t)\right]\bigg\},\ \ \ \ \ {\rm if}\ z_j=0.
\end{eqnarray}
{We will prove in the verification theorem that $\vartheta_{j,{\bm z}}^{*,{\bm\pi}}(t)$ given by \eqref{eq:werst-controlj} indeed identifies the worst-case measure {corresponding to ${\bm\pi}$}.
Substituting the expression~\eqref{eq:werst-controlj} into the master HJB equation \eqref{eq:HJB} yields
\begin{eqnarray}\label{eq:HJB2}
&&\sup_{{\bm\pi}\in{\cal U}_t}\left\{U(v)B'_{\bm z}(t) +H_{\bm z}^{{\bm\pi},{\bm\vartheta}^{*,{\bm\pi}}}(t,v)\right\}=0
\end{eqnarray}
with terminal condition $w_{\bm z}(T,v) = U(v)$. In the above expression, for ${\bm\pi}\in{\cal U}_t$, the Hamiltonian $H_{\bm z}^{{\bm\pi},{\bm\vartheta}_{\bm z}^{*,{\bm\pi}}}(t)$ is defined by \eqref{eq:H} with ${\bm\vartheta}$ replaced by ${\bm\vartheta}_{\bm z}^{*,{\bm\pi}}=(\vartheta_{i,{\bm z}}^{*,{\bm\pi}})_{i=1,\ldots,M}$ given by \eqref{eq:werst-controlj}. In the sequel of the paper, to lighten notation, we use $H_{\bm z}^{{\bm\pi},{\bm\vartheta}^{*}}$ in place of $H_{\bm z}^{{\bm\pi},{\bm\vartheta}_{\bm z}^{*,{\bm\pi}}}$.

\subsection{Optimal Feedback Functions under Worst Case Measure} \label{sec:worstopt}
This section derives an explicit expression for the optimal feedback functions. These are associated with the
HJB equation~\eqref{eq:HJB2}. Our objective is to find the optimal admissible feedback function ${\bm\pi}_{\bm z}^*(t)=(\pi_{i,{\bm z}}^*(t))_{i=1,\ldots,M}$, $(t,{\bm z})\in[0,T]\times{\cal S}$,
where $\pi_{i,{\bm z}}^*(t)$ is the feedback function yielding the optimal fraction of wealth invested in the $i$-th bond when the default state is ${\bm z}\in{\cal S}$. {From the
definition of admissibility, for each $(t,{\bm z})\in[0,T]\times{\cal S}$ it must hold that}
\begin{eqnarray}\label{eq:cond-optimum}
1 + \Gamma_{j,{\bm z}}^{{\bm\pi}^*}(t) >0,\ \ \ \ \ \ j=1,\ldots,M,
\end{eqnarray}
where $\Gamma_{j,{\bm z}}^{{\bm\pi}}(t)$ is given by \eqref{eq:linear}. From \eqref{eq:wealth}, it can be seen that the condition \eqref{eq:cond-optimum} guarantees that the wealth process remains positive after any occurrence of a default event.

We solve the system of first order conditions $\frac{\partial H_{\bm z}^{{\bm\pi},{\bm\vartheta}^{*}}(t,v) }{\partial\pi_i}=0$, $i=1,\ldots,M$, by using \eqref{eq:H} and \eqref{eq:werst-controlj} and obtain
\begin{eqnarray}\label{eq:foc-eq}
&&\sum_{j=1}^M(1-z_j)h_{j,{\bm z}}^{\Px}(t)B_{{\bm z}^j}(t)(1-z_i)G_{i,j,{\bm z}}(t)\big(1+\Gamma_{j,{\bm z}}^{{\bm\pi}}(t)\big)^{\gamma-1}\vartheta_{j,{\bm z}}^{*,{\bm\pi}}\nonumber\\
&&\qquad\qquad=B_{\bm z}(t)\sum_{j=1}^M(1-z_i)G_{i,j,{\bm z}}(t)(1-z_j)h_{j,{\bm z}}(t).
\end{eqnarray}
For any integer $0\leq m\leq M$, and default state ${\bm z}={\bm0}^{j_1,\ldots,j_m}$, Definition \ref{def:control-U} implies that
the feedback function $\pi_{j_1,j_1,\ldots,j_m}^*(t)=\cdots=\pi_{j_m,j_1,\ldots,j_m}^*(t)=0$.
Hence, { for $0\leq m\leq M-1$}, the
optimal strategies in the bonds which have not yet defaulted, ${\bm\pi}_{j_1,\ldots,j_m}^{*}:=(\pi^*_{j,j_1,\ldots,j_m})_{j\notin\{j_1,\ldots,j_m\}}$, is obtained from
Eq.~\eqref{eq:foc-eq} and given by
\begin{eqnarray}\label{eq:foc-eq-zj1-jm}
&&\sum_{j\notin\{j_1,\ldots,j_m\}} h_{j,j_1,\ldots,j_m}^{\Px}(t)B_{j_1,\ldots,j_m,j}(t)G_{i,j,j_1,\ldots,j_m}(t)\big(1+\Gamma_{j,j_1,\ldots,j_m}^{{\bm\pi}^*}(t)\big)^{\gamma-1}
\vartheta_{j,j_1,\ldots,j_m}^{*,{\bm\pi}^*}\nonumber\\
&&\qquad\qquad=B_{j_1,\ldots,j_m}(t)\sum_{j\notin\{j_1,\ldots,j_m\}}G_{i,j,j_1,\ldots,j_{m}}(t)h_{j,j_1,\ldots,j_m}(t),\ \ \forall\ i\notin\{j_1,\ldots,j_m\}.
\end{eqnarray}

{\small\bf Note: In the the rest of the paper, to lighten notation we omit the dependence of the above quantities on $j_1,\ldots,j_m$, i.e. on the obligors which have defaulted.
}

Our next step is to rewrite the above system in matrix form. Let $\{j_{m+1},\ldots,j_{M}\}:=\{1,\ldots,M\}\setminus\{j_1,\ldots,j_m\}$. Define the following matrices
\begin{eqnarray}\label{eq:matrix}
{\bm G}(t) :=\left[
                                  \begin{array}{cccc}
                                    G_{j_{m+1},j_{m+1}}(t) & G_{j_{m+1},j_{m+2}}(t) &\ldots & G_{j_{m+1},j_{M}}(t)\\
                                    G_{j_{m+2},j_{m+1}}(t) & G_{j_{m+2},j_{m+2}}(t) &\ldots & G_{j_{m+2},j_{M}}(t)\\
                                    \vdots & \vdots & \ddots & \vdots \\
                                    G_{j_{M},j_{m+1}}(t) & G_{j_M,j_{m+2}}(t) &\ldots & G_{j_M,j_{M}}(t)\\
                                  \end{array}
                                \right]_{(M-m)\times(M-m)},
\end{eqnarray}
and
\begin{eqnarray}\label{eq:matrix2}
{\bm A}(t) :=\left[
                                  \begin{array}{cccc}
                                    h_{j_{m+1}}^{\Px}(t)B_{j_{m+1}}(t) & 0 &\ldots &0\\
                                    0 & h_{j_{m+2}}^{\Px}(t)B_{j_{m+2}}(t) &\ldots &0\\
                                    \vdots & \vdots & \ddots & \vdots \\
                                    0 & 0 &\ldots &h_{j_{M}}^{\Px}(t)B_{j_M}(t)\\
                                  \end{array}
                                \right]_{(M-m)\times(M-m)}.
\end{eqnarray}
The matrix ${\bm G}(t)$ can be interpreted as a bond depreciation matrix. Each entry of this matrix gives the
 depreciation of a bond underwritten by an alive obligor in case another obligor defaults.
The matrix ${\bm A}(t)$, instead, can be interpreted as a matrix of default risk adjustments. Each diagonal entry scales the reference default intensity of an
alive obligor by a factor equal to the value function in an augmented default state where a new obligor defaults. We make the following assumption:
\begin{itemize}
  \item[{\bf(A1)}] For $t \in [0,T]$, the matrix ${\bm G}(t)$ has full rank.
\end{itemize}

Such an assumption means that at any time $t$ there exist no redundant bond securities. In other words, each bond security cannot be replicated via a linear combination of the others. Clearly, it is always satisfied in the absence of default contagion because in this case, for each time $t$, the matrix~\eqref{eq:matrix} would become diagonal with nonzero entries.

Next, for $(y,x)\in\R_+^2$, we define the following function
\begin{eqnarray}\label{eq:fcnY}
{\cal Y}_y(x) &:=& xe^{-yx^{\frac{\gamma}{\gamma-1}}}.
\end{eqnarray}
For fixed $y>0$, the positive function $x\too{\cal Y}_y(x)$ is smooth and increasing on $x\in\R_+$, since $\gamma\in(0,1)$. This implies that it admits an inverse function  $x\too{\cal Y}_y^{-1}(x)$, $x\in\R_+$, which is also smooth and increasing. Let $\delta:=\frac{\gamma}{1-\gamma}$. Then
\begin{eqnarray*}
{\cal Y}_y(x) =  xe^{-yx^{-\delta}}= xe^{-(y^{-1/\delta}x)^{-\delta}} = y^{1/\delta}y^{-1/\delta}xe^{-(y^{-1/\delta}x)^{-\delta}}
=y^{1/\delta}{\cal Y}_1(y^{-1/\delta}x).
\end{eqnarray*}
Hence, for fixed $x\in\R_+$ the positive function $y\too{\cal Y}_y(x)$ is also smooth on $y\in\R_+$, and further it holds that
\begin{eqnarray*}
{\cal Y}_y\big(y^{1/\delta}{\cal Y}_1^{-1}(y^{-1/\delta}x)\big) &=& y^{1/\delta}{\cal Y}_1\big(y^{-1/\delta}y^{1/\delta}{\cal Y}_1^{-1}(y^{-1/\delta}x)\big)= y^{1/\delta}{\cal Y}_1\big({\cal Y}_1^{-1}(y^{-1/\delta}x)\big)=x.
\end{eqnarray*}
Hence, we obtain the following useful relation
\begin{eqnarray}\label{eq:relation-inverse}
{\cal Y}_y^{-1}(x) &=& y^{1/\delta}{\cal Y}_{1}^{-1}\big(y^{-1/\delta}x\big).
\end{eqnarray}
This shows that for fixed $x\in\R_+$, the positive inverse function $y\too {\cal Y}_y^{-1}(x)$, $y\in\R_+$, is also smooth. We next discuss the analytic properties of the derivative of the function $x\too{\cal Y}_y^{-1}(x)$. Since ${\cal Y}_y\big({\cal Y}_y^{-1}(x)\big)=x$, application of the chain rule leads to
\begin{eqnarray}\label{eq:deri-inverse}
\frac{\partial{\cal Y}_y^{-1}(x)}{\partial x} &=& \frac{1}{\frac{\partial{\cal Y}_y(z)}{\partial z}\big|_{z={\cal Y}_y^{-1}(x)}}.
\end{eqnarray}
Hence, we deduce that $x\too\frac{\partial{\cal Y}_y^{-1}(x)}{\partial x}$, $x\in\R_+$, is continuous and $y\too\frac{\partial{\cal Y}_y^{-1}(x)}{\partial x}$, $y\in\R_+$, is also continuous.
Using \eqref{eq:werst-controlj} and \eqref{eq:fcnY}, it holds that
\begin{eqnarray}\label{eq:equation-Y}
{\cal Y}_{j}^{{\bm\pi}^*}(t) &:=& e^{\mu_{j}(t)B(t)}{\cal Y}_{\mu_{j}(t)B_{j}(t)}\big((1+\Gamma_{j}^{{\bm\pi}^*}(t))^{\gamma-1}\big)
={e^{\mu_{j}(t)B(t)}(1+\Gamma_{j}^{{\bm\pi}^*}(t))^{\gamma-1}e^{-\mu_{j}(t)B_j(t)(1+\Gamma_{j}^{{\bm\pi}^*}(t))^{\gamma}}}\nonumber\\
&=&{(1+\Gamma_{j}^{{\bm\pi}^*}(t))^{\gamma-1}e^{-\mu_{j}(t)\big[B_j(t)(1+\Gamma_{j}^{{\bm\pi}^*}(t))^{\gamma}-B(t)\big]}}
=\big(1+\Gamma_{j}^{{\bm\pi}^*}(t)\big)^{\gamma-1}\vartheta_{j}^{*,{\bm\pi}^*}(t).
\end{eqnarray}
Further, define the following matrices:
\begin{eqnarray}\label{eq:matrix3}
\boldsymbol{\mathcal{Y}}^{{\bm\pi}^*}(t) :=\left[
                                  \begin{array}{c}
                                    {\cal Y}_{j_{m+1}}^{{\bm\pi}^*}(t)\\
                                    {\cal Y}_{j_{m+2}}^{{\bm\pi}^*}(t)\\
                                    \vdots \\
                                    {\cal Y}_{j_{M}}^{{\bm\pi}^*}(t)\\
                                  \end{array}
                                \right]_{(M-m)\times1},\ \ \ \ {\rm and}\ \ \
                                {\bm B}(t) :=B(t)\left[
                                  \begin{array}{c}
                                    h_{j_{m+1}}(t)\\
                                    h_{j_{m+2}}(t)\\
                                    \vdots \\
                                    h_{j_{M}}(t)\\
                                  \end{array}
                                \right]_{(M-m)\times1}.
\end{eqnarray}
Then, we can rewrite \eqref{eq:foc-eq-zj1-jm} in the matrix form given by
\begin{eqnarray}\label{eq:matrix-eq}
{\bm G}(t){\bm A}(t)\boldsymbol{\mathcal{Y}}^{{\bm\pi}^*}(t)={\bm G}(t){\bm B}(t),\ \ \ \ \ \ \ \ t\in[0,T].
\end{eqnarray}
This leads to the following lemma.
\begin{lemma}\label{lem:Y-star}
Under Assumption {\bf(A1)}, we have $\boldsymbol{\mathcal{Y}}^{{\bm\pi}^*}(t) =  {\bm A}^{-1}(t){\bm B}(t)$, for all $t\in[0,T]$,
i.e. for each $j\in\{j_{m+1},\ldots,j_{M}\}$,
\begin{eqnarray}\label{eq:sol-Y-star}
{\cal Y}_{j}^{{\bm\pi}^*}(t)
= \frac{B(t)h_{j}(t)}{B_{j}(t)h^{\Px}_{j}(t)}. 
\end{eqnarray}
We recall that we are omitting the subscripts $j_1,\ldots,j_m$ to lighten notation.
\end{lemma}

We next use the above lemma to obtain the optimal feedback functions. From \eqref{eq:fcnY}, it can be seen that the positive smooth function $x\too{\cal Y}_{\mu_j(t)B_{j}(t)}(x)$ is increasing on $x\in\R_+$, being  $\gamma\in(0,1)$. Hence, the corresponding positive smooth inverse function $x\too{\cal Y}_{\mu_j(t)B_{j}(t)}^{-1}(x)$, $x\in\R_+$, is also increasing. Further, it follows from~\eqref{eq:fcnY} and~\eqref{eq:relation-inverse} that $\lim_{x\downarrow0}{\cal Y}_{\mu_j(t)B_{j}(t)}^{-1}(x)=0$, and $\lim_{x\uparrow+\infty}{\cal Y}_{\mu_j(t)B_{j}(t)}^{-1}(x)=+\infty$. Using \eqref{eq:sol-Y-star}, we obtain, for $j\in\{j_{m+1},\ldots,j_M\}$,
\begin{eqnarray}\label{eq:1+Gamma}
\big(1 + \Gamma_{j}^{{\bm\pi}^*}(t)\big)^{\gamma-1} &=& {\cal Y}_{\mu_j(t)B_{j}(t)}^{-1}\left( \frac{h_j(t)B(t)}{h_{j}^{\Px}(t)B_{j}(t)}e^{-\mu_j(t)B(t)}\right).
\end{eqnarray}

Next, define the following $(M-m)$-dimensional column vector of optimal feedback functions by ${\bm\pi}^*(t)=(\pi_j^*(t))_{j\in\{j_{m+1},\ldots,j_{M}\}}$,
and the following $(M-m)$-dimensional column vector
\begin{eqnarray}\label{eq:matrix-Y-invese}
\hat{\boldsymbol{\mathcal{Y}}}(t)&:=&\left[
                                             \begin{array}{c}
                                               \Big[{\cal Y}_{\mu_{j_{m+1}}(t)B_{j_{m+1}}(t)}^{-1}\big( \frac{h_{j_{m+1}}(t)B(t)}{h_{j_{m+1}}^{\Px}(t)B_{j_{m+1}}(t)}e^{-\mu_{j_{m+1}}(t)B(t)}\big)\Big]^{\frac{1}{\gamma-1}}-1\\
                                               \Big[{\cal Y}_{\mu_{j_{m+2}}(t)B_{j_{m+2}}(t)}^{-1}\big( \frac{h_{j_{m+2}}(t)B(t)}{h_{j_{m+2}}^{\Px}(t)B_{j_{m+2}}(t)}e^{-\mu_{j_{m+2}}(t)B(t)}\big)\Big]^{\frac{1}{\gamma-1}}-1\\
                                               \vdots\\
                                               \Big[{\cal Y}_{\mu_{j_{M}}(t)B_{j_{M}}(t)}^{-1}\big( \frac{h_{j_{M}}(t)B(t)}{h_{j_M}^{\Px}(t)B_{j_{M}}(t)}e^{-\mu_{j_{M}}(t)B(t)}\big)\Big]^{\frac{1}{\gamma-1}}-1\\
                                             \end{array}
                                           \right]_{(M-m)\times1}.
\end{eqnarray}
We then have the following main result.
\begin{proposition}\label{prop:optimum}
Under Assumption {\bf(A1)}, the vector of optimal feedback functions is
\begin{eqnarray}\label{eq:optima-stra}
{\bm\pi}^*(t) &=& \big({\bm G}^{-1}(t)\big)^{\top}\hat{\boldsymbol{\mathcal{Y}}}(t),\ \ \ \ \ \ t\in[0,T].
\end{eqnarray}
\end{proposition}
\noindent{\it Proof.} Plugging the expression for $\Gamma_{j,{\bm z}}^{{\bm\pi}^*}(t)$ with $j\in\{j_{m+1},\ldots,j_M\}$ given by Eq.~\eqref{eq:linear} inside Eq.~\eqref{eq:1+Gamma}, we obtain, for $t\in[0,T]$,
\[
\sum_{i=1}^M\pi_i^*G_{i,j}(t) = \left[{\cal Y}_{\mu_j(t)B_{j}(t)}^{-1}\left( \frac{h_j(t)B(t)}{h_{j}^{\Px}(t)B_{j}(t)}e^{-\mu_j(t)B(t)}\right)\right]^{\frac{1}{\gamma-1}}-1,\ \ \ \ \forall\ j\in\{j_{m+1},\ldots,j_M\}.
\]
We can then rewrite the above equations in a matrix-vector form, and recover ${\bm\pi}^*(t)$, $t\in[0,T]$ as the solution of a system of linear equations
${\bm G}^{\top}(t)\ {\bm\pi}^*(t)=\hat{\boldsymbol{\mathcal{Y}}}(t)$, where $\top$ denotes the transpose of the matrix. The result then follows using the invertibility assumption on ${\bm G}$. \hfill$\Box$

We will prove in the verification theorem that the $(M-m)$-dimensional vector ${\bm\pi}^*(t)$, $t\in[0,T]$, given by \eqref{eq:optima-stra} is indeed the optimal feedback function at time $t$
in the default state where names $j_{m+1},\ldots,j_M$ are alive and $j_1,\ldots,j_m$ defaulted.

\section{HJB Equations} \label{sec:HJB}
This section is devoted to analyze the HJB equation~\eqref{eq:HJB2}. Our goal is to establish existence and uniqueness of a classical solution.

Throughout the section, we set ${\bm\vartheta}_{\bm z}^*(t)={\bm\vartheta}_{\bm z}^{*,{\bm\pi}^*}(t)$ for notational convenience. Using \eqref{eq:H} and the expression for $\vartheta_{j,{\bm z}}^{*,{\bm\pi}}(t)$ given in Eq.~\eqref{eq:werst-controlj}, it follows that the Hamiltonian associated with the optimal feedback function and the worst-case measure is given by
\begin{eqnarray}\label{eq:H2}
H_{\bm z}^{{\bm\pi}^*,{\bm\vartheta}^*}(t,v)
&=& U(v) H_{\bm z}^{{\bm\pi}^*}(t),
\end{eqnarray}
where, for $(t,{\bm z})\in[0,T]\times{\cal S}$, the function
\begin{eqnarray}\label{eq:Hamiton-without-v}
H_{\bm z}^{{\bm\pi}^*}(t) &:=&\gamma B_{\bm z}(t)\left[r -\sum_{j=1}^M\Gamma_{j,{\bm z}}^{{\bm\pi}^*}(t)(1-z_j)h_{j,{\bm z}}(t)\right]+ \sum_{j=1}^M \frac{(1-z_j)h_{j,{\bm z}}^{\Px}(t)}{\mu_{j,{\bm z}}(t)}\big(1-\vartheta^*_{j,{\bm z}}\big).
\end{eqnarray}
Then, the HJB equation~\eqref{eq:HJB2} is equivalent to the following equation:
\begin{eqnarray}\label{eq:HJB-nov}
B_{\bm z}'(t) + H_{\bm z}^{{\bm\pi}^*}(t) = 0,\ \ \ \ t\in[0,T),\ \ {\rm and}\ \ B_{\bm z}(T)=1.
\end{eqnarray}

Next, we analyze existence and uniqueness of the solution to the default-state dependent Eq.~\eqref{eq:HJB-nov}.
We proceed inductively and prove that for some positive default-state dependent constants $\underline{\theta}_{\bm z}<\bar{\theta}_{\bm z}$, $B_{\bm z}(t) \in[\underline{\theta}_{\bm z},\bar{\theta}_{\bm z}]$, $t\in[0,T]$, is the unique global positive solution to the HJB equation \eqref{eq:HJB-nov} associated with the default state ${\bm z}$.


\begin{itemize}
  \item {\sf Base step}: $m = M$. The default state is ${\bm z}={\bf0}^{j_1,\ldots,j_M}={\bf 1}$. 
      By definition of admissible strategies, the optimal feedback function $\pi_{j}^*(t)=\pi_{j,j_1,\ldots,j_M}^*(t)=0$ for all $j=1,\ldots,M$. From~\eqref{eq:linear}, it follows that $\Gamma_j^{{\bm\pi}^*}(t)=\Gamma_{j,j_1,\ldots,j_M}^{{\bm\pi}^*}(t)=0$ for all $j=1,\ldots,M$.
      Using Eq.~\eqref{eq:Hamiton-without-v}, the HJB equation~\eqref{eq:HJB-nov} is reduced to $B_{j_1,\ldots,j_M}'(t)+\gamma rB_{j_1,\ldots,j_M}(t)=0$ on $t\in[0,T)$ and $B_{j_1,\ldots,j_M}(T)=1$. The unique solution is then given by
\begin{eqnarray}\label{eq:sol-HJBM}
 B_{j_1,\ldots,j_M}(t) = e^{\gamma r(T-t)}\in[1,e^{\gamma rT}],\ \ \ \ \ \ \ t\in[0,T].
\end{eqnarray}
Hence, our statement holds.
\item {The default state is ${\bm z}={\bm0}^{j_1,\ldots,j_{m}}$ with $m \leq M-1$. By the induction hypothesis, there exists $\underline{\theta}_{j_1,\ldots,j_m}<\bar{\theta}_{j_1,\ldots,j_m}$ such that $B_j(t)=B_{j_1,\ldots,j_{m},j}(t)\in[\underline{\theta}_{j_1,\ldots,j_m},\bar{\theta}_{j_1,\ldots,j_m}]$, $t\in[0,T]$, is the unique global positive solution
    to the HJB equation \eqref{eq:HJB-nov} associated with the default state ${\bm z}^{j}={\bm0}^{j_1,\ldots,j_{m},j}$, $j\notin\{j_1,\ldots,j_m\}$.
    Given this inductive assumption, we show the existence of a unique global positive solution to the HJB equation \eqref{eq:HJB-nov}, when the default state is ${\bm z}$.}

 First, by definition of admissibility, the optimal feedback function $\pi_{j}^*={\pi}_{j,j_1,\ldots,j_m}^*=0$ for all $j\in\{j_1,\ldots,j_m\}$, and
 hence for all $j\notin\{j_1,\ldots,j_m\}$, it holds that $\Gamma_{j,j_1,\ldots,j_m}^{{\bm\pi}^*}(t)=\sum_{i\notin\{j_1,\ldots,j_{m}\}}\pi_{i}^{*}(t)G_{i,j}(t)$. For all $j\notin\{j_1,\ldots,j_m\}$, we further have
\begin{eqnarray}
\vartheta_{j}^{*}(t)=\vartheta_{j,j_1,\ldots,j_m}^{*}(t)= \exp\Big\{-\mu_{j}(t)\big[B_{j}(t)\big(1+\Gamma_{j}^{{\bm\pi}^*}(t)\big)^{\gamma} - B(t)\big]\Big\},
\label{eq:varthjexpr}
\end{eqnarray}
where we recall that $B(t)=B_{j_1,\ldots,j_m}(t)$, since we are omitting the dependence on the defaulted obligors to lighten notation. Then the HJB equation~\eqref{eq:HJB-nov} is reduced to
\begin{eqnarray}\label{eq:hjbm1}
0 &=& B'(t) + \gamma B(t)\left(r - \sum_{j\notin\{j_1,\ldots,j_{m}\}}\Gamma_{j}^{{\bm\pi}^*}(t){h_{j}(t)}\right)\nonumber\\
&& + \sum_{j\notin\{j_1,\ldots,j_{m}\}}\frac{h_{j}^{\Px}(t)}{\mu_{j}(t)}\Big\{1-e^{-\mu_{j}(t)[B_{j}(t)(1+\Gamma_{j}^{{\bm\pi}^*}(t))^{\gamma} - B(t)]}\Big\},
\end{eqnarray}
with terminal condition $B(T)=1$.

Next, we derive an equivalent representation for Eq.~\eqref{eq:hjbm1}, which turns out to be more convenient for the analysis of the solution. First, using \eqref{eq:1+Gamma} we obtain
\begin{eqnarray}\label{eq:Gammajm}
\Gamma_{j}^{{\bm\pi}^*}(t) &=& \left[{\cal Y}_{\mu_j(t)B_{j}(t)}^{-1}\left( \frac{h_{j}(t)}{h_{j}^{\Px}(t)}\frac{B(t)}{B_{j}(t)}e^{-\mu_j(t)B(t)}\right)\right]^{\frac{1}{\gamma-1}}-1,
\end{eqnarray}
and from \eqref{eq:equation-Y}, \eqref{eq:1+Gamma} and Lemma \ref{lem:Y-star}, it follows that
\begin{eqnarray}\label{eq:vartheta=}
\vartheta_{j}^{*}(t) &=& \frac{{\cal Y}_{j}^{{\bm\pi}^*}(t)}{\big(1+\Gamma_{j}^{{\bm\pi}^*}(t)\big)^{\gamma-1}}
={\frac{h_{j}(t)}{h_{j}^{\Px}(t)}} \frac{\frac{B(t)}{B_{j}(t)}}{{\cal Y}_{\mu_j(t)B_{j}(t)}^{-1}\Big( \frac{h_{j}(t)B(t)}{h_{j}^{\Px}(t)B_{j}(t)}e^{-\mu_j(t)B(t)}\Big)}.
\end{eqnarray}
Thus, the Hamiltonian $H^{{\bm\pi}^*}(t)$ given by \eqref{eq:Hamiton-without-v} can be rewritten in the following form (notice that, as stated earlier, we are omitting the dependence on ${\bm z}$):
\begin{eqnarray}\label{eq:Hpsi-star}
H^{{\bm\pi}^*}(t) &=& B(t)\left[\gamma\bigg(r + \sum_{j\notin\{j_1,\ldots,j_m\}}h_j(t)\bigg) - \sum_{j\notin\{j_1,\ldots,j_m\}} C_j\big(t,\mu_j(t)B_j(t);B(t)\big)\right]\nonumber\\
&& + \sum_{j\notin\{j_1,\ldots,j_m\}}\frac{h_j^{\Px}(t)}{\mu_{j}(t)},
\end{eqnarray}
where, the function $C_j(t,y;x)$, $j\notin\{j_1,\ldots,j_m\}$, on $(t,y,x)\in[0,T]\times\R_+^2$ is defined as
\begin{eqnarray}\label{eq:Cj}
C_j(t,y;x) &:=& \gamma {h_j(t)}\left[{\cal Y}_{y}^{-1}\left(\frac{\mu_j(t)h_{j}(t)}{yh_{j}^{\Px}(t)}{\cal J}\big(\mu_j(t);x\big)\right)\right]^{\frac{1}{\gamma-1}}\nonumber\\
&&+\frac{h_j(t)}{y}\left[{\cal Y}_{y}^{-1}\left(\frac{\mu_j(t)h_{j}(t)}{yh_{j}^{\Px}(t)}{\cal J}\big(\mu_j(t);x\big)\right)\right]^{-1},
\end{eqnarray}
with the function ${\cal J}(a;x):=xe^{-ax}$ for $(a,x)\in\R_+^2$. Then we obtain the following equation which is equivalent to the original HJB equation \eqref{eq:hjbm1}:
\begin{eqnarray}\label{eq:hjbm2}
0 &=& B'(t) + {B(t)}\left[\gamma\bigg(r + \sum_{j\notin\{j_1,\ldots,j_m\}}{h_j(t)}\bigg) - \sum_{j\notin\{j_1,\ldots,j_m\}} C_j\big(t,\mu_j(t)B_j(t);B(t)\big)\right]\nonumber\\
&& + \sum_{j\notin\{j_1,\ldots,j_m\}}\frac{h_j^{\Px}(t)}{\mu_{j}(t)}.
\end{eqnarray}
Given any positive continuous function $f(t)$ on $t\in[0,T]$ which admits a strictly positive lower bound, set
\begin{equation}
m_T^f:=\inf_{t\in[0,T]}f(t), \qquad \qquad M_T^f:=\sup_{t\in[0,T]}f(t).
\label{eq:bounds}
\end{equation}
Then $0< m_T^f\leq M_T^f<+\infty$. Existence of a classical solution to the HJB equation is then proved in two main steps. We first show that if a solution exists, then it must be bounded. We then use the established lower and upper bounds to show existence and uniqueness of a solution to the HJB equation.

\begin{proposition}\label{prop:bound-sol-hjb}
If Eq.~\eqref{eq:hjbm2} admits a unique global solution $B(t)$, $t\in[0,T]$, then there exist two positive constants $\underline{\theta}<\bar{\theta}$ so that $B(t)\in[\underline{\theta},\bar{\theta}]$ for all $t\in[0,T]$.
\end{proposition}

\noindent{\it Proof.}\quad First we notice that for fixed $a\in\R_+$, the smooth function ${\cal J}(a;x)=xe^{-ax}$ with $x\in\R_+$ admits $\sup_{x\in\R_+}{\cal J}(a;x)={\cal J}(a;a^{-1})=a^{-1}e^{-1}$ and $\lim_{x\downarrow0}{\cal J}(a,x)=\lim_{x\uparrow+\infty}{\cal J}(a,x)=0$. Recall that the the positive inverse function $x\too{\cal Y}_{\mu_j(t)B_j(t)}^{-1}(x)$ is $C^1$ and increasing. 
Then, for fixed  $(t,y)\in[0,T]\times(0,+\infty)$, and for all $x\in\R_+$,
\begin{eqnarray*}
&&\left[{\cal Y}_{\mu_j(t)B_j(t)}^{-1}\left(\frac{h_{j}(t)}{h_{j}^{\Px}(t)B_j(t)}{\cal J}\big(\mu_j(t);x\big)\right)\right]^{-1}
\geq\left[{\cal Y}_{\mu_j(t)B_j(t)}^{-1}\left(\frac{h_{j}(t)}{h_{j}^{\Px}(t)B_j(t)}{\cal J}\big(\mu_j(t);\mu_j^{-1}(t)\big)\right)\right]^{-1}\nonumber\\
&&\qquad\quad\geq\left[{\cal Y}_{\mu_j(t)B_j(t)}^{-1}\left(e^{-1}\sup_{t\in[0,T]}\Big[ \frac{h_{j}(t)}{h_{j}^{\Px}(t)\mu_j(t)B_j(t)}\Big]\right)\right]^{-1}=:\kappa_T\big(\mu_j(t)B_j(t)\big),
\end{eqnarray*}
where we used the fact that $\frac{h_{j}(t)}{h_{j}^{\Px}(t)B_j(t)}{\cal J}\big(\mu_j(t);\mu_j^{-1}(t)\big)=e^{-1} \frac{h_{j}(t)}{h_{j}^{\Px}(t)\mu_j(t)B_j(t)}$. Moreover,
the quantity $\kappa_T$ is finite since, recalling the notation introduced in~\eqref{eq:bounds}, we have
\[
0<\frac{m_T^{{h_{j}}/{h_{j}^{\Px}}}}{M_T^{\mu_j} M_T^{B_j}}\leq \sup_{t\in[0,T]}\left[{\frac{h_{j}(t)}{h_{j}^{\Px}(t)}} \frac{1}{\mu_j(t)B_j(t)}\right]\leq\frac{M_T^{{h_{j}}/{h_{j}^{\Px}}}}{m_T^{\mu_j} m_T^{B_j}}<+\infty.
\]
Using the above given lower bound, we obtain the following lower bound for all $x\in\R_+$,
\begin{eqnarray}\label{eq:Cj2}
C_j\big(t,\mu_j(t)B_j(t);x\big) &\geq& \gamma {h_j(t)}\left(\kappa_T \big(\mu_j(t)B_j(t)\big)\right)^{\frac{1}{1-\gamma}}+\frac{h_j(t)}{\mu_j(t)B_j(t)}\kappa_T\big(\mu_j(t)B_j(t)\big).
\end{eqnarray}
Define the function
\begin{eqnarray}\label{eq:ine-barDm}
D_m\big(t,\mu_j(t)B_j(t),x\big)&:=&\gamma\bigg(r + \sum_{j\notin\{j_1,\ldots,j_m\}}{h_j(t)}\bigg) - \sum_{j\notin\{j_1,\ldots,j_m\}} C_j\big(t,\mu_j(t)B_j(t);x\big).
\end{eqnarray}
It then holds that
\begin{eqnarray}\label{eq:ine-barDmbounds}
D_m\big(t,\mu_j(t)B_j(t),x\big)
&\leq& \gamma\bigg(r + \sum_{j\notin\{j_1,\ldots,j_m\}}{h_j(t)}\bigg) - \gamma \sum_{j\notin\{j_1,\ldots,j_m\}} {h_j(t)} \left(\kappa_T\big(\mu_j(t)B_j(t)\big) \right)^{\frac{1}{1-\gamma}}\nonumber\\
&&-\sum_{j\notin\{j_1,\ldots,j_m\}}\frac{h_j(t)}{\mu_j(t)B_j(t)}\kappa_T\big(\mu_j(t)B_j(t)\big)=:\bar{D}_m(t).
\end{eqnarray}
Notice that the {strictly positive} function $y\too\kappa_T(y)$ is continuous in $y>0$ since $y\too{\cal Y}_y^{-1}(x)$ is continuous for fixed $x\in\R_+$ using \eqref{eq:relation-inverse}. Then
\begin{eqnarray*}
M_T^{\bar{D}_m} &\leq& \gamma\bigg(r + \sum_{j\notin\{j_1,\ldots,j_m\}}M_T^{h_j}\bigg)- \gamma \sum_{j\notin\{j_1,\ldots,j_m\}} m_T^{h_j}\left\{\inf_{y\in[m_T^{\mu_j}m_T^{B_j},M_T^{\mu_j}M_T^{B_j}]}\kappa_T(y)\right\}^{\frac{1}{1-\gamma}}\nonumber\\
&&-\sum_{j\notin\{j_1,\ldots,j_m\}}\frac{m_T^{h_j^{\Px}}m_T^{{h_{j}}/{h_{j}^{\Px}}}}{M_T^{\mu_j}M_T^{B_j}}
\left\{\inf_{y\in[m_T^{\mu_j}m_T^{B_j},M_T^{\mu_j}M_T^{B_j}]}\kappa_T(y)\right\}<+\infty,
\end{eqnarray*}
since $\gamma\in(0,1)$. Using the integral representation of the solution to Eq.~\eqref{eq:hjbm2} and the inequality \eqref{eq:ine-barDm}, it follows that
\begin{eqnarray}\label{eq:bar-theta}
B(t) &=& e^{\int_t^TD_m(u,\mu_j(u)B_j(u),B(u))\D u} +  \sum_{j\notin\{j_1,\ldots,j_m\}}\int_t^T\frac{h_j^{\Px}(s)}{\mu_{j}(s)}e^{\int_t^sD_m(u,\mu_j(u)B_j(u),B(u))\D u}\D s\nonumber\\
&\leq& e^{M_T^{\bar{D}_m}(T-t)} +  \sum_{j\notin\{j_1,\ldots,j_m\}}\int_t^T\frac{h_j^{\Px}(s)}{\mu_{j}(s)}e^{M_T^{\bar{D}_m}(s-t)}\D s\nonumber\\
&\leq&  e^{M_T^{\bar{D}_m}(T-t)}\Bigg[1+ (M-m)(T-t)\frac{M_T^{h_j^{\Px}}}{m_T^{\mu_{j}}}\Bigg]\leq e^{TM_T^{\bar{D}_m}}\Bigg[1+ TM\frac{M_T^{h_j^{\Px}}}{m_T^{\mu_{j}}}\Bigg]=:\bar{\theta},
\end{eqnarray}
where the constant {$\bar{\theta} \in\R_+$}. Above, $M-m$ is the number of obligors which are alive. It is also clear from the first equality in the above array of equations that $B(t)>0$ for all $t\in[0,T]$. This is because $h_j^{\Px}(t)$ and $\mu_j(t)$ are all strictly positive on $t\in[0,T]$.

Next, we use the above established upper bound $\bar{\theta}>0$ to conclude the existence of a positive lower bound for the solution $B(t)$, $t\in[0,T]$. 
Recall the definition of $\Gamma_{j}^{{\bm\pi}}(t)$ given in Eq.~\eqref{eq:linear}. We choose an admissible control $\hat{\bm\pi}=(\hat{\pi}_i(t,B(t));\ i\in\{1,\ldots,M\})$ which satisfies the following relation for $i\notin\{j_1,\ldots,j_m\}$:
\begin{eqnarray}\label{eq:Gammahat}
\Gamma_{j}^{\hat{{\bm\pi}}}(t)=\Gamma_{j}^{\hat{{\bm\pi}}}(t,B(t)):=\sum_{i\notin\{j_1,\ldots,j_m\}}\hat{\pi}_i(t,B(t))G_{i,j}(t)
=\left(\frac{B(t)}{B_j(t)}\right)^{\frac{1}{\gamma}}-1,\ \ j\notin\{j_1,\ldots,j_m\}.
\end{eqnarray}
We define $\hat{\pi}_i(t,B(t)))=0$ for $i\in\{j_1,\ldots,j_m\}$. Notice that the existence of such an admissible control is guaranteed by Assumption {\bf (A1)} on the invertibility of the matrix $\bm{G}(t)$, $t\in[0,T]$, given in Eq.~\eqref{eq:matrix}. In fact, this admissible control is given by
$\hat{\bm\pi}=[\hat{\pi}_i;\ i\notin\{j_1,\ldots,j_m\}]^{\top}=({\bm G}^{-1}(t))^{\top}[(B(t)/B_j(t))^{1/\gamma}-1;\ j \notin \{j_{1},..,j_m\}]^{\top}$.
Rearranging terms in Eq.~\eqref{eq:Gammahat}, we obtain
\begin{eqnarray*}
B_{j}(t)\big(1+\Gamma_{j}^{\hat{\bm\pi}}(t)\big)^{\gamma} - B(t) =0,\ \ \ \ j\notin\{j_1,\ldots,j_m\}.
\end{eqnarray*}
The above equation directly implies that
\begin{eqnarray}\label{eq:hat0}
\sum_{j\notin\{j_1,\ldots,j_{m}\}}\frac{h_{j}^{\Px}(t)}{\mu_{j}(t)}\Big\{1-e^{-\mu_{j}(t)\big[B_{j}(t)(1+\Gamma_{j}^{\hat{\bm\pi}}(t))^{\gamma} - B(t)\big]}\Big\}=0.
\end{eqnarray}
Using that $B(t)>0$ for all $t\in[0,T]$, and recalling the expression for $\vartheta^*_j(t)$ given in Eq.~\eqref{eq:varthjexpr}, the Hamiltonian \eqref{eq:Hamiton-without-v} satisfies
\begin{eqnarray}\label{eq:ine-H}
H^{{\bm\pi}^*}(t)& =&\gamma B(t)\left(r - \sum_{j\notin\{j_1,\ldots,j_{m}\}}\Gamma_{j}^{{\bm\pi}^*}(t){h_j(t)}\right) + \sum_{j\notin\{j_1,\ldots,j_{m}\}}\frac{h_{j}^{\Px}(t)}{\mu_{j}(t)}\Big\{1-e^{-\mu_{j}(t)\big[B_{j}(t)(1+\Gamma_{j}^{{\bm\pi}^*}(t))^{\gamma} - B(t)\big]}\Big\}\nonumber\\
&\geq&H^{\hat{{\bm\pi}}}(t)\nonumber\\
&=&{\gamma B(t)\left(r - \sum_{j\notin\{j_1,\ldots,j_{m}\}}\Gamma_{j}^{\hat{\bm\pi}}(t){h_j(t)}\right) + \sum_{j\notin\{j_1,\ldots,j_{m}\}}\frac{h_{j}^{\Px}(t)}{\mu_{j}(t)}\Big\{1-e^{-\mu_{j}(t)\big[B_{j}(t)(1+\Gamma_{j}^{\hat{\bm\pi}}(t))^{\gamma} - B(t)\big]}\Big\}}\nonumber\\
&=&\gamma B(t)\left(r - \sum_{j\notin\{j_1,\ldots,j_{m}\}}\Gamma_{j}^{\hat{\bm\pi}}(t){h_j(t)}\right)
=\gamma B(t)\left\{r - \sum_{j\notin\{j_1,\ldots,j_{m}\}}\left[\left(\frac{B(t)}{B_j(t)}\right)^{\frac{1}{\gamma}}-1\right]{ h_j(t)}\right\}\nonumber\\
&\geq&\gamma B(t)\left[r +\sum_{j\notin\{j_1,\ldots,j_{m}\}}{h_j(t)}- \sum_{j\notin\{j_1,\ldots,j_{m}\}}\left(\frac{\bar{\theta}}{B_j(t)}\right)^{\frac{1}{\gamma}}{h_j(t)}\right]\nonumber\\
&\geq&\gamma B(t)\left[r+\sum_{j\notin\{j_1,\ldots,j_{m}\}}m_T^{h_j}- \sum_{j\notin\{j_1,\ldots,j_{m}\}}\left(\frac{\bar{\theta}}{m_T^{B_j}}\right)^{\frac{1}{\gamma}}M_T^{h_{j}}\right]  =:B(t)K_T.
\end{eqnarray}
Above, the third equality is obtained using~\eqref{eq:hat0}. The first inequality follows because the Hamiltonian achieves its maximum value at the optimum ${\bm\pi}^*$.
The second inequality follows from the upper bound for $B(t)$ established in Eq.~\eqref{eq:bar-theta}. The last inequality follows from the the fact that, by definition~\eqref{eq:bounds}, we have $m_T^{h_j} \leq h_j(t) \leq M_T^{h_{j}}$ and $B_j(t) \leq M_T^{B_{j}}$ for all $t\in[0,T]$.

Let $b\in(0,1]$ be an arbitrary constant. Consider the following ODE:
\begin{eqnarray*}
v'(t) + v(t)K_T = 0,\ \ \ \ t\in[0,T),\ \ \ \ \ v(T)=b.
\end{eqnarray*}
Using the comparison theorem of ODEs along with the inequality \eqref{eq:ine-H}, it follows that the solution to Eq.~\eqref{eq:hjbm1} is lower bounded by the solution of the above ODE, i.e., $B(t)\geq v(t) = b e^{K_T(T-t)}$ for $t\in[0,T]$. Further, define the positive constant
\begin{eqnarray}\label{eq:underline-theta}
\underline{\theta}:=\left\{ \begin{array}{cc}
                       b, & \mbox{if}\ \ K_T\geq0,\\
                       be^{K_TT}, & \mbox{if}\ \ K_T<0.
                     \end{array}\right.
\end{eqnarray}
Then we obtain $B(t)\geq \underline{\theta}$ for all $t\in[0,T]$. This completes the proof of the proposition. \hfill$\Box$

Using the above established lower and upper bounds, we can prove the main theorem which ensures existence and uniqueness of a solution to our HJB equation.
\begin{theorem}\label{thm:hjb}
There exists a unique solution $B(t)$, $t\in[0,T]$, to the HJB equation~\eqref{eq:hjbm2} satisfying $B(t)\in[\underline{\theta},\bar{\theta}]$ for all $t\in[0,T]$. {We recall that the positive constants $\underline{\theta}$ and $\bar{\theta}$ have been given in Proposition \ref{prop:bound-sol-hjb},
see equations~\eqref{eq:underline-theta} and \eqref{eq:bar-theta} therein.}
\end{theorem}

\noindent{\it Proof.}\quad
We first consider the following truncated HJB equation:
\begin{eqnarray}\label{eq:trunc-hjbm2}
0 &=& B_{\theta}'(t) + \sum_{j\notin\{j_1,\ldots,j_m\}}C_j^{(\theta)}\big(t,\mu_j(t)B_j(t);B_{\theta}(t)\big) + \sum_{j\notin\{j_1,\ldots,j_m\}}\frac{h_j^{\Px}(t)}{\mu_{j}(t)},
\end{eqnarray}
with $B_{\theta}(T)=1$. For all $j\notin\{j_1,\ldots,j_m\}$ and $(t,x)\in[0,T]\times\R_+$, we define the function
\begin{eqnarray*}
C_j^{(\theta)}\big(t,\mu_j(t)B_j(t);x\big) &:=& x\gamma\bigg(r + \sum_{j\notin\{j_1,\ldots,j_m\}}h_j(t)\bigg)- \sum_{j\notin\{j_1,\ldots,j_m\}} \big(\underline{\theta}\vee x\wedge\bar{\theta}\big)C_j\big(t,\mu_j(t)B_j(t);\underline{\theta}\vee x\wedge\bar{\theta}\big).
\end{eqnarray*}
Using \eqref{eq:relation-inverse} and \eqref{eq:deri-inverse}, and recalling the expression~\eqref{eq:Cj}, it can be easily verified that the function $x\too C_j^{(\theta)}\big(t,\mu_j(t)B_j(t);x\big)$, $x\in\R_+$, is Lipschitz continuous uniformly in $t\in[0,T]$. This implies existence and uniqueness of the solution to Eq.~\eqref{eq:trunc-hjbm2}, since the function $t\too\sum_{j\notin\{j_1,\ldots,j_m\}}\frac{h_j^{\Px}(t)}{\mu_{j}(t)}$ is continuous and bounded on $t\in[0,T]$. By Proposition~\ref{prop:bound-sol-hjb}, $B_{\theta}(t)\in[\underline{\theta},\bar{\theta}]$ for $t\in[0,T]$. This yields that $C_j^{(\theta)}\big(t,\mu_j(t)B_j(t);B_{\theta}(t)\big)=C_j\big(t,\mu_j(t)B_j(t);B_{\theta}(t)\big)$. Then the uniqueness of the solution of Eq.~ \eqref{eq:trunc-hjbm2} implies that $B(t)=B_{\theta}(t)$ is the unique solution of the HJB equation~\eqref{eq:hjbm2}. \hfill$\Box$

\end{itemize}

\section{Verification Theorem} \label{sec:verification}

In this section, we show that the optimal feedback function is given by Eq.~\eqref{eq:optima-stra}. 
Notice that the optimal feedback function immediately yields the optimal bond investment strategy in light of Eq.~\eqref{eq:pit-}.
We also show that the value function associated with the control problem is given by the product $U(v) B_{\bm z}^{*}(t)$, $(v,t,{\bm z})\in\R_+\times[0,T]\times{\cal S}$, where $B_{\bm z}^{*}(t)$, $(t,{\bm z})\in[0,T]\times{\cal S}$, is the unique classical solution of Eq.~\eqref{eq:HJB-nov}.

\begin{theorem}\label{thm:verifi-thm}
For $t\in[0,T]$, let the default state ${\bm Z}(t-)={\bm0}^{j_1,\ldots,j_m}$, where $0\leq m\leq M$. Here, $j_1,\ldots,j_m$ denote $m$ distinct obligors on which risky bonds are underwritten. We then have
\begin{itemize}
  \item If $m=M$ (i.e., all obligors have defaulted), the time $t$-optimal strategy in risky bonds is given by $\tilde{\pi}_1^*(t)=\cdots=\tilde{\pi}_M^*(t)=0$.
      The corresponding value function is given by $w_{\bm 1}(t,v)=U(v)e^{\gamma r(T-t)}$ for $(t,v)\in[0,T]\times\R_+$.
  \item If $0\leq m\leq M-1$, let $(\pi_{j,j_1,\ldots,j_m}^{*}(t),B^*_{j_1,\ldots,j_m}(t))_{j\notin\{j_1,\ldots,j_m\}}$, $t\in[0,T]$, be given by \eqref{eq:optima-stra} and by the unique positive bounded solution to Eq.~\eqref{eq:hjbm2} respectively. Then the following holds:
      \begin{enumerate}
      \item The time $t$-optimal strategy in each risky bond is given by $\tilde{\pi}^*_j(t)=0$ for $j\in\{j_1,\ldots,j_m\}$, and $\tilde{\pi}_{j}^*(t)=\pi_{j,j_1,\ldots,j_m}^{*}(t)$ for $j\notin\{j_1,\ldots,j_m\}$.
      \item The time $t$-worst-case measure {corresponding to the optimal feedback function ${\bm\pi}^*$} $\vartheta_{j,j_1,\ldots,j_m}^{*,{\bm\pi}^*}(t):=\vartheta_{j,{\bm0}^{j_1,\ldots,j_m}}^{*,{\bm\pi}^*}(t)$ is given by \eqref{eq:werst-controlj}, where ${\bm\pi}^*=({\pi}_{j,j_1,\ldots,j_m}^*(t), t\in[0,T];\ j\notin\{j_1,\ldots,j_m\})$ has been specified above.
      \item  The value function associated with the robust optimization criterion~\eqref{eq:robust-problem} is given by $w_{j_1,\ldots,j_m}(t,v):=w_{{\bf0}^{j_1,\ldots,j_m}}(t,v)=U(v)B^*_{j_1,\ldots,j_m}(t)$ for $(t,v)\in[0,T]\times\R_+$.
    \end{enumerate}
\end{itemize}
\end{theorem}

\noindent{\it Proof.}\quad Recall the Hamiltonian given by~\eqref{eq:H}. Then, given the default state ${\bm Z}(t-)={\bm z}={\bm0}^{j_1,\ldots,j_m}$,  $0\leq m\leq M-1$, $(j,n)\notin\{j_1,\ldots,j_m\} \times \{j_1,\ldots,j_m\}$, and for a fixed admissible feedback function ${\bm\pi}$ satisfying \eqref{eq:cond-optimum}, it holds that
\begin{eqnarray}\label{eq:der-foc-varthetaj}
H_{j,n}^{{\bm\pi},{\bm\vartheta}}(t,v):=\frac{\partial^2 H^{{\bm\pi},{\bm\vartheta}}(t,v)}{\partial\vartheta_j\partial\vartheta_n}=0,\ \ {\rm if}\ j\neq n,\ {\rm and}\ \ H_{j,j}^{{\bm\pi},{\bm\vartheta}}(t,v)=U(v)\frac{h_j^{\Px}(t)}{\mu_j(t)}\vartheta_j^{-1}>0,
\end{eqnarray}
since $\vartheta_j>0$. {Hence $H^{{\bm\pi},{\bm\vartheta}}(t,v)$ is concave in ${\bm\vartheta}$}, and ${\bm\vartheta}^{*,{\bm\pi}}=(\vartheta_j^{*,{\bm\pi}};\ j\notin\{j_1,\ldots,j_m\})$ given by \eqref{eq:werst-controlj} is the worst-case measure {corresponding to ${\bm\pi}$}, {i.e.,
\begin{align}\label{eq:ine1}
H^{{\bm\pi},{\bm\vartheta}^{*,{\bm\pi}}}(t,v)\leq H^{{\bm\pi},{\bm\vartheta}}(t,v),\ \ \ \ \ \ {\rm for\ all}\ ({\bm\pi},{\bm\vartheta})\in{\cal U}_0\times{\cal V}_0.
\end{align}}
Fix ${\bm\vartheta}^{*,{\bm\pi}}$. Then, for the default state ${\bm z}={\bf0}^{j_1,\ldots,j_m}$ with $0\leq m\leq M-1$,  $(i,k)\notin\{j_1,\ldots,j_m\}^2$, and ${\bm\pi}$ satisfying \eqref{eq:cond-optimum}, one has
\begin{eqnarray}\label{eq:der-foc-psi}
H_{i,k}^{{\bm\pi}}(t,v)&:=&\frac{\partial^2 H^{{\bm\pi},{\bm\vartheta}^{*,{\bm\pi}}}(t,v) }{\partial\pi_i\partial\pi_k}
=\sum_{j\notin\{j_1,\ldots,j_m\}}\ell_{j}(t,v)G_{i,j}(t)G_{k,j}(t),
\end{eqnarray}
where for all $j\notin\{j_1,\ldots,j_m\}$ and $(t,v)\in[0,T]\times(0,+\infty)$, we define
\begin{eqnarray*}
\ell_{j}(t,v) &:=&-\gamma U(v)h_{j}^{\Px}(t)B_{j}(t)\big(1+\Gamma_{j}^{{\bm\pi}}(t)\big)^{\gamma-2}\vartheta_{j}^{*,{\bm\pi}}\Big[(1-\gamma)+\gamma \mu_j(t) B_j(t)\big(1+\Gamma_{j}^{{\bm\pi}}(t)\big)^{\gamma}\Big]<0.
\end{eqnarray*}
Notice that $\ell_{j}(t,v)$ is negative in the whole domain, using \eqref{eq:cond-optimum} and the fact that $\gamma\in(0,1)$. Hence, the Hessian matrix of the Hamiltonian $H^{{\bm\pi},{\bm\vartheta}^{*,{\bm\pi}}}(t,v)$ in ${\bm\pi}$ is given by
\begin{eqnarray}\label{eq:Hessian-matrix}
{\bm H}^{{\bm\pi}}(t,v) &=& {\bm G}(t) {\bm\ell}(t,v) {\bm G}^{\top}(t),\ \ \ \ \ \   (t,v)\in[0,T]\times(0,+\infty),
\end{eqnarray}
where the $(M-m)\times(M-m)$-dimensional matrix ${\bm G}(t)$ is given by \eqref{eq:matrix}, and the $(M-m)\times(M-m)$-dimensional matrix
\begin{eqnarray}\label{eq:L}
{\bm\ell}(t,v) &:=& \left[
                                  \begin{array}{cccc}
                                    \ell_{j_{m+1}}(t,v) & 0 &\ldots &0\\
                                    0 & \ell_{j_{m+2}}(t,v) &\ldots &0\\
                                    \vdots & \vdots & \ddots & \vdots \\
                                    0 & 0 &\ldots &\ell_{j_{M}}(t,v)\\
                                  \end{array}
                                \right]_{(M-m)\times(M-m)}.
\end{eqnarray}
Using \eqref{eq:Hessian-matrix}, for every non-zero row vector ${\bm x}$ consisting of $M-m$ real components, we obtain
\begin{eqnarray}\label{eq:less0}
{\bm x}{\bm H}^{{\bm\pi}}(t,v){\bm x}^{\top} &=& {\bm x}{\bm G}(t) {\bm \ell}(t,v) {\bm G}^{\top}(t){\bm x}^{\top}=\sum_{j\notin\{j_1,\ldots,j_m\}}\ell_j(t,v)\left(\sum_{k\notin\{j_1,\ldots,j_m\}}x_k G_{k,j}(t)\right)^2<0,
\end{eqnarray}
since $\ell_j(t,v)<0$ for all $j\notin\{j_1,\ldots,j_m\}$. This shows that for fixed $(t,v)\in[0,T]\times(0,\infty)$, the Hessian ${\bm H}^{{\bm\pi}}(t,v)$ is negative definite for all feedback functions ${\bm\pi}$ satisfying \eqref{eq:cond-optimum}. Hence, ${\bm\pi}^*=(\pi_{j}^*)_{j\notin\{j_1,\ldots,j_m\}}$ obtained from the first order conditions~\eqref{eq:foc-eq-zj1-jm} is the optimum, the value of ${\bm\pi}$ at which $H^{{\bm\pi},{\bm\vartheta}^{*,{\bm\pi}}}(t,v)$ achieves the maximum, {i.e.,
\begin{align}\label{eq:ine2}
H^{{\bm\pi},{\bm\vartheta}^{*,{\bm\pi}}}(t,v)\leq H^{{\bm\pi}^*,{\bm\vartheta}^{*,{\bm\pi}^*}}(t,v),\ \ \ \ \ \ {\rm for\ all}\ {\bm\pi}\in{\cal U}_0.
\end{align}
By the inequality~\eqref{eq:ine1} and take ${\bm\pi}={\bm\pi}^*$ therein, we obtain $H^{{\bm\pi}^*,{\bm\vartheta}^{*,{\bm\pi}^*}}(t,v)\leq H^{{\bm\pi}^*,{\bm\vartheta}}(t,v)$ for all ${\bm\vartheta}\in{\cal V}_0$. Further, using the inequality \eqref{eq:ine2}, we obtain $H^{{\bm\pi},{\bm\vartheta}^{*,{\bm\pi}}}(t,v)\leq H^{{\bm\pi}^*,{\bm\vartheta}^{*,{\bm\pi}^*}}(t,v)\leq H^{{\bm\pi}^*,{\bm\vartheta}}(t,v)$ for all $({\bm\pi},{\bm\vartheta})\in{\cal U}_0\times{\cal V}_0$. This implies that ${\bm\pi}^*$ is the optimal feedback function and ${\bm\vartheta}^{*,{\bm\pi}^*}$ is the worse-case measure corresponding to ${\bm\pi}^*$.}


{Recall the notation ${\bm\vartheta}^*:={\bm\vartheta}^{*,{\bm\pi}^*}$ introduced at the beginning of section~\ref{sec:HJB}.}
For $u\in[t,T]$, define the process $Y_u^{\tilde{\bm\pi}^*} := U(V_u^{\tilde{\bm\pi}^*})B_{{\bm Z}(u)}^*(u)$, where {$\tilde{\bm\pi}^*=(\tilde{\pi}^*_i)_{i=1,\ldots,M}$ with $\tilde{\pi}^*_i(u)={\pi}^*_{i,{\bm Z}(u-)}(u)$, $u\in[t,T]$,} and $V_u^{\tilde{\bm\pi}^*}$ is the optimally controlled wealth process satisfying the dynamics \eqref{eq:wealth}.
Applying It\^o's formula, for $u\in[t,T]$ we obtain
\begin{eqnarray}\label{eq:itoY}
Y_{u}^{\tilde{\bm\pi}^*} = Y_{t}^{\tilde{\bm\pi}^*} +\int_t^{u}R_{{\bm Z}(s)}^{\tilde{\bm\pi}^*}(s,V_s^{\tilde{\bm\pi}^*})\D s + {M}_{u}^{\tilde{\bm\pi}^*} - {M}_{t}^{\tilde{\bm\pi}^*},
\end{eqnarray}
where, for $(u,v)\in[t,T]\times\R_+$, $R_{\bm z}^{\tilde{\bm\pi}^*}(s,v) := U(v) B_{\bm z}^{*'}(s) + {\cal L}^{{\bm\pi}^*,{\bm\vartheta}^*}(U(v)B_{\bm z}^*(s))$.
The operator ${\cal L}^{{\bm\pi}^*,{\bm\vartheta}^*}:= {\cal L}_c^{{\bm\pi}^*}+{\cal L}_J^{{\bm\pi}^*,{\bm\vartheta}^*}$, where ${\cal L}_c^{{\bm\pi}^*}$ and ${\cal L}_J^{{\bm\pi}^*,{\bm\vartheta}^*}$ are defined by \eqref{eq:operatorsL} with $( {\bm \pi},{\bm\vartheta})$ replaced by the optimum $({\bm\pi}^*,{\bm\vartheta}^*)$. Moreover, the $\tilde{\Px}$-(local) martingale ${M}_u^{\tilde{\bm\pi}^*}$, $u\in[t,T]$, is given by
\begin{eqnarray}\label{eq:localmart-P}
{M}_u^{\tilde{\bm\pi}^*} &:=& \sum_{j=1}^M\int_0^u U(V_{s-}^{\tilde{\bm\pi}^*})\Big[B^*_{{\bm Z}^{j}(s-)}(s)\big(1+\Gamma_{j,{\bm Z}(s-)}^{{\bm\pi}^*}(s)\big)^{\gamma}-B^*_{{\bm Z}(s-)}(s)\Big]\D\xi_j^{\tilde{\Px}}(s).
\end{eqnarray}

Notice that $B_{\bm z}^*(t)$ satisfies \eqref{eq:HJB-nov}. Then
\[
R_{\bm z}^{\tilde{\bm\pi}^*}(s,v) = - U(v)\sum_{j=1}^M\frac{(1-z_j)\big[\vartheta^*_{j,{\bm z}}(s)\log(\vartheta^*_{j,{\bm z}}(s))-\vartheta^*_{j,{\bm z}}(s)+1\big]h_{j,{\bm z}}^{\Px}(s)}{\mu_{j,{\bm z}}(s)},
\]
where we set $\vartheta^*_{j,{\bm z}}(s):=\vartheta^{*,{\bm\pi}^*}_{j,{\bm z}}(s)$, given by \eqref{eq:werst-controlj} with $({\bm\pi},B_{{\bm z}^j}(t),B_{\bm z}(t))$ replaced by $({\bm\pi}^*,B^*_{{\bm z}^j}(t),B^*_{\bm z}(t))$. Recall the notation $\Ex^{\tilde{\Px}}_t[\cdot]:=\Ex^{\tilde{\Px}}[\cdot|\G_t]$ introduced earlier. Then
\begin{eqnarray*}
&& \Ex_t^{\tilde{\Px}}\left[Y_{u}^{\tilde{\bm\pi}^*}+\sum_{j=1}^M\int_t^uU\big(V_s^{\tilde{\bm\pi}^*}\big)\frac{(1-Z_j(s))\big[\vartheta^*_{j,{\bm Z}(s)}(s)\log(\vartheta^*_{j,{\bm Z}(s)}(s))-\vartheta^*_{j,{\bm Z}(s)}(s)+1\big]h_{j,{\bm Z}(s)}^{\Px}(s)}{\mu_{j,{\bm Z}(s)}(s)}\D s\right]\nonumber\\
 &&\qquad\qquad = U\big(V_t^{\tilde{\bm\pi}^*}\big)B^*_{{\bm Z}(t)}(t) + \Ex_t^{\tilde{\Px}}\left[{M}_{u}^{\tilde{\bm\pi}^*} -{M}_{t}^{\tilde{\bm\pi}^*}\right].
\end{eqnarray*}

We next take $u=T\wedge\tau_{a,b}$, where $\tau_{a,b}:= \inf \{s\geq t;\ V_s^{\tilde{\bm\pi}^*}\geq b^{-1},\ {\rm or}\ V_s^{\tilde{\bm\pi}^*}\leq a\}$, with $0<a<V_t^{\tilde{\bm\pi}^*}=v<b^{-1}<+\infty$. Notice that for each ${\bm z} \in {\cal S}$, both $B_{\bm z}^*(s)$ and $\Gamma_{i,{\bm z}}^{{\bm\pi}^*}(s)$,
$i\in\{1,\ldots,M\}$, are bounded on the closed time interval $[0,T]$. Then, for $0<a<b^{-1}<+\infty$, it holds that $\Ex_t^{\tilde{\Px}}\big[{M}_{T\wedge\tau_{a,b}}^{\tilde{\bm\pi}^*} - {M}_{t}^{\tilde{\bm\pi}^*}\big]=0$.
Thus,  we obtain
\begin{eqnarray}\label{eq:YleqV}
&&\Ex_t^{\tilde{\Px}}\left[Y_{T\wedge\tau_{a,b}}^{\tilde{\bm\pi}^*}\right]=U\big(V_t^{\tilde{\bm\pi}^*}\big)B^*_{{\bm Z}(t)}(t)\\
 &&\quad- \sum_{j=1}^M\Ex_t^{\tilde{\Px}}\left[\int_t^{T\wedge\tau_{a,b}}U\big(V_s^{\tilde{\bm\pi}^*}\big)\frac{(1-Z_j(s))\big[\vartheta^*_{j,{\bm Z}(s)}(s)\log(\vartheta^*_{j,{\bm Z}(s)}(s))-\vartheta^*_{j,{\bf Z}(s)}(s)+1\big]h_{j,{\bm Z}(s)}^{\Px}(s)}{\mu_{j,{\bm Z}(s)}(s)}\D s\right].\nonumber
\end{eqnarray}

Next, we want to prove that
\begin{eqnarray}\label{eq:limitab}
\lim_{a,b\too0}\Ex_t^{\tilde{\Px}}\big[Y_{T\wedge\tau_{a,b}}^{\tilde{\bm\pi}^*}\big]=\Ex_t^{\tilde{\Px}}\big[Y_{T}^{\tilde{\bm\pi}^*}\big].
\end{eqnarray}
Let $C_T > 0$ be a generic positive constant depending on $T$ that may be different for each inequality below. Since for each ${\bm z} \in {\cal S}$, $B^*_{\bm z}(t)$ is bounded on $t\in[0,T]$ and $\gamma\in(0,1)$, by employing H\"older{'}s inequality, it follows that
\begin{eqnarray*}
&&\Ex_t^{\tilde{\Px}}\left[Y_{T\wedge\tau_{a,b}}^{\tilde{\bm\pi}^*}\right] = \frac{1}{\gamma}\Ex_t^{\tilde{\Px}}\left[\big(V_{T\wedge\tau_{a,b}}^{\tilde{\bm\pi}^*}\big)^\gamma B^*_{{\bm Z}(T\wedge\tau_{a,b})}(T\wedge\tau_{a,b})\right]\leq C_T  \Ex_t^{\Px}\left[\left(V_{T\wedge\tau_{a,b}}^{\tilde{\bm\pi}^*}\right)^2\right].
\end{eqnarray*}
Moreover, according to Corollary 7.1.5 in \cite{ChowTeicher78}, in order to prove \eqref{eq:limitab}, it suffices to prove that
there exists a constant $C_T > 0$ so that
\begin{eqnarray}\label{eq:2nd-moment-V}
\Ex^{\tilde{\Px}}_t\left[\sup_{u\in[t,T]}\left|V_{u}^{\tilde{\bm\pi}^*}-V_{t}^{\tilde{\bm\pi}^*}\right|^{2}\right]
\leq C_T\left[1+\left|V_{t}^{\tilde{\bm\pi}^*}\right|^2\right].
\end{eqnarray}

In order to establish the estimate \eqref{eq:2nd-moment-V}, we first recall the dynamics of the wealth process $V_u^{\tilde{\bm\pi}^*}$ given by \eqref{eq:wealth}. Writing it under $\tilde{\Px}$, we obtain $\D V_u^{\tilde{\bm\pi}^*} = V_u^{\tilde{\bm\pi}^*}\varpi_{{\bm Z}(u)}(u)\D u+ V_{u-}^{\tilde{\bm\pi}^*}\sum_{j=1}^M\Gamma_{j,{\bm Z}(u-)}^{{\bm\pi}^*}(u)\D {\xi}_j^{\tilde{\Px}}(u)$,
where, for $j\in\{1,\ldots,M\}$, ${\bm z}\in{\cal S}$, and $u\in[t,T]$, $\varpi_{{\bm z}}(u):= r + \sum_{j=1}^M(1-z_j)\Gamma_{j,{\bm z}}^{{\bm\pi}^*}(u)\big(\vartheta_{j,{\bm z}}^*(u)-\frac{h_{j,{\bm z}}(u)}{h_{j,{\bm z}}^{\Px}(u)} \big) h_{j,{\bm z}}(u)$.
Using \eqref{eq:werst-controlj} and Proposition~\ref{prop:optimum}, it follows that, $\forall\ {\bm z}\in{\cal S}$,
\begin{eqnarray}\label{eq:condsmu}
\max\left\{\sup_{t\in[0,T]}\left|\varpi_{{\bm z}}(t)\right|,\sup_{t\in[0,T]}\sum_{j=1}^M\left|\Gamma_{j,{\bm z}}^{{\bm\pi}^*}(t)\right|\right\}<+\infty.
\end{eqnarray}
Using H\"older{'}s inequality, we get using \eqref{eq:condsmu} that for $u\in[t,T]$,
\begin{eqnarray*}
&& \Ex_t^{\tilde{\Px}}\left[\sup_{u\in[t,T]}\left|\int_t^uV_s^{\tilde{\bm\pi}^*} \varpi_{{\bm Z}(s)}(s) \D s\right|^2\right] \leq (T-t)\Ex_t^{\tilde{\Px}}\left[\int_t^T \left|V_s^{\tilde{\bm\pi}^*}\right|^2 \left|\varpi_{{\bm Z}(s)}(s)\right|^2 \D s\right]\nonumber\\
&&\qquad\leq 2(T-t)\Ex_t^{\tilde{\Px}}\left[\int_t^T \left(\left|V_s^{\tilde{\bm\pi}^*}-V_t^{\tilde{\bm\pi}^*}\right|^2 + \left|V_t^{\tilde{\bm\pi}^*}\right|^2\right) \left|\varpi_{{\bm Z}(s)}(s)\right|^2 \D s\right]\nonumber\\
&&\qquad\leq C_T \left\{\Ex_t^{\tilde{\Px}}\left[\int_t^T \left|V_s^{\tilde{\bm\pi}^*}-V_t^{\tilde{\bm\pi}^*}\right|^2\D s\right] + \left|V_t^{\tilde{\bm\pi}^*}\right|^2\right\}.
\end{eqnarray*}
Since $t\too h^{\Px}_{j,{\bm z}}(t)$ is continuous, it is bounded on $t\in[0,T]$. From Burkh\"older-Davis-Gundy inequality  (see \cite{Protter}, Theorem IV.48, pag. 193), it follows that
\begin{eqnarray*}
&& \Ex_t^{\tilde{\Px}}\left[\sup_{u\in[t,T]}\left|\sum_{j=1}^M\int_t^uV_{s-}^{\tilde{\bm\pi}^*} \Gamma_{j,{\bm Z}(s-)}^{{\bm\pi}^*}(s)\D{\xi}_j^{\tilde{\Px}}(s)\right|^2\right]
\leq C_T \Ex_t^{\tilde{\Px}}\left[\sum_{j=1}^M\int_t^T\left|V_{s-}^{\tilde{\bm\pi}^*}\right|^2 \left|\Gamma_{j,{\bm Z}(s-)}^{{\bm\pi}^*}(s)\right|^2\D Z_j(s)\right]\nonumber\\
&&\qquad\qquad = C_T \sum_{j=1}^M\Ex_t^{\tilde{\Px}}\left[\int_t^T\left|V_s^{\tilde{\bm\pi}^*}\right|^2 \left|\Gamma_{j,{\bm Z}(s)}^{{\bm\pi}^*}(s)\right|^2\vartheta_{j,{\bm Z}(s)}^*(s)h^{\Px}_{j,{\bm Z}(s)}(s)\D s\right]\nonumber\\
&&\qquad\qquad
\leq C_T \left\{\Ex_t^{\tilde{\Px}}\left[\int_t^T \left|V_s^{\tilde{\bm\pi}^*}-V_t^{\tilde{\bm\pi}^*}\right|^2\D s\right] + \left|V_t^{\tilde{\bm\pi}^*}\right|^2\right\}.
\end{eqnarray*}
Then the moment estimate \eqref{eq:2nd-moment-V} follows from the Grownwall's lemma. This shows the limiting equality~\eqref{eq:limitab}.
Similarly, using \eqref{eq:werst-controlj}, for each ${\bm z}\in{\cal S}$ and $j\in\{1,\ldots,M\}$, it holds that
\[
\sup_{u\in[t,T]}\left|\frac{\big[\vartheta^*_{j,{\bm z}}(u)\log(\vartheta^*_{j,{\bm z}}(u))-\vartheta^*_{j,{\bm z}}(u)+1\big]h_{j,{\bm z}}^{\Px}(u)}{\mu_{j,{\bm z}}(u)}\right|<+\infty.
\]
We also have that for all $j\in\{1,\ldots,M\}$,
\begin{eqnarray*}
&&\lim_{a,b\too0}\Ex_t^{\tilde{\Px}}\left[\int_t^{T\wedge\tau_{a,b}}U\big(V_s^{\tilde{\bm\pi}^*}\big)\frac{(1-Z_j(s))\big[\vartheta^*_{j,{\bm Z}(s)}(s)\log(\vartheta^*_{j,{\bm Z}(s)}(s))-\vartheta^*_{j,{\bm Z}(s)}(s)+1\big]h_{j,{\bm Z}(s)}^{\Px}(s)}{\mu_{j,{\bm Z}(s)}(s)}\D s\right]\nonumber\\
&&\qquad=\Ex_t^{\tilde{\Px}}\left[\int_t^{T}U\big(V_s^{\tilde{\bm\pi}^*}\big)\frac{(1-Z_j(s))\big[\vartheta^*_{j,{\bm Z}(s)}(s)\log(\vartheta^*_{j,{\bm Z}(s)}(s))-\vartheta^*_{j,{\bm Z}(s)}(s)+1\big]h_{j,{\bm Z}(s)}^{\Px}(s)}{\mu_{j,{\bm Z}(s)}(s)}\D s\right].
\end{eqnarray*}
Then, by \eqref{eq:YleqV} and using the relation~\eqref{eq:weights}, along with the terminal condition $B^*_{\bm z}(T)=1$ for all ${\bm z}\in{\cal S}$, we obtain
\begin{eqnarray*}
&&\Ex_t^{\tilde{\Px}}\left[Y_{T}^{\tilde{\bm\pi}^*}+\sum_{j=1}^M\int_t^T\frac{(1-Z_j(s))\big[\vartheta^*_{j,{\bm Z}(s)}(s)\log(\vartheta^*_{j,{\bm Z}(s)}(s))-\vartheta^*_{j,{\bm Z}(s)}(s)+1\big]h_{j,{\bm Z}(s)}^{\Px}(s)}{\Upsilon_{j,{\bm Z}(s)}(s,V_s^{\tilde{\bm\pi}^*})}\D s\right]\nonumber\\ &&\qquad=\Ex_t^{\tilde{\Px}}\left[U\big(V_T^{\tilde{\bm\pi}^*}\big)+\sum_{j=1}^M\int_t^T\frac{(1-Z_j(s))\big[\vartheta^*_{j,{\bm Z}(s)}(s)\log(\vartheta^*_{j,{\bm Z}(s)}(s))-\vartheta^*_{j,{\bm Z}(s)}(s)+1\big]h_{j,{\bm Z}(s)}^{\Px}(s)}{\Upsilon_{j,{\bm Z}(s)}(s,V_s^{\tilde{\bm\pi}^*})}\D s\right]\nonumber\\
&&\qquad = U\big(V_t^{\tilde{\bm\pi}^*}\big)B^*_{{\bm Z}(t)}(t).
\end{eqnarray*}
This shows that the value function defined by \eqref{eq:robust-problem} and associated with the robust optimization criterion admits the decomposition $w_{\bm z}(t,v)=U(v)B^*_{{\bm z}}(t)$. This completes the proof of the verification theorem. \hfill$\Box$

\section{Numerical Analysis} \label{sec:numanalysis}
We perform a numerical study to assess the impact of robustness on feedback and value functions.
We develop an efficient implementation to solve the coupled system of HJB equations and recover the optimal controls.
This study is for illustrative purposes only. In particular, parameter values are chosen in an ad hoc manner in order to exemplify typical qualitative behavior of the model. We consider two obligors, i.e, set $M=2$, with $j_1=1$ and $j_2=2$.

We describe the fixed point procedure used to recover the value function and the optimal feedback function in Section \ref{sec:fixedpoint}.
We provide a comparative statics in Section \ref{sec:comparatstat}.

\subsection{Fixed Point Algorithm} \label{sec:fixedpoint}
We solve for the coupled value function and optimal feedback functions by first computing the fixed point solution $(C_j, B)$, $j=1,2$ of Eq.~\eqref{eq:hjbm2} with coefficient~\eqref{eq:Cj}. The solution $B(t)$ solves the fixed point equation $B(t) = g(t,B(t))$, where
\begin{eqnarray}\label{eq:hjbm200}
g(t,B(t))&:=& B'(t) + {B(t)}\left[1+\gamma\bigg(r + \sum_{j\notin\{j_1,\ldots,j_m\}}{h_j(t)}\bigg) - \sum_{j\notin\{j_1,\ldots,j_m\}} C_j\big(t,\mu_j(t)B_j(t);B(t)\big)\right]\nonumber\\
&& + \sum_{j\notin\{j_1,\ldots,j_m\}}\frac{h_j^{\Px}(t)}{\mu_{j}(t)},\ \ \ \ \ m\in\{0,1,2\},
\end{eqnarray}
and is guaranteed by Theorem~\ref{thm:hjb}. We then plug this solution in the system~\eqref{eq:optima-stra} and recover the optimal feedback functions. Concretely, we proceed backwards as follows:
\begin{enumerate}
\item[{\sf(I)}] ${\bm z} = (1,0)$. Since the name $j_1=1$ has defaulted, Eq.~\eqref{eq:hjbm2} becomes
\begin{eqnarray}
0 &=& B_{10}'(t) + B_{10}(t)\left[\gamma\big(r + h_{2,10}(t)\big) - C_{2,10}\big(t,\mu_{2,10}(t)B_{11}(t);B_{10}(t)\big)\right]+\frac{h_{2,10}^{\Px}(t)}{\mu_{2,10}(t)}
\label{eq:B10SP}
\end{eqnarray}
with terminal condition $B_{10}(T)=1$. Moreover, the function $C_{2,10}$ is obtained from Eq.~\eqref{eq:Cj} under this default state and given by
\begin{eqnarray}
C_{2,10}(t,y;x) &:=& \gamma h_{2,10}(t)\left[{\cal Y}_{y}^{-1}\left(\frac{\mu_{2,10}(t)h_{2,10}(t)}{yh_{2,10}^{\Px}(t)}{\cal J}\big(\mu_{2,10}(t);x\big)\right)\right]^{\frac{1}{\gamma-1}}\nonumber\\
&&+\frac{h_{2,10}(t)}{y}\left[{\cal Y}_y^{-1}\left(\frac{\mu_{2,10}(t)h_{2,10}(t)}{yh_{2,10}^{\Px}(t)}{\cal J}\big(\mu_{2,10}(t);x\big)\right)\right]^{-1}.
\label{eq:C210SP}
\end{eqnarray}
We use the following procedure to solve the coupled system given by the equations~\eqref{eq:B10SP} and~\eqref{eq:C210SP} above. Suppose we are at the $n$-th iteration step of the procedure. Let
$C^{(n-1)}_{2,10} := \big(C^{(n-1)}_{2,10}(t) \; ; \; t\in[0,T]\big)$ and $B_{10}^{(n-1)} := \big(B^{(n-1)}_{10}(t) \; ; \; t\in[0,T]\big)$ be respectively the $(n-1)$-th order approximation of the
function $C_{2,10}$ and of the function $B_{10}$ recovered in the $(n-1)$-th iteration. Using $C^{(n-1)}_{2,10}$, we solve the nonlinear equation~\eqref{eq:B10SP}. The corresponding solution yields the $n$-th order approximation of {$B_{10}$},
which we denote by  $B^{(n)}_{10}$. Then, for each $t\in[0,T]$, we use $B^{(n)}_{10}(t)$ to compute $C_{2,10}^{(n)}={ C_{2,10}(t,\mu_{2,10}(t)B_{11}(t);B_{10}^{(n)}(t))}$. We continue iterating until convergence is achieved. Let $B_{10}^*(t)$ be the function when the procedure stops.
We compute the optimal fraction of wealth invested in the risky bond ``2'' using Eq.~\eqref{eq:optima-stra}. In this specific case, it reduces to
$\pi^*_{2,10}(t)={G}^{-1}_{2,2,10}(t)\hat{{\mathcal{Y}}}_{10}(t)$, where {$G_{2,2,10}(t) = \frac{R_2}{F_{2,10}(t)}-1$}, and $\hat{{\mathcal{Y}}}_{10}(t):=
\big[{\cal Y}_{\mu_{2,10}(t)B_{11}(t)}^{-1}\big(\frac{h_{2,10}(t)B^*_{10}(t)}{h_{2,10}^{\Px}(t)B_{11}(t)}e^{-\mu_{2,10}(t) B^*_{10}(t)}\big)\big]^{\frac{1}{\gamma-1}}-1$.
We recall that $B_{11}(t)$ is explicitly given by Eq.~\eqref{eq:sol-HJBM}.\\

\item[{\sf(II)}] ${\bm z} = (0,1)$. This case is completely symmetric to the one for the default state ${\bm z} = (1,0)$. Hence, we omit the description.\\

\item[{\sf(III)}] ${\bm{z}} = (0,0)$. In this case, both names are alive. Then Eq.~\eqref{eq:hjbm2} becomes
\begin{eqnarray}
0 &=& B_{00}'(t) + B_{00}(t)\bigg[\gamma\big(r + h_{1,00}(t) + h_{2,00}(t)\big) - C_{1,00}\big(t,\mu_{1,00}(t)B_{10}(t);B_{00}(t)\big)\nonumber\\
&& - C_{2,00}\big(t,\mu_{2,00}(t)B_{01}(t);B_{00}(t)\big) \bigg] + \frac{h_{1,00}^{\Px}(t)}{\mu_{1,00}(t)} + \frac{h_{2,00}^{\Px}(t)}{\mu_{2,00}(t)},
\end{eqnarray}
with terminal condition $B(T)=1$. Moreover, the functions $C_{1,00}$ and $C_{2,00}$ are obtained from Eq.~\eqref{eq:Cj} and given by
\begin{eqnarray}
C_{i,00}(t,y;x) &:=& \gamma h_{i,00}(t)\left[{\cal Y}_y^{-1}\left(\frac{\mu_{i,00}(t)h_{i,00}(t)}{yh_{i,00}^{\Px}(t)}{\cal J}\big(\mu_{i,00}(t);x\big)\right)\right]^{\frac{1}{\gamma-1}}\nonumber\\
&&+\frac{h_{i,00}(t)}{y}\left[{\cal Y}_y^{-1}\left(\frac{\mu_{i,00}(t)h_{i,00}(t)}{yh_{i,00}^{\Px}(t)}{\cal J}\big(\mu_{i,00}(t);x\big)\right)\right]^{-1}, \ \ \ i = 1,2. 
\end{eqnarray}
Using the same fixed point procedure described in {\sf(I)}, where we successively refine the $n$-th order approximations of $B_{00}^{(n)}$ and of the pairs $(C_{1,00}^{(n)},C_{2,00}^{(n)})$ until
achieving convergence, we estimate $B_{00}$. Denote the corresponding estimate by $B_{00}^*$. We then use it to compute the optimal feedback function. Using Eq.~\eqref{eq:optima-stra}, we obtain that the optimal feedback control function giving the fraction of wealth invested in risky bonds is given by ${\bm\pi}^*(t)=\big({\bm G}^{-1}(t)\big)^{\top}\hat{\boldsymbol{\mathcal{Y}}}(t)$, $t\in[0,T]$,
where
\begin{eqnarray*}
\hat{\boldsymbol{\mathcal{Y}}}(t)=\left[
                                             \begin{array}{c}
                                               \Big[{\cal Y}_{\mu_{1,00}(t)B^*_{10}(t)}^{-1}\big( \frac{h_{1,00}(t)B^*_{00}(t)}{h_{1,00}^{\Px}(t)B^*_{10}(t)}e^{-\mu_{1,00}(t)B^*_{00}(t)}\big)\Big]^{\frac{1}{\gamma-1}}-1\\
                                               \Big[{\cal Y}_{\mu_{2,00}(t)B^*_{01}(t)}^{-1}\big(\frac{h_{2,00}(t)B^*_{00}(t)}{h_{2,00}^{\Px}(t)B^*_{01}(t)}e^{-\mu_{2,00}(t)B^*_{00}(t)}\big)\Big]^{\frac{1}{\gamma-1}}-1
                                             \end{array}
                                           \right],\ {\bm G}(t)=\left[
                                  \begin{array}{cccc}
                                    G_{1,1,00}(t) & G_{1,2,00}(t) \\
                                    G_{2,1,00}(t) & G_{2,2,00}(t)
                                  \end{array}
                                \right].
\end{eqnarray*}
The components of this matrix are given by, for $t\in[0,T]$,
\[
G_{1,1,00}(t) = \frac{R_1}{F_{1,00}(t)}-1,\ G_{1,2,00}(t)=\frac{F_{1,01}(t)}{F_{1,00}(t)}-1,\ G_{2,1,00}(t)=\frac{F_{2,10}(t)}{F_{2,00}(t)}-1,\
G_{2,2,00}(t)=\frac{R_2}{F_{2,00}(t)}-1.
\]
\end{enumerate}

\subsection{Comparative Statics Analysis} \label{sec:comparatstat}
Throughout the analysis, whenever the following parameters are kept fixed and unless otherwise specified, we use the following benchmark values.
We consider the same contractual parameters for the two risky bonds. Their loss rates are $L_1=L_2=0.3$, and the coupon rates $\nu_1 = \nu_2 = 0.6$. We set the investment horizon to
$T=1$, and the maturities of the two bonds to $T_1 = T_2 = 3$. We choose $r=0.05$, and $\gamma = 0.5$.
The reference default  intensities are set to $h_{1,00}^{\Px} = 0.5$, $h_{2,00}^{\Px} = 0.5$, $h_{1,01}^{\Px} = h^{\Px}_{2,10} = 1$. The penalty parameters are set to $\mu_{1,00} = \mu_{2,00} = \mu_{1,01} = \mu_{2,10} = 0.5$. The risk neutral default intensities are set to
$h_{1,00} = h_{2,00} = 1$ and $h_{1,01} = h_{2,10} = 2$. We set the investment time to $t=0$.

Notice that in the verification theorem, we have proven that {\sf(I)} the vector of optimal wealth fractions $\tilde{\bm\pi}^*$, is independent of the
wealth variable $v$ and {\sf(II)} the robust value function $w_{{\bm z}}(t,v) = \frac{v^{\gamma}}{\gamma} B^*_{{\bm z}}(t)$. In light of this decomposition result, in the forthcoming section we will plot the time component $B^*_{{\bm z}}(t)$ of the robust value function, given that the additional term $\frac{v^{\gamma}}{\gamma}$ would not have any informative role in the sensitivity analysis. Moreover, we will not specify the wealth level $v$ in the plots, given that the fractional strategies are independent of it.

\subsubsection{Impact of Credit Risk}
Figure \ref{figstrat2} shows that the investor increases his position in bonds if, ceteris paribus, the reference default intensities increase. When this happens,
the rate of bond returns increases because the investor receives higher compensation for bearing default risk. The top panels indicate that the investor faces a
trade-off between receiving compensation for being exposed to default risk and bearing the consequences deriving from the bond's default: when the reference default intensity of obligor ``1'' is lower than that of name ``2'', the investor allocates a higher fraction of wealth to bond 1. However, as this exceeds the value of the reference default intensity of obligor ``2'', his risk aversion dominates and leads the investor to invest more in the safer bond ``2'' and to reduce the fraction allocated to the riskier bond ``1''.

The bottom panels of figure~\ref{figstrat2}
indicate that the investor achieves higher utility by increasing the size of his bond position in reaction to an increase in the reference default intensity.

%
}

\begin{figure}[ht!]
\begin{center}
\hspace{-.7cm} \includegraphics[width=7.2cm]{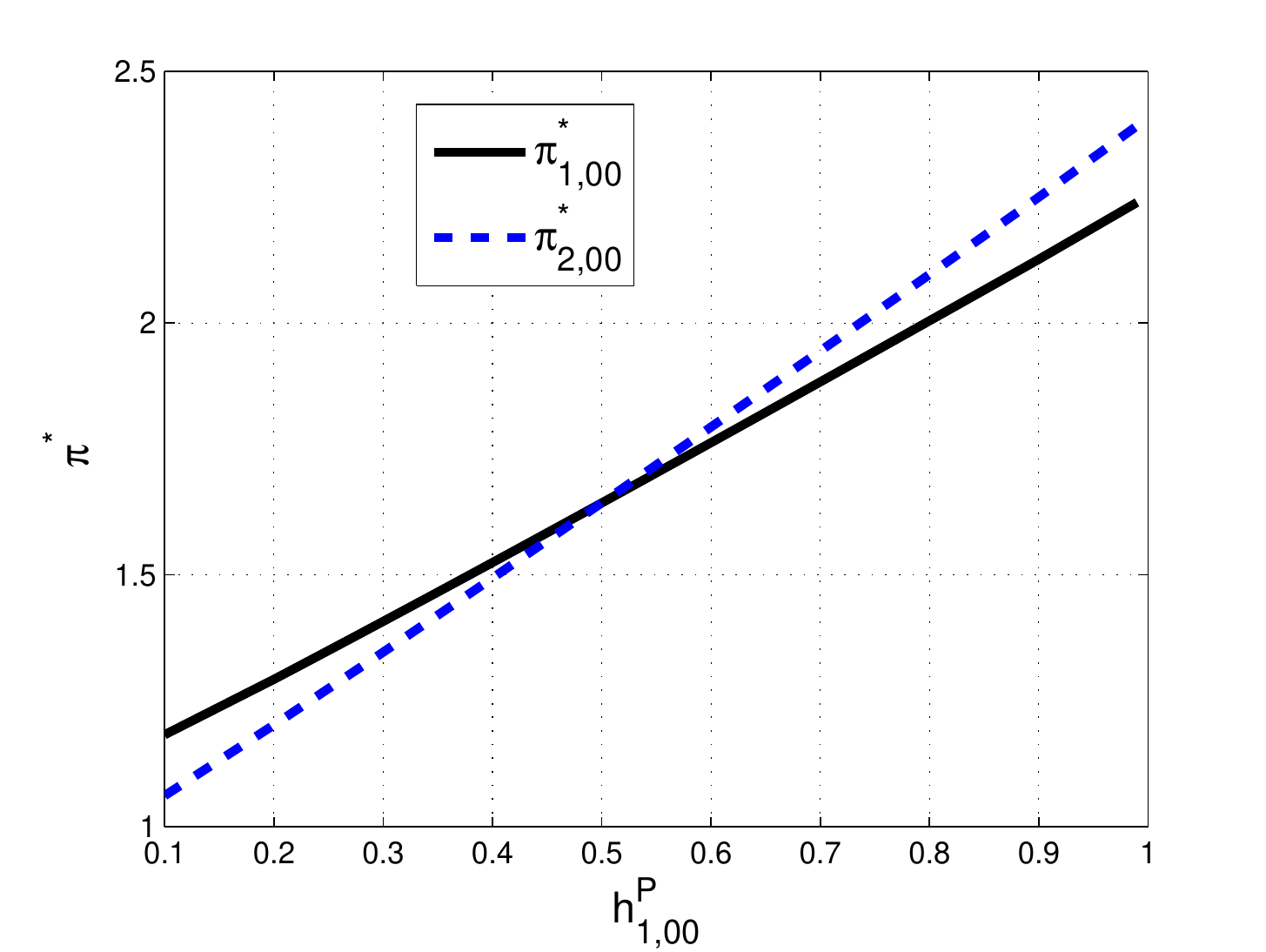}
\hspace{-.7cm} \includegraphics[width=7.2cm]{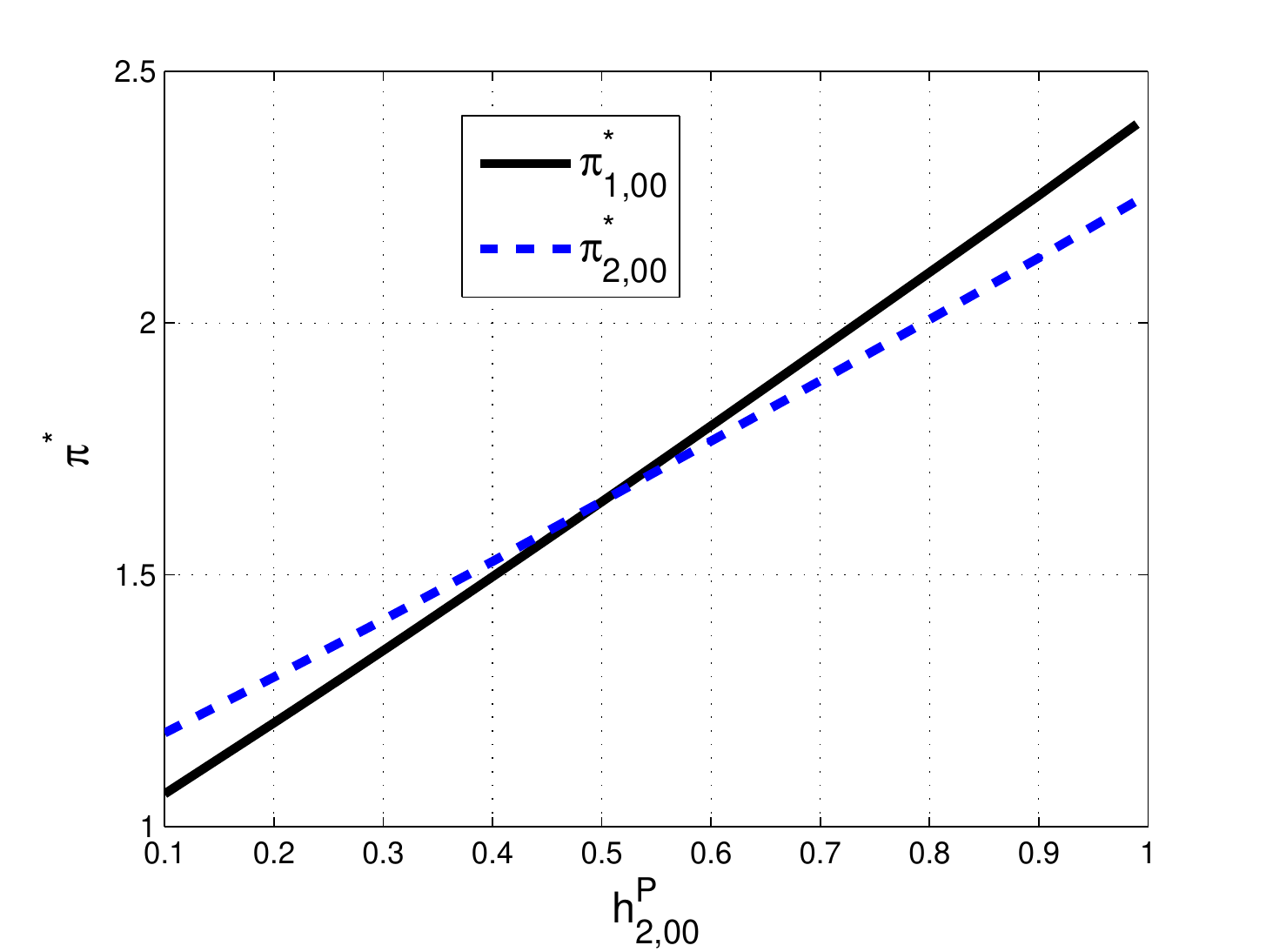}
\hspace{-.7cm} \includegraphics[width=7.2cm]{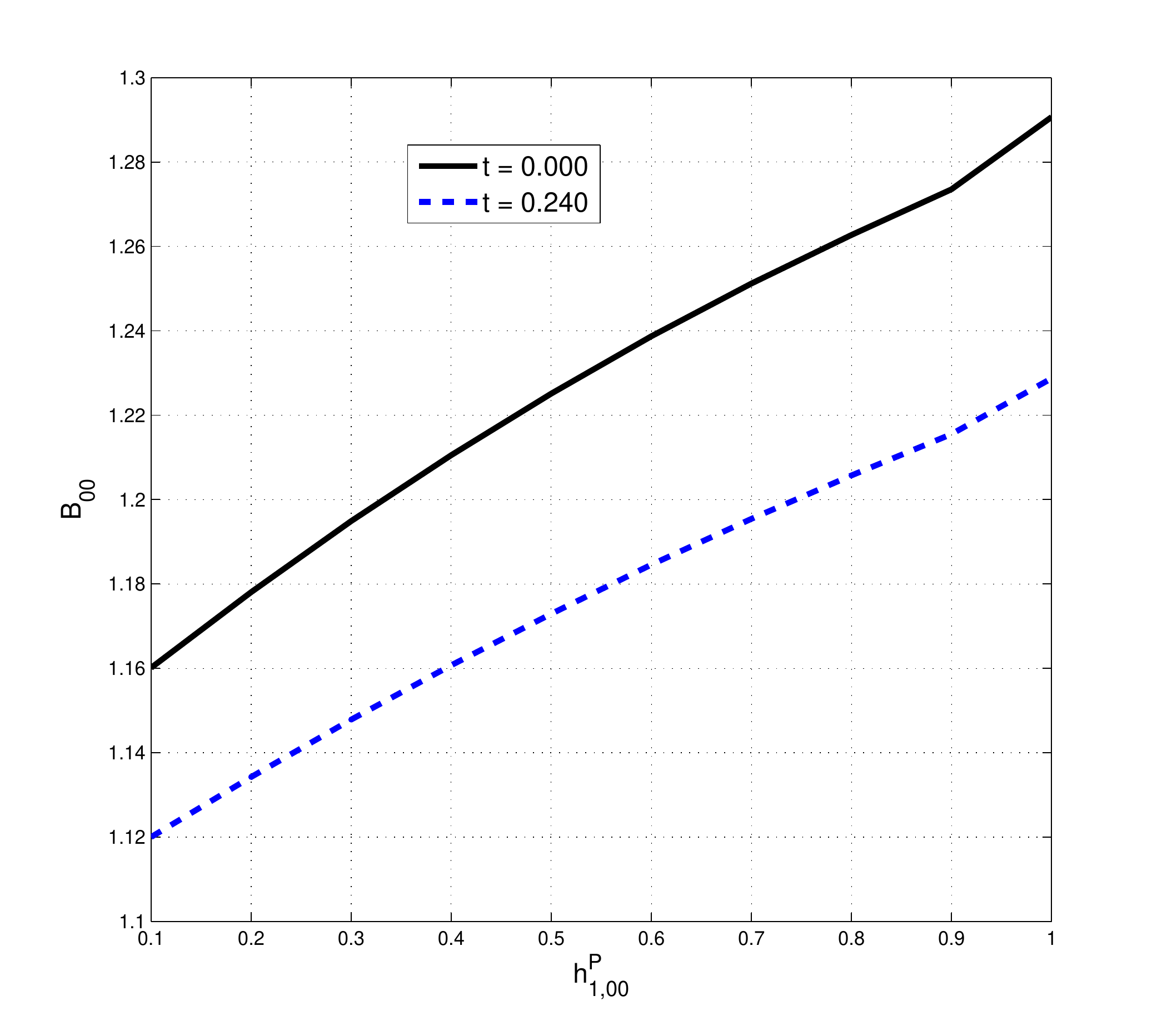}
\hspace{-.7cm} \includegraphics[width=7.2cm]{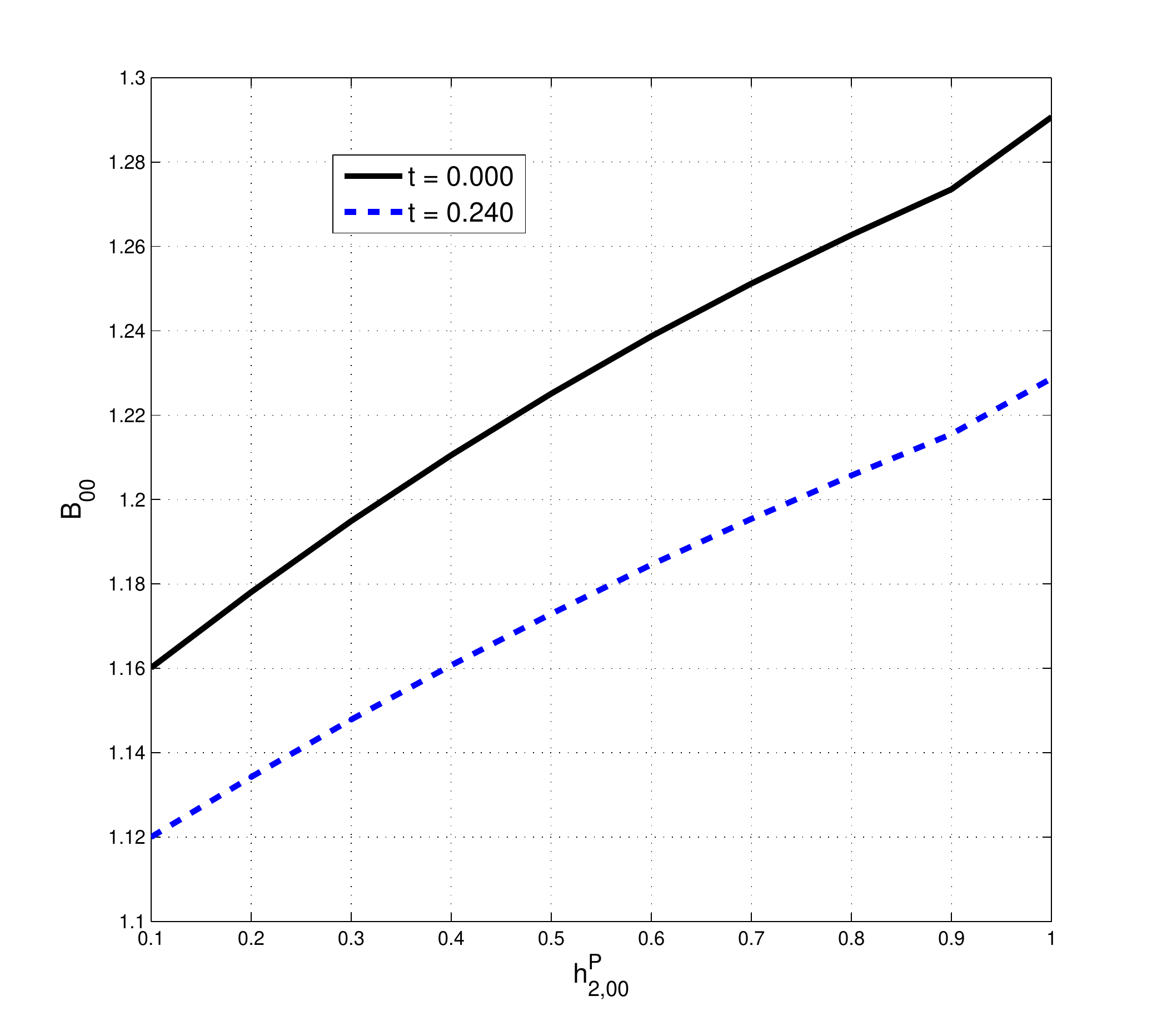}
\end{center}
\caption{The top panels report the dependence of the optimal feedback functions on the default intensities $h_{1,00}^{\Px}$ and $h_{2,00}^{\Px}$. The bottom panels give the same dependence
for the time component $B_{00}$ of the robust value function. We set the risk-neutral default intensities $h_{1,00} = 2 h_{1,00}^{\Px}$ and $h_{2,00} = 2 h_{2,00}^{\Px}$. }
\label{figstrat2}
\end{figure}

\subsubsection{Impact of Robustness}
This section analyzes the sensitivity of the feedback functions and of the value functions with respect to the robustness parameters.

\begin{figure}[ht!]
\begin{center}
\hspace{-.7cm} \includegraphics[width=7.2cm]{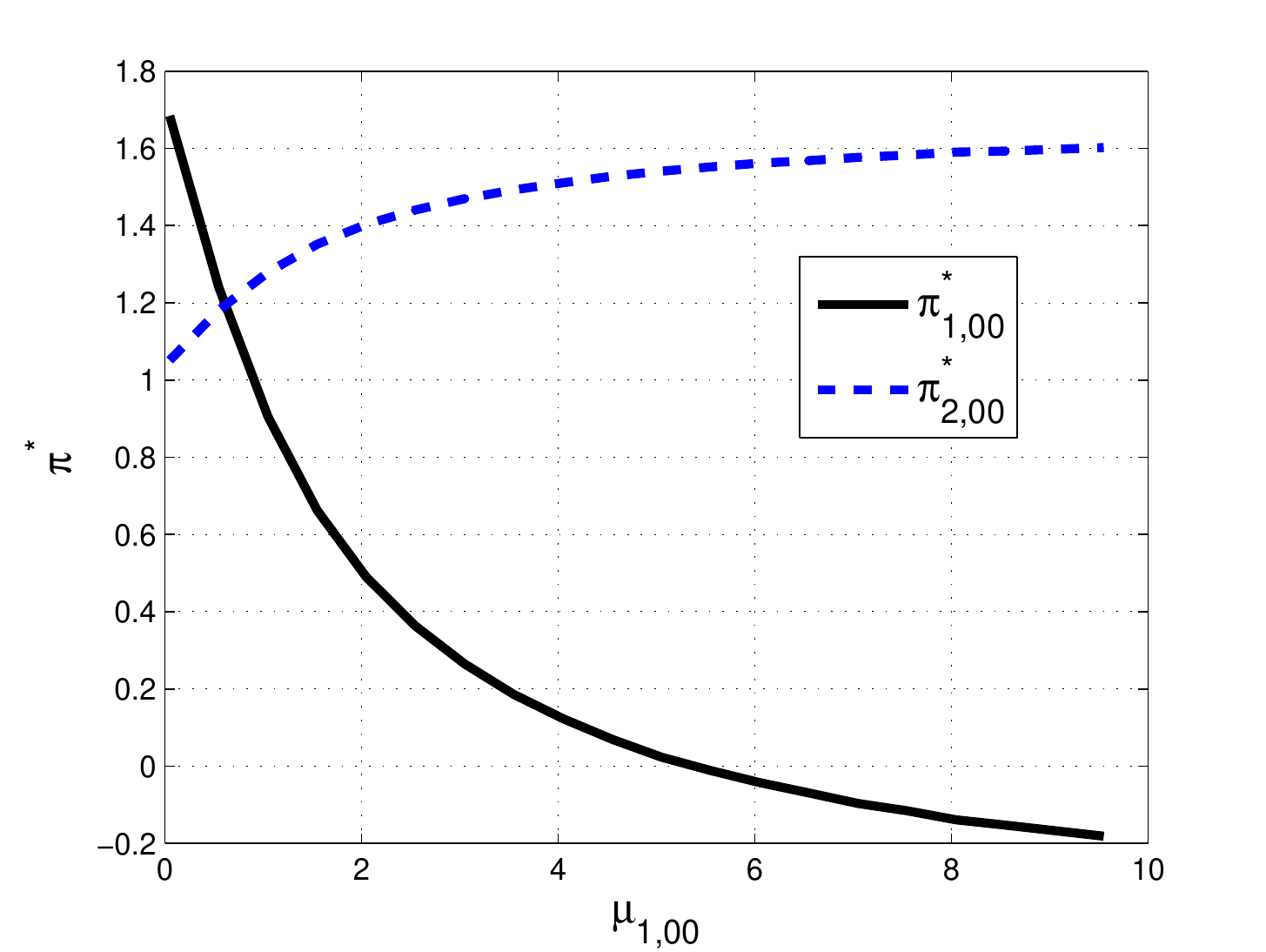}
\hspace{-.7cm} \includegraphics[width=7.2cm]{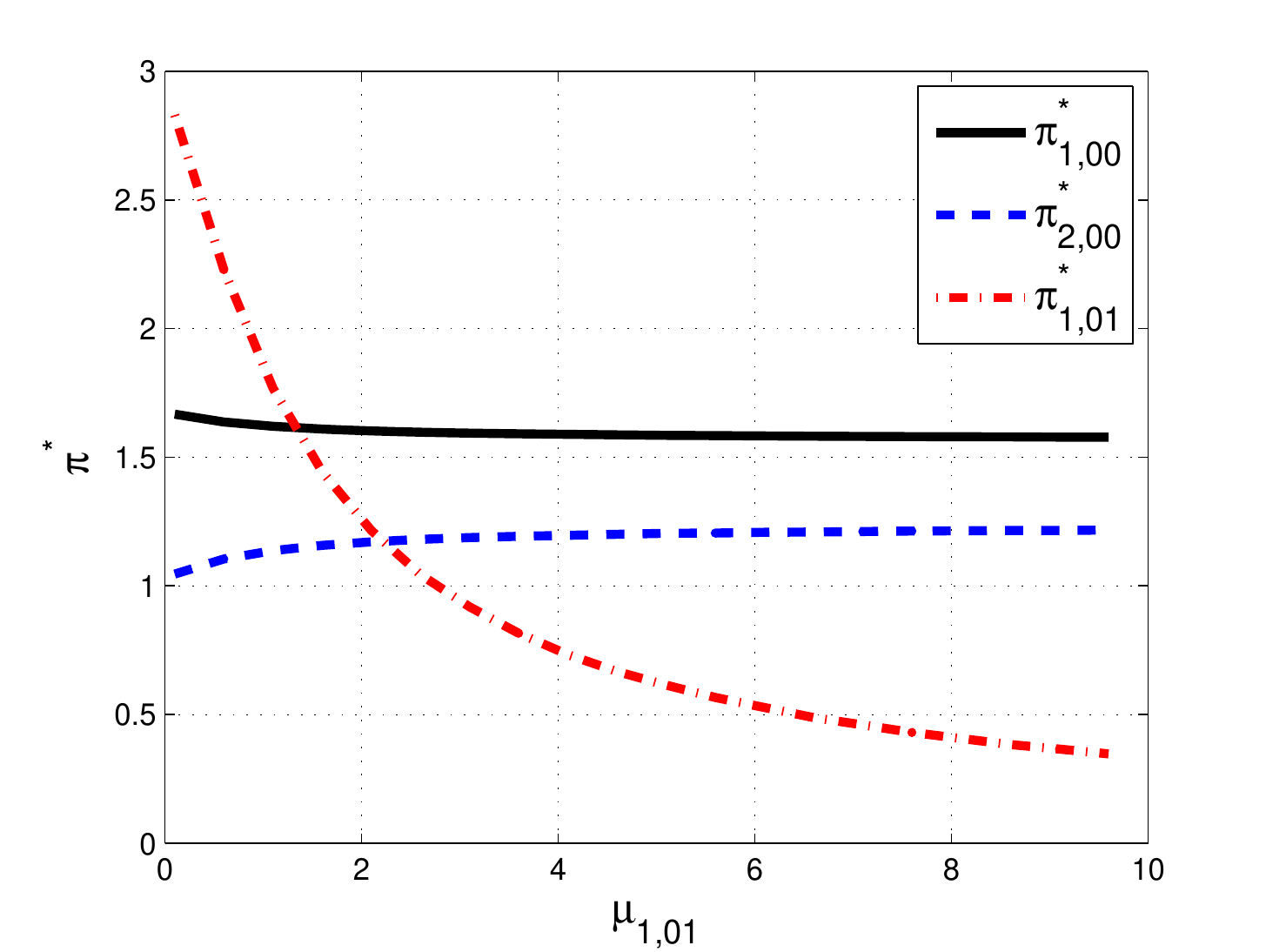} 
\hspace{-.7cm} \includegraphics[width=7.2cm]{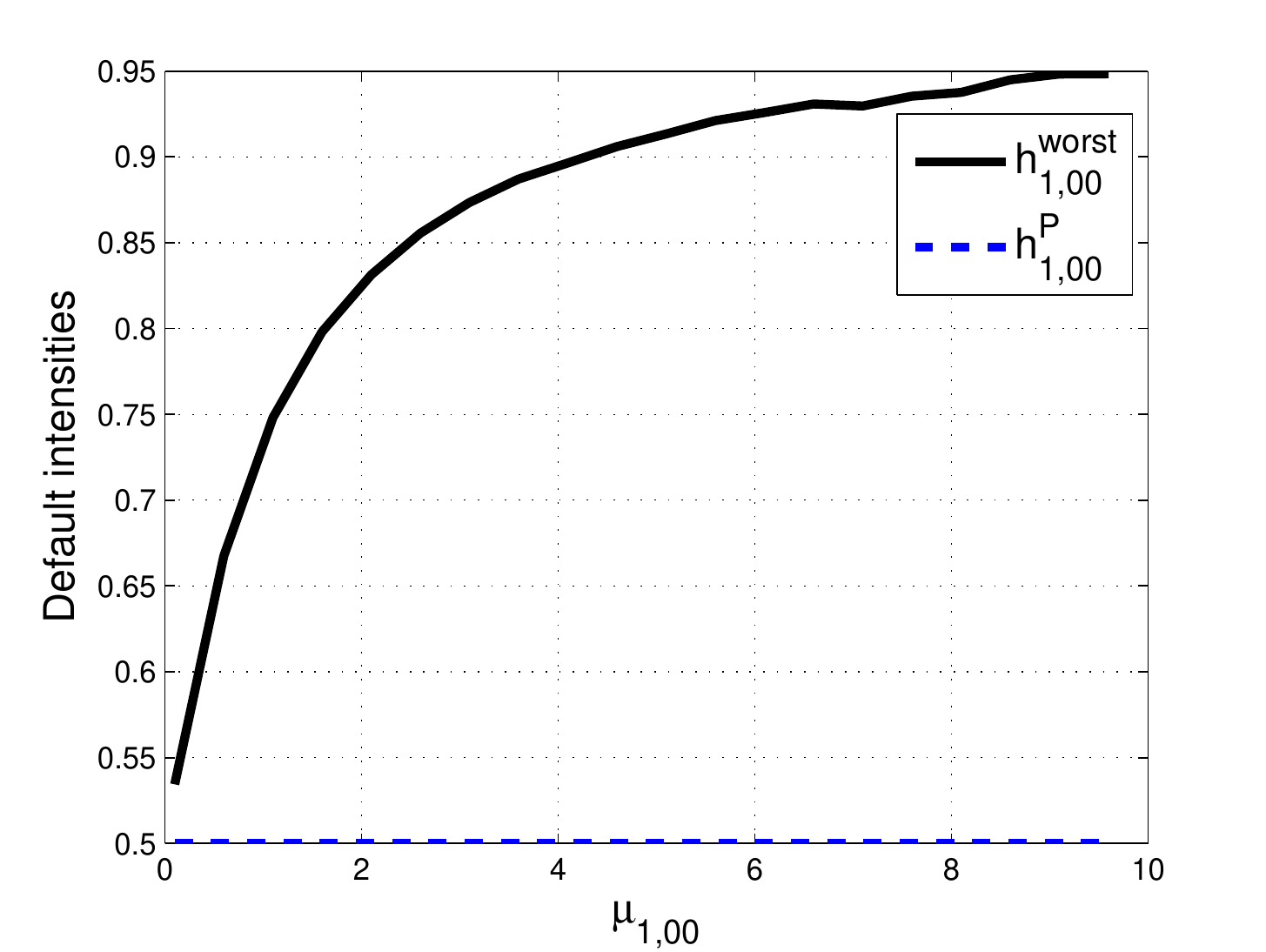}
\hspace{-.7cm} \includegraphics[width=7.2cm]{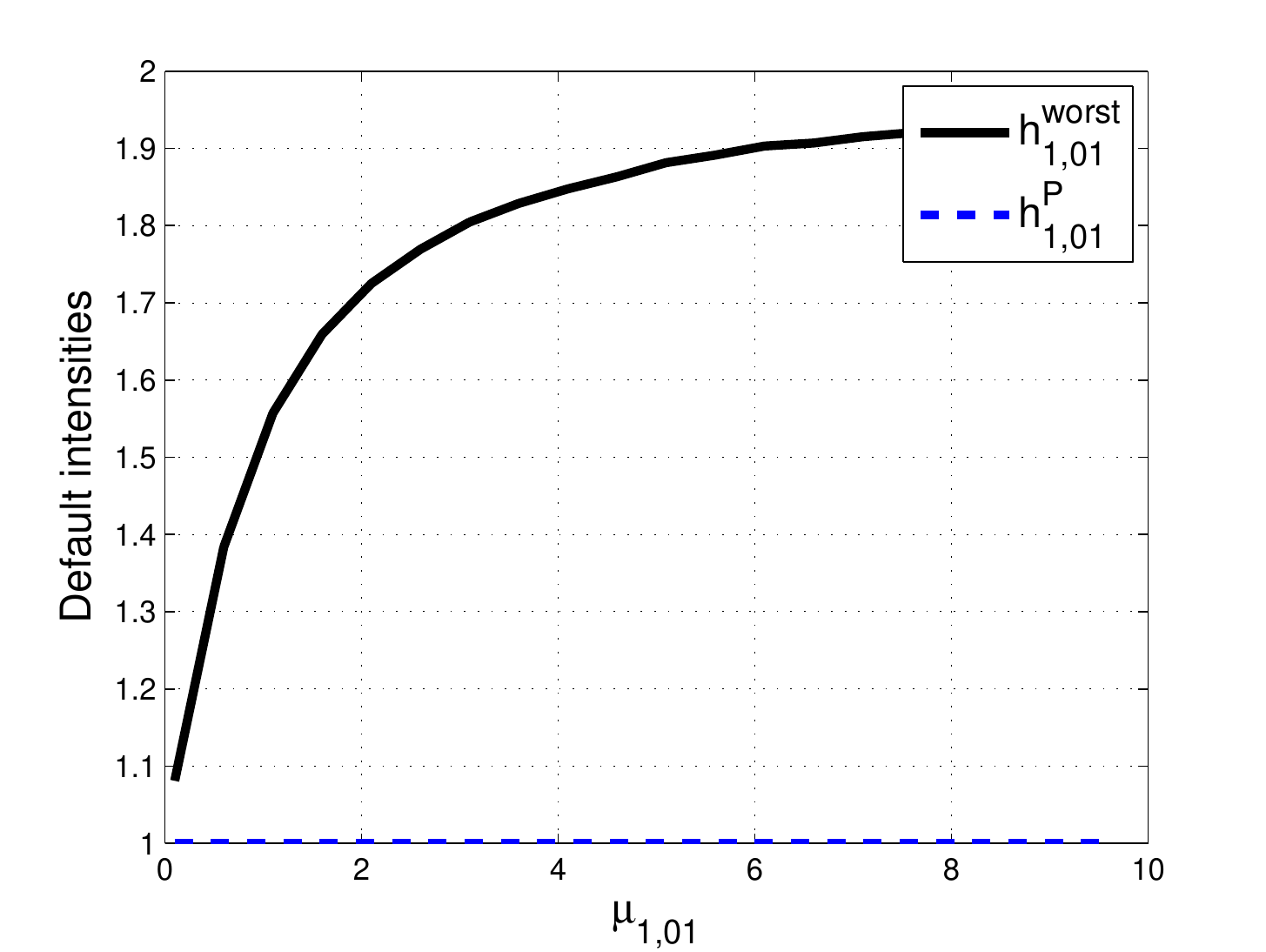}
\end{center}
\caption{
The top panels report the dependence of the feedback functions on the penalty parameters $\mu_{1,00}$ and $\mu_{1,01}$. The bottom panels give the dependence of the worst-case
default intensities on the penalty parameters.}
\label{figstrat}
\end{figure}

We find that robustness reduces the demand for risky bonds. The investor allocates a higher fraction of his wealth to the risky bond if he is more confident about the reference model.
As the penalty for misspecification of the reference default intensity of name ``1'' becomes lower (large values of $\mu_{1,00}$), the investor decreases the fraction of wealth
allocated to bond ``1'' and invests the saved proceeds in the bond ``2'' (see top left panel of figure \ref{figstrat}). This can be understood together with
the graph of the worst-case default intensity. When the investor is more tolerant about deviations from the reference model, his worst-case default intensity is higher
(see bottom panels of figure~\ref{figstrat}). Then the risk-averse investor decreases the size of his long bond position because he considers a worst-case scenario where
default is more likely to occur that what estimated by his reference model. 
Increasing the tolerance against misspecifications of the reference default intensity $h_{1,01}^{\Px}$ has the highest impact on the investment strategy in bond ``1'' when the
state is (0,1). However, it also affects the strategy of the investor in the state $(0,0)$ when both names are alive, pushing him to decrease the size of his long position in bond ``1'' and to
increase the corresponding position in bond 2 (see also right panels of figure \ref{figstrat}).


%

\begin{figure}[ht!]
\begin{center}
\hspace{-.7cm} \includegraphics[width=7.2cm]{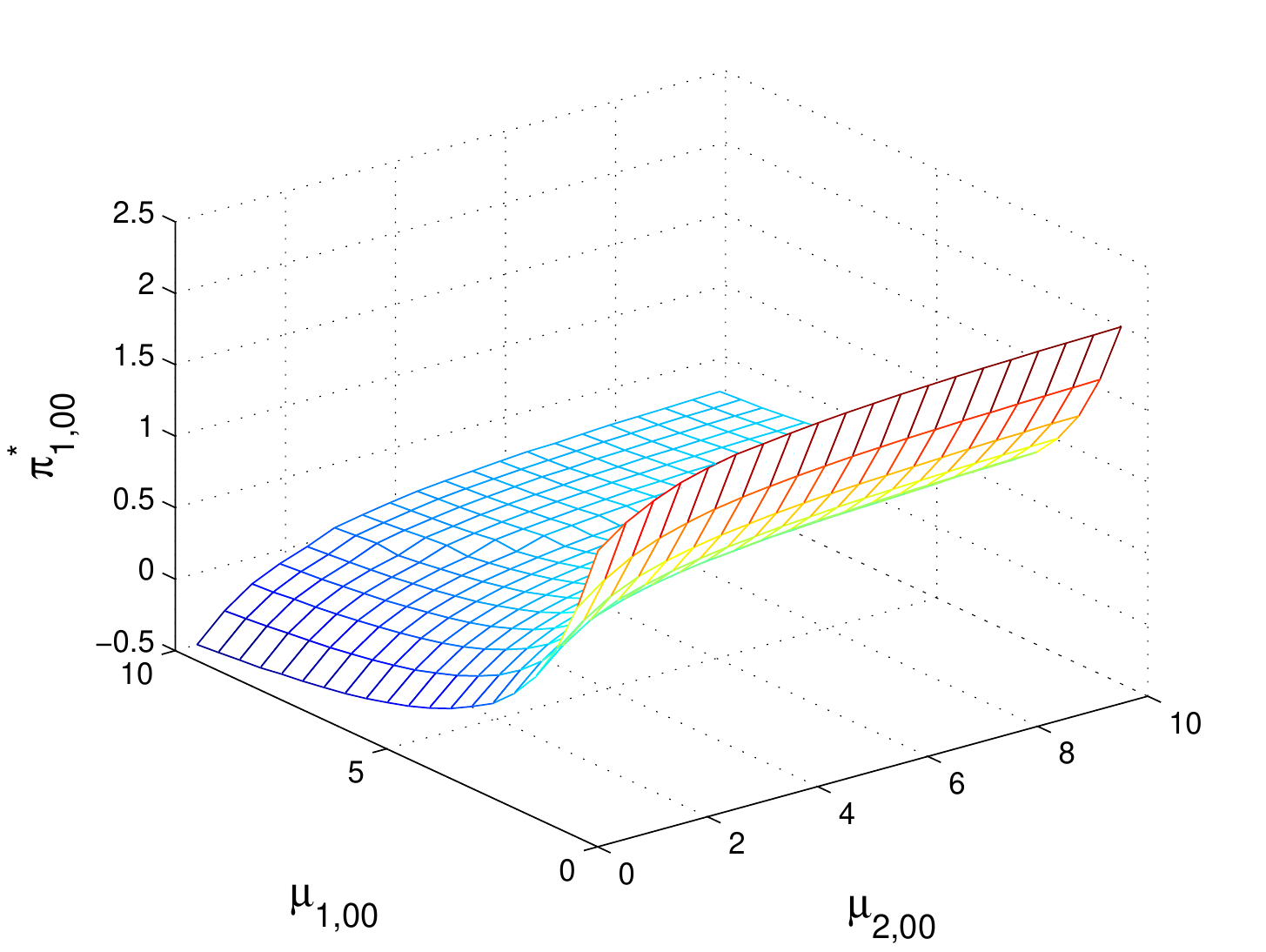}
\hspace{-.7cm} \includegraphics[width=7.2cm]{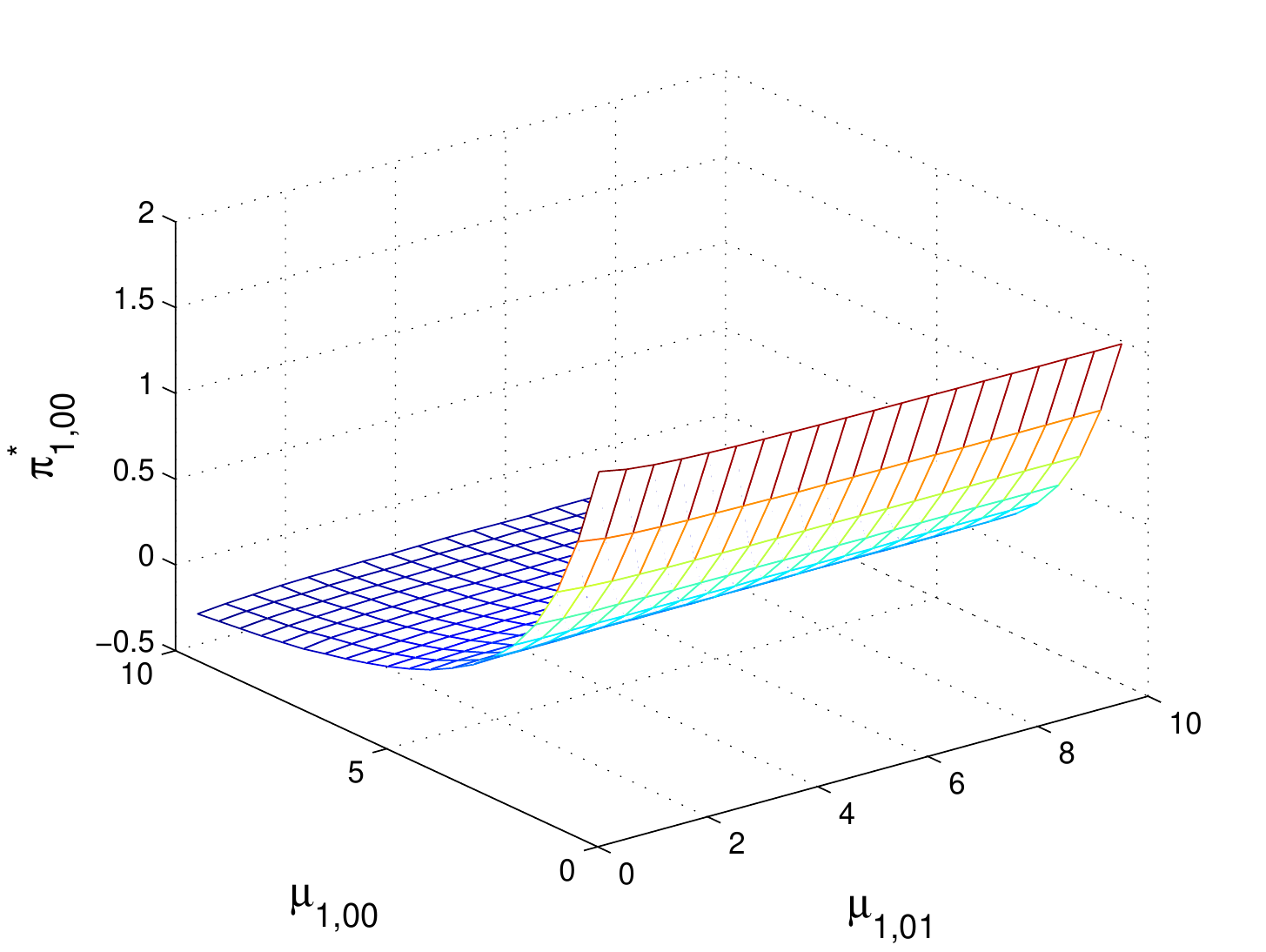}
\end{center}
\caption{
The left panel reports the dependence of the optimal feedback function on the penalty parameters $(\mu_{1,00}, \mu_{2,00})$. The right panel give the joint dependence of the same
feedback function on the penalty parameters $(\mu_{1,00}, \mu_{1,01})$.}
\label{figstrat1}
\end{figure}

Changes of penalty for mispecification of the reference default intensity of obligor ``1'' lead the investor to revise his investment strategy in bond 1, but only mildly affects
his investment strategy in bond ``2'' (see left panel of figure~\ref{figstrat1}). Moreover, the investor is less sensitive to penalty against misspecification of default risk in a future default state, i.e. the state when obligor ``2'' has defaulted, than to penalty against misspecification of credit risk in the current state, i.e. the state when both obligors are alive
(see right panel of figure~\ref{figstrat1}). This is because the default risk of name ``2'', $h_{2,00}^{\Px}$, is relatively low. Hence, the probability that the investor will find himself in the state $01$ where the default intensity $h_{1,01}^{\Px}$ matters is not too high. We expect stronger dependence of the strategy to penalty for mispecification of default intensities in future default states, if the credit risk in the current state were higher.


\begin{figure}[ht!]
\begin{center}
\hspace{-.7cm} \includegraphics[width=7.2cm]{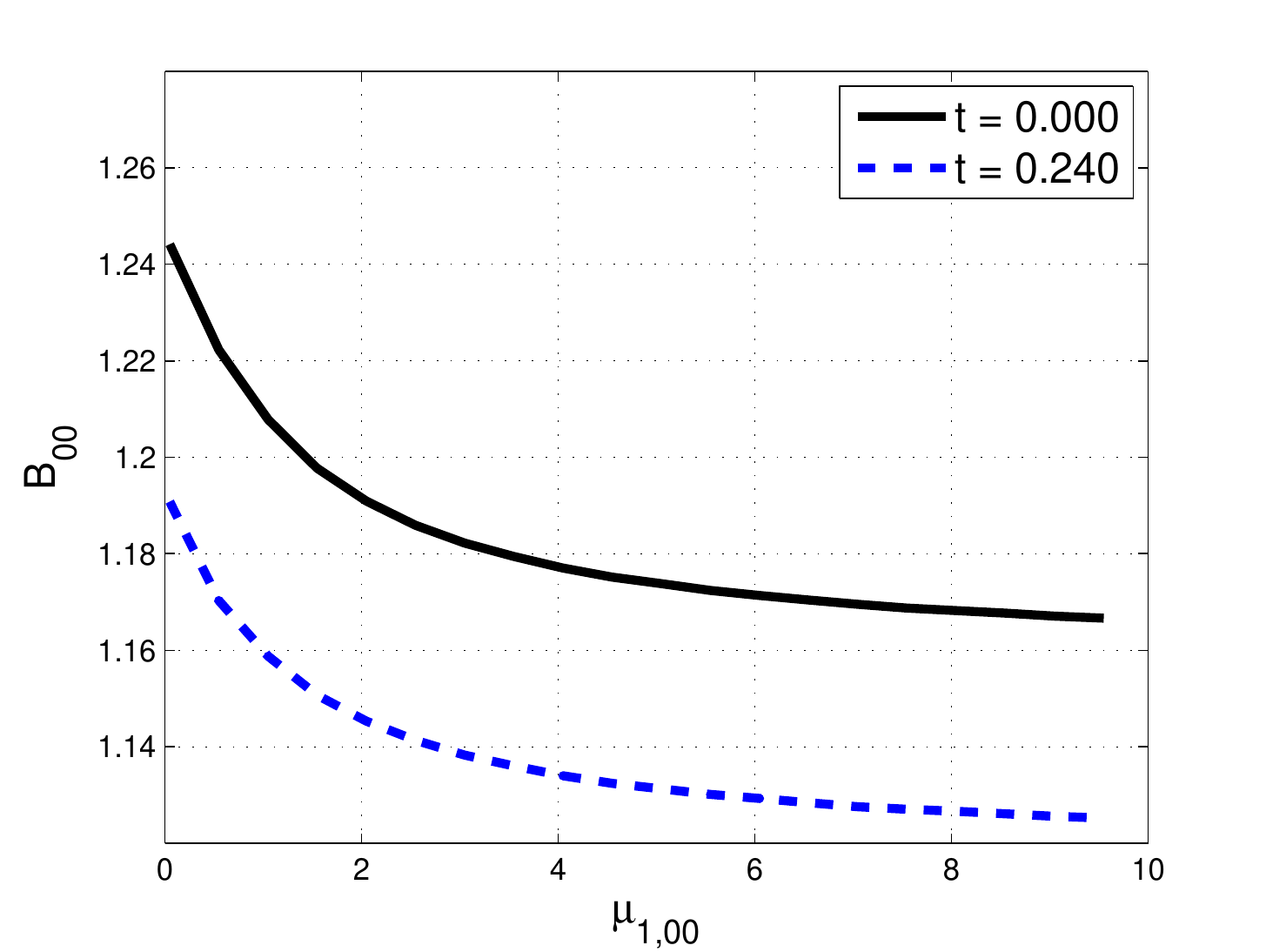}
\hspace{-.7cm} \includegraphics[width=7.2cm]{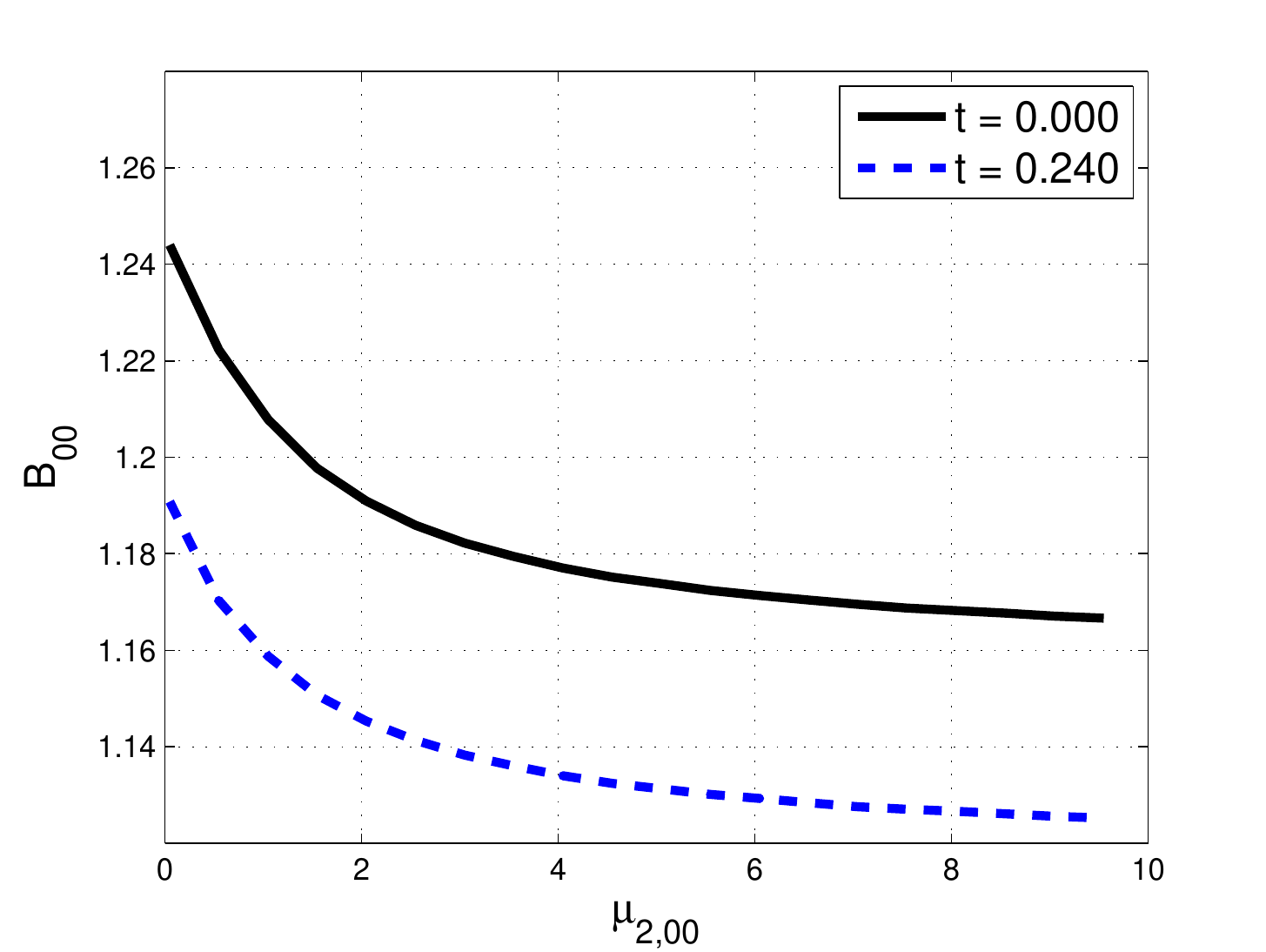}
\end{center}
\caption{
The left panel reports the dependence of the time component of the robust value function on the penalty parameter $\mu_{1,00}$. The right panel reports the dependence on the penalty
parameter $\mu_{2,00}$.}
\label{robustvfn}
\end{figure}

Figure \ref{robustvfn} shows that model uncertainty reduces the utility achievable by the investor. This finding is consistent with \cite{Glasserman}, who also find that the
robust value function is bounded above by the nonrobust value function (corresponding to the parameter setting $\mu_{1,00}=0$ and $\mu_{2,00}=0$ in our case), see section 5.1 therein.
Together with figure~\ref{figstrat}, this indicates that by allocating a smaller fraction of wealth to the risky bond, the robust investor incurs a loss of utility.  He would have
achieved higher returns from a larger position in the risky bonds, had he been very confident on the reference model of default intensities.
As expected, when the planning horizon is higher, the investor achieves higher utility because he has more investment opportunities at his disposal.
Notice that the dependence of the expected utility on the robustness parameters $\mu_{1,00}$ and $\mu_{2,00}$ is the same. This is expected given that the default characteristics of the
two names as well as the contractual parameters of the bonds underwritten by them are the same in our numerical setup.

\section{Conclusion} \label{sec:concl}
It is well understood that historical estimation of default risk is challenging and often prone to estimation errors. This is because the available
dataset is limited due to the rarity of default events. Nevertheless, current literature on optimal credit portfolios has so far assumed the reference credit
model to be known with certainty. Since optimal strategies crucially depend on the ratio of risk-neutral and historical default intensities (the so-called default risk premium)
as well as on contagion effects, it is important to account for misspecifications of the credit model when designing optimal decision rules.

This paper has introduced a novel dynamic framework, where an investor can choose optimal investment strategies while protecting himself against misspecification of the reference credit model. We have obtained an explicit characterization of the optimal feedback function yielding the strategy in risky bonds.
The latter has been shown to be coupled with the value function of the robust control problem, which we have shown to correspond with the unique classical solution of the corresponding HJB equation.

The introduced framework is rich enough to accommodate several features of default risk, but at the same time tractable since both optimal feedback functions and value
functions can be recovered, respectively, as a matrix-vector product, and as the solution of an ordinary differential equation.

%
%
%
\appendix

%

\section{Proofs of Section \ref{sec:CDSdyn}} \label{sec:appHJB}
We give the following proofs:

We first have the following lemma on the uniqueness of the risk-neutral probability measure.
\begin{lemma}\label{lem:uniqe-Q}
Let the matrix ${\bm \Phi}_{\bm z}(t)=[(1-z_j)G_{i,j,{\bm z}}(t)]_{i,j=1,\ldots,M}$ with $(t,{\bm z})\in[0,T]\times{\cal S}$. Then the following are equivalent: {\sf(I)} The square matrix ${\bm \Phi}_{\bm Z(t)}(t)$ is invertible a.s.; {\sf(II)} The risk-neutral default intensity $(h_{j,{\bm Z}(t)}(t))_{j=1,\ldots,M}$ is unique.
\end{lemma}

\noindent{\it Proof.}\quad Let $\hat{\Qx}\sim\Px$ be a risk-neutral probability measure with the corresponding risk-neutral default intensity $(\hat{h}_{i,{\bm Z}(t)}(t))_{i=1,\ldots,M}$ with $t\in[0,T]$. Then under $\hat{\Qx}$, $\hat{\xi}_j(t):=Z_j(t)-\int_0^{t\wedge\tau_j}\hat{h}_{j,{\bm Z}(u)}(u)\D u$ is a martingale for each $j=1,\ldots,M$. Using \eqref{eq:P+D}, under $\hat{\Qx}$, we then have
\begin{eqnarray*}\label{eq:P+D2}
\frac{\D\big(P_i(t)+D_i(t)\big)}{P_i(t-)} = r \D t+\sum_{j=1}^M G_{i,j,{\bm Z}(t)}(t)(1-Z_j(t))\big(\hat{h}_{j,{\bm Z}(t)}(t)-h_{j,{\bm Z}(t)}(t)\big)\D t
+\sum_{j=1}^M G_{i,j,{\bm Z}(t-)}(t)\D\hat{\xi}_j(t).
\end{eqnarray*}
Hence the discounted prices are (local) $\hat{\Qx}$-martingales if and only if $\sum_{j=1}^M G_{i,j,{\bm Z}(t)}(t)(1-Z_j(t))\big(\hat{h}_{j,{\bm Z}(t)}(t)-h_{j,{\bm Z}(t)}(t)\big)=0$, a.s. for all $i=1,\ldots,M$. The system of linear equations admits a unique solution $\hat{h}_{j,{\bm Z}(t)}(t)-h_{j,{\bm Z}(t)}(t)=0$ with $j=1,\ldots,M$ if and only if the matrix ${\bm \Phi}_{\bm Z(t)}(t)$ has a full rank a.s.. \hfill$\Box$

We next give the lemma on the risk-neutral dynamics of the $i$-th risky bond price process, which is a key result to prove Lemma~\ref{lem:P+D}.
\begin{lemma}\label{thm:mul-cds-price}
The risk-neutral dynamics of the $i$-th risky bond price process is given by
\begin{eqnarray}\label{eq:mul-cds-dynamics-p0}
\D P_i(t) &=& \left[ rP_i(t) -(1-Z_i(t))\big(C_i + R_i h_{i,{\bm Z}(t)}(t)\big) \right] \D t-P_i(t-)\D \xi_i(t)\nonumber\\
&&+P_i(t-)\sum_{j\neq i}G_{i,j,{\bm Z}(t-)}(t)\D \xi_j(t),\ \ \ \ \ {t\in[0,T]},
\end{eqnarray}
where, for $i,j=1,\ldots,M$, the functions $G_{i,j,{\bm z}}(t)$, $(t,{\bm z})\in[0,T]\times{\cal S}$, are given by \eqref{eq:Phi}.
\end{lemma}

In order to prove Lemma~\ref{thm:mul-cds-price}, we need the following auxiliary lemma:
\begin{lemma}\label{lem:mul-FK-Eqns}
The pre-default price $F_{i,{\bm Z}(t)}(t)$, $t\in[0,T_i]$, of the $i$-th risky bond admits the decomposition:
\begin{eqnarray}\label{eq:mul-Dykin}
F_{i,{\bm Z}(t)}(t)&=&F_{i,{\bm Z}(0)}(0)+\sum_{j=1}^M\int_0^t\left[F_{i,{\bm Z}^j(u-)}(u)-F_{i,{\bm Z}(u-)}(u)\right]\D \xi_j(u)\\
&&+\int_0^t\Big[r(1-Z_i(u))F_{i,{\bm Z}(u)}(u) + rZ_i(u) \big(C_iF_{i,{\bm Z}(u)}^{b}(u)
+F_{i,{\bm Z}(u)}^{c}(u)\big)-C_i(1-Z_i(u))\Big]\D u,\nonumber
\end{eqnarray}
where $F_{i,{\bm z}}^{b}(t)$ and $F_{i,{\bm z}}^{c}(t)$ are defined in Eq.~\eqref{eq:Ffun}.
\end{lemma}

\noindent{\it Proof.}\quad
{Define the operator ${\A} g_{{\bm z}}(t)= \sum_{j=1}^M (1-z_j)h_{j,{\bm z}}(t)[g_{{\bm z}^{j}}(t)-g_{{\bm z}}(t)]$ acting on
 an arbitrary measurable function $g_{\bm z}(\cdot)$ with ${\bm z}\in{\cal S}$.}
Using Feynman-Kac's formula, we obtain that $F_{i,{\bm z}}^{a}(t)$, $F_{i,{\bm z}}^{b}(t)$ and $F_{i,{\bm z}}^{c}(t)$ satisfy
\begin{eqnarray}\label{eq:mul-FK12}
&&\left(\frac{\partial}{\partial t} + \A\right)F_{i,{\bm z}}^{a}(t) = r (1-z_i) F_{i,{\bm z}}^{a}(t),\ \ \ \ \
F_{i,{\bm z}}^{a}(T_i) = z_i,\nonumber\\
&&\left(\frac{\partial}{\partial t} + \A\right)F_{i,{\bm z}}^{b}(t) + (1-z_i) = r F_{i,{\bm z}}^{b}(t),\ \ \ \ \
F_{i,{\bm z}}^{b}(T_i) = 0,\\
&&\left(\frac{\partial}{\partial t} + \A\right)F_{i,{\bm z}}^{c}(t)= r F_{i,{\bm z}}^{c}(t),\ \ \ \ \
F_{i,{\bm z}}^{c}(T_i) = 1-z_i.\nonumber
\end{eqnarray}
Then the $i$-th pre-default price function $F_{i,{\bm z}}(t)$ given by \eqref{eq:pre-default-cds-price} satisfies
\begin{eqnarray}\label{eq:PDE-F}
\left(\frac{\partial}{\partial t} +\A\right)F_{i,{\bm z}}(t)
&=&R_i\left(\frac{\partial}{\partial t} + \A\right)F_{i,{\bm z}}^a(t)+C_i\left(\frac{\partial}{\partial t} + \A\right)F^b_{i,{\bm z}}(t)+\left(\frac{\partial}{\partial t} + \A\right)F^c_{i,{\bm z}}(t)\nonumber\\
&=&rR_i(1-z_i)F_{i,{\bm z}}^a(t) + rC_i F_{i,{\bm z}}^b(t) -C_i(1-z_i)+ r F_{i,{\bm z}}^{c}(t)\nonumber\\
&=&r(1-z_i)F_{i,{\bm z}}(t) + rz_i \left(C_iF_{i,{\bm z}}^{b}(t)+F_{i,{\bm z}}^{c}(t)\right)
-C_i(1-z_i).
\end{eqnarray}
Using It\^o's formula, we have
\begin{eqnarray*}
F_{i,{\bm Z}(t)}(t)&=&F_{i,{\bm Z}(0)}(0)
+ \int_0^t\left(\frac{\partial}{\partial{u}}+\A\right)F_{i,{\bm Z}(u)}(u)\D u+\sum_{j=1}^N\int_0^t\left[F_{i,{\bm Z}^j(u-)}(u)-F_{i,{\bm Z}(u-)}(u)\right]\D \xi_j(u)\nonumber\\
&=&F_{i,{\bm Z}(0)}(0)+\sum_{j=1}^M\int_0^t\left[F_{i,{\bm Z}^j(u-)}(u)-F_{i,{\bm Z}(u-)}(u)\right]\D  \xi_j(u)\nonumber\\
&&+\int_0^t\Big[r(1-Z_i(u))F_{i,{\bm Z}(u)}(u) + rZ_i(u) \big(C_iF_{i,{\bm Z}(u)}^{b}(u)
+F_{i,{\bm Z}(u)}^{c}(u)\big)-C_i(1-Z_i(u))\Big]\D u
\end{eqnarray*}
which corresponds to the equality in \eqref{eq:mul-Dykin}. \hfill$\Box$


\noindent{\it Proof of Lemma~\ref{thm:mul-cds-price}.}\quad Using~\eqref{eq:cds-price} and It\^o's formula, it follows that
\begin{eqnarray}\label{eq:mul-Ito-cds2}
\D P_i(t) &=& (1-Z_i(t-)) \D F_{i,{\bm Z}(t)}(t) -F_{i,{\bm Z}(t-)}(t) \D Z_i(t) +
\Delta(1-Z_i(t))\Delta F_{i,{\bm Z}(t)}(t)\nonumber\\
&=&(1-Z_i(t-)) \D F_{i,{\bm Z}(t)}(t) -F_{i,{\bm Z}(t-)}(t) \D Z_i(t) -\Delta F_{i,{\bm Z}(t)}(t)\D Z_i(t)\nonumber\\
&=&(1-Z_i(t-)) \D F_{i,{\bm Z}(t)}(t) -F_{i,{\bm Z}(t-)}(t) \D Z_i(t)-\left[F_{i,{\bm Z}^i(t-)}(t)-F_{i,{\bm Z}(t-)}(t)\right]\D Z_i(t),
\end{eqnarray}
where we used the equality $\Delta F_{i,{\bm Z}(t)}(t)\D Z_i(t)=\left[F_{i,{\bm Z}^i(t-)}(t)-F_{i,{\bm Z}(t-)}(t)\right]\D Z_i(t)$
which follows from the fact that our default model excludes the occurrence of simultaneous defaults. Using Eq.~\eqref{eq:mul-Dykin} in Lemma \ref{lem:mul-FK-Eqns}, we obtain
\begin{eqnarray*}
&& (1-Z_i(t-)) \D F_{i,{\bm Z}(t)}(t)\nonumber\\
&&\qquad  =(1-Z_i(t))\Big[r(1-Z_i(t))F_{i,{\bm Z}(t)}(t) + rZ_i(t) \big(C_iF_{i,{\bm Z}(t)}^{b}(t)
+F_{i,{\bm Z}(t)}^{c}(t)\big)-C_i(1-Z_i(t))\Big]\nonumber\\
&&\qquad\quad+(1-Z_i(t-))\sum_{j=1}^M\left[F_{i,{\bm Z}^j(t-)}(t)-F_{i,{\bm Z}(t-)}(t)\right]\D \xi_j(t)\nonumber\\
&&\qquad  =r(1-Z_i(t))F_{i,{\bm Z}(t)}(t) -C_i(1-Z_i(t))+(1-Z_i(t-))\sum_{j=1}^M\left[F_{i,{\bm Z}^j(t-)}(t)-F_{i,{\bm Z}(t-)}(t)\right]\D \xi_j(t).
\end{eqnarray*}
It follows from \eqref{eq:mul-Ito-cds2} that
\begin{eqnarray}\label{eq:mul-sde-cds2}
\D P_i(t) &=&(1-Z_i(t))\big[rF_{i,{\bm Z}(t)}(t)-C_i\big]\D t - F_{i,{\bm Z}(t-)}(t)\D Z_i(t)\\
&&+(1-Z_i(t-))\sum_{j=1}^M\left[F_{i,{\bm Z}^j(t-)}(t)-F_{i,{\bm Z}(t-)}(t)\right]\D \xi_j(t)-\left[F_{i,{\bm Z}^i(t-)}(t)-F_{i,{\bm Z}(t-)}(t)\right]\D Z_i(t).\nonumber
\end{eqnarray}
From Proposition~ \ref{lem:solution-Phi}-{\sf(I)} below, we know that $F_{i,{\bm z}^i}(t)=R_i$ for all $(t,{\bm z})\in[0,T_i]\times{\cal S}$. Using this
along with the fact that $P_i(t)=(1-Z_i(t))F_{i,{\bm Z}(t)}(t)$ and $\xi_j(t)=Z_j(t)-\int_0^t(1-Z_j(s))h_{j,{\bm Z}(s)}(s)\D s$, we obtain the desired result. \hfill$\Box$

\section{Explicit Recursive Representation of Price Functions }\label{appen:express-pre-fcn}

Recalling the definition of ${\bm z}^j$ given in~\eqref{eq:zjdef}, we will write
${\bm z}^{j}={\bm0}^{j_1,\ldots,j_m,j}$ if ${\bm z}={\bm0}^{j_1,\ldots,j_m}$, and $j\notin\{j_1,\ldots,j_m\}$. Then
\begin{proposition}\label{lem:solution-Phi}
Let $i=1,\ldots,M$. Then $ F_{i,{\bm z}}(T_i)=R_iz_i+1-z_i$ for all ${\bm z}\in{\cal S}$, and on $(t,{\bm z})\in[0,T_i)\times{\cal S}$, it holds that
\begin{itemize}
  \item[{\sf(I)}] If  $m=M$, or there exists an integer $l=1,\ldots,m$ so that $j_l=i$ for $m=1,\ldots,M-1$, then
  \[
   F_{i,j_1,\ldots,j_m}(t):=F_{i,{\bm0}^{j_1,\ldots,j_m}}(t)=R_i.
  \]
  \item[{\sf(II)}] If $m=M-1$, then for $i=j_M$,
  \begin{eqnarray}\label{eq:sol-PDE-F1}
F_{i,j_1,\ldots,j_{M-1}}(t)
&=&\exp\left\{-\int_t^{T_i}\Big(r+h_{i,j_1,\ldots,j_{M-1}}(s)\Big)\D s\right\}\\
&&+\int_t^{T_i}\Big(C_i+R_ih_{i,j_1,\ldots,j_{M-1}}(u)\Big)\exp\left\{-\int_t^{u}\Big(r+h_{i,j_1,\ldots,j_{M-1}}(s)\Big)\D s\right\}\D u.\nonumber
\end{eqnarray}
  \item[{\sf(III)}] If $i\notin\{j_1,\ldots,j_m\}$ with $m=0,1,\ldots,M-2$, then
  \begin{eqnarray}\label{eq:solution-Phi}
&& F_{i,j_1,\ldots,j_m}(t) = \exp\left\{-\int_t^{T_i}\bigg(r+\sum_{j\notin\{j_1,\ldots,j_m\}}h_{j,j_1,\ldots,j_m}(s)\bigg)\D s\right\}\nonumber\\
&&\qquad+\int_t^{T_i}\Big(C_i+R_ih_{i,j_1,\ldots,j_m}(u)\Big)\exp\left\{-\int_t^{u}\bigg(r+\sum_{j\notin\{j_1,\ldots,j_m\}}h_{j,j_1,\ldots,j_m}(s)\bigg)\D s\right\}\D u\\
&&\qquad+\int_t^{T_i}\sum_{j\notin\{j_1,\ldots,j_m,i\}}h_{j,j_1,\ldots,j_m}(u) F_{i,j_1,\ldots,j_{m},j}(u)\exp\left\{-\int_t^{u}\bigg(r+\sum_{j\notin\{j_1,\ldots,j_m\}}h_{j,j_1,\ldots,j_m}(s)\bigg)\D s\right\}\D u.\nonumber
  \end{eqnarray}
\end{itemize}
\end{proposition}

\noindent{\it Proof.}\quad
Using \eqref{eq:PDE-F}, $F_{i,{\bm z}}(t)$, $(t,{\bm z})\in[0,T_i)\times{\cal S}$, admits
\begin{eqnarray}\label{eq:PDE-F0}
\left(\frac{\partial}{\partial t} +\A\right)F_{i,{\bm z}}(t)
&=&r(1-z_i)F_{i,{\bm z}}(t) + rz_i \left(C_iF_{i,{\bm z}}^{b}(t)+F_{i,{\bm z}}^{c}(t)\right)
-C_i(1-z_i),
\end{eqnarray}
and $F_{i,{\bm z}}(T_i)=R_iz_i+1-z_i$ for all ${\bm z}\in{\cal S}$. Hence Eq.~\eqref{eq:PDE-F0} can be rewritten as
\begin{eqnarray}\label{eq:PDE-F1}
\frac{\partial}{\partial t}F_{i,{\bm z}}(t)
&=&r(1-z_i)F_{i,{\bm z}}(t) + rz_i \left(C_iF_{i,{\bm z}}^{b}(t)+F_{i,{\bm z}}^{c}(t)\right)
-C_i(1-z_i)\nonumber\\
&&-\sum_{j=1}^M \big[F_{i,{\bm z}^j}(t) - F_{i,{\bm z}}(t)\big](1-z_j)h_{j,{\bm z}}(t),
\end{eqnarray}
and $F_{i,{\bm z}}(T_i)=R_iz_i+1-z_i$ for all ${\bm z}\in{\cal S}$.

The conclusion {\sf(I)} can be followed from the definition \eqref{eq:pre-default-cds-price} of the pre-default function directly. Next, we consider {\sf(II)}. In this case, the only $j_M$-name is alive. In terms of \eqref{eq:PDE-F1}, we have that for $i\notin\{j_1,\ldots,j_{M-1}\}$ (and hence $z_{i}=z_{j_M}=0$), $F_{i,j_1,\ldots,j_{M-1}}(t):=F_{i,{\bm0}^{j_1,\ldots,j_{M-1}}}(t)$, $t\in[0,T_i)$, satisfies
\begin{eqnarray}\label{eq:PDE-FN-1}
\frac{\partial}{\partial t} F_{i,j_1,\ldots,j_{M-1}}(t)
&=&rF_{i,j_1,\ldots,j_{M-1}}(t) -C_i-\big[F_{i,{\bm1}}(t) - F_{i,j_1,\ldots,j_{M-1}}(t)\big]h_{i,j_1,\ldots,j_{M-1}}(t)\nonumber\\
&=& \big(r+h_{i,j_1,\ldots,j_{M-1}}(t)\big)F_{i,j_1,\ldots,j_{M-1}}(t)-\big(C_i+R_ih_{i,j_1,\ldots,j_{M-1}}(t)\big),
\end{eqnarray}
and $F_{i,j_1,\ldots,j_{M-1}}(T_i)=1$. Here we used {\sf(I)}, i.e., $F_{i,{\bm1}}(t)=R_i$ for all $t\in[0,T_i]$. Then the solution to Eq.~\eqref{eq:PDE-F1}~admits \eqref{eq:sol-PDE-F1} for $i=j_M$.

Finally, we consider the proof of {\sf(III)}. In this case, the all $j\notin\{j_1,\ldots,j_m\}$ names are alive, i.e., $z_{j}=0$ for all $j\notin\{j_1,\ldots,j_m\}$. We assume that for all $j\notin\{j_1,\ldots,j_m\}$,  $F_{i,j_1,\ldots,j_{m},j}(t)$ satisfies Eq.~\eqref{eq:PDE-F1} at the default state ${\bm z}={\bm0}^{j_1,\ldots,j_m,j}$. From Eq.~\eqref{eq:PDE-F1}, it follows that for $i\notin\{j_1,\ldots,j_m\}$, $F_{i,j_1,\ldots,j_{m}}(t):=F_{i,{\bm0}^{j_1,\ldots,j_{m}}}(t)$, $t\in[0,T_i)$, satisfies
\begin{eqnarray}\label{eq:PDE-Fm1}
\frac{\partial}{\partial t}F_{i,j_1,\ldots,j_m}(t)
&=&\left(r+\sum_{j\notin\{j_1,\ldots,j_m\}}h_{j,j_1,\ldots,j_m}(t)\right)F_{i,j_1,\ldots,j_m}(t)-\big(C_i+R_ih_{i,j_1,\ldots,j_m}(t)\big)\nonumber\\
&&-\sum_{j\notin\{j_1,\ldots,j_m,i\}}h_{j,j_1,\ldots,j_m}(t) F_{i,j_1,\ldots,j_{m},j}(t)
\end{eqnarray}
with $F_{i,j_1,\ldots,j_{m}}(T_i)=1$, where we used $F_{i,j_1,\ldots,j_m,i}(t)=R_i$ obtained in {\sf(I)}. Then the closed-form solution to Eq.~\eqref{eq:PDE-Fm1} is given by \eqref{eq:solution-Phi}.
\hfill$\Box$

We next provide a lower bound for the price function $F_{i,j_1,\ldots,j_m}(t)$ assuming $C_i\geq rR_i$ for $i=1,\ldots,M$. This condition is in line with empirical evidence. Bond coupon rates are typically set at the prevailing market rate (proxied by $r$ in our case) when issued. Since $R_i < 1$, this assumption is clearly satisfied. In particular, if $R_i=0$, i.e. there is zero recovery rate on default of obligor $i$, the assumption is trivially satisfied given that the coupon rate $C_i\geq0$.
\begin{lemma}\label{lem:upperb}
Let $i=1,\ldots,M$, and $j_1,\ldots,j_m \in \{1,\ldots,M\} \setminus \{i\}$. Then $F_{i,j_1,\ldots,j_m}(t) > R_i$ for all $t\in[0,T_i)$ if $C_i\geq rR_i$.
\end{lemma}

\noindent{\it Proof.}\quad
From \eqref{eq:sol-PDE-F1}, and the fact $C_i\geq rR_i$ with $R_i\in[0,1)$, it follows that, for all $t\in[0,T_i)$,
\begin{eqnarray}\label{ineq:sol-PDE-F1}
F_{i,j_1,\ldots,j_{M-1}}(t)
&\geq&\exp\left\{-\int_t^{T_i}\Big(r+h_{i,j_1,\ldots,j_{M-1}}(s)\Big)\D s\right\}\nonumber\\
&&+R_i\int_t^{T_i}\Big(r+h_{i,j_1,\ldots,j_{M-1}}(u)\Big)\exp\left\{-\int_t^{u}\Big(r+h_{i,j_1,\ldots,j_{M-1}}(s)\Big)\D s\right\}\D u\nonumber\\
&=& R_i + (1-R_i)\exp\left\{-\int_t^{T_i}\Big(r+h_{i,j_1,\ldots,j_{M-1}}(s)\Big)\D s\right\}
>R_i.
\end{eqnarray}

Next, assume that, for $m=0,1,\ldots,M-2$, and $i\notin\{j_1,\ldots,j_m\}$, it holds that $F_{i,j_1,\ldots,j_{m},j}(t)> R_i$ for all $j\notin\{j_1,\ldots,j_m\}$. We want to prove $F_{i,j_1,\ldots,j_{m}}(t)> R_i$. Using \eqref{eq:solution-Phi}, we obtain
\begin{eqnarray}\label{ineq:solution-Phi}
&& F_{i,j_1,\ldots,j_m}(t) \geq \exp\left\{-\int_t^{T_i}\bigg(r+\sum_{j\notin\{j_1,\ldots,j_m\}}h_{j,j_1,\ldots,j_m}(s)\bigg)\D s\right\}\nonumber\\
&&\qquad+R_i\int_t^{T_i}\Big(r+h_{i,j_1,\ldots,j_m}(u)\Big)\exp\left\{-\int_t^{u}\bigg(r+\sum_{j\notin\{j_1,\ldots,j_m\}}h_{j,j_1,\ldots,j_m}(s)\bigg)\D s\right\}\D u\nonumber\\
&&\qquad+\int_t^{T_i}R_i\sum_{j\notin\{j_1,\ldots,j_m,i\}}h_{j,j_1,\ldots,j_m}(u)\exp\left\{-\int_t^{u}\bigg(r+\sum_{j\notin\{j_1,\ldots,j_m\}}
h_{j,j_1,\ldots,j_m}(s)\bigg)\D s\right\}\D u\nonumber\\
&&\qquad\qquad= R_i + (1-R_i)\exp\left\{-\int_t^{T_i}\bigg(r+\sum_{j\notin\{j_1,\ldots,j_m\}}h_{j,j_1,\ldots,j_m}(s)\bigg)\D s\right\}> R_i,
\end{eqnarray}
using the fact $R_i\in[0,1)$ again. Thus we prove recursively that, for $i\notin\{j_1,\ldots,j_m\}$, the pre-default price function $F_{i,j_1,\ldots,j_m}>R_i$ for all $t\in[0,T_i)$. \hfill$\Box$\\

\end{document}